\documentclass[12pt]{article}
\usepackage{amssymb,epsfig,times,graphicx}
\usepackage{palatino}
\usepackage{amsmath}
\usepackage{amsthm}
\usepackage{graphicx}
\usepackage{tikz}
\usepackage[all]{xy}
\numberwithin{equation}{section}
\newtheorem{thm}{Theorem}[section]     

\newtheorem{prop}[thm]{Proposition}

\newtheorem{defi}[thm]{Definition}

\newtheorem{lemma}[thm]{Lemma}

\newtheorem{cor}[thm]{Corollary}

\newtheorem{rmk}[thm]{Remark}

\newcommand{\beq}{\begin{equation}}
\newcommand{\eeq}{\end{equation}}

\def\d{\partial}
\def\n{\noindent}
\def\f{\frac}

\begin{document}
\title{$F$-manifolds, multi-flat structures and Painlev\'e transcendents}
\author{Alessandro Arsie* and Paolo Lorenzoni**\\
\\
{\small *Department of Mathematics and Statistics}\\
{\small The University of Toledo,}
{\small 2801 W. Bancroft St., 43606 Toledo, OH, USA}\\
{\small **Dipartimento di Matematica e Applicazioni}\\
{\small Universit\`a di Milano-Bicocca,}
{\small Via Roberto Cozzi 55, I-20125 Milano, Italy}\\
{\small *alessandro.arsie@utoledo.edu,  **paolo.lorenzoni@unimib.it}}
\date{}
\maketitle

\begin{abstract}
In this paper we study $F$-manifolds equipped with multiple flat connections (and multiple $F$-products), that are required to be compatible in a suitable sense. Multi-flat $F$-manifolds are the analogue for $F$-manifolds of Frobenius manifolds with multi-Hamiltonian structures. 

In the semisimple case we show that a necessary condition for the existence of such multiple flat connections can be expressed in terms of the integrability (in the sense of the Frobenius Theorem) of a distribution of vector fields that are related to the eventual identities for the multiple products involved. These vector fields moreover satisfy the commutation relations of the centerless Virasoro algebra. We prove that the distributions associated to bi-flat and tri-flat $F$-manifolds are integrable, while in other cases they are maximally non-integrable. Using this fact we show that in general there can not be semisimple multi-flat structures with more than three flat connections.

When the relevant distributions are integrable, coupling the invariants of the foliations they determine with Tsarev's conditions, we construct bi-flat  $F$-manifolds in dimension $2$ and $3$, and tri-flat $F$-manifolds in dimensions $3$ and $4$. In particular  we
 obtain a parametrization of three-dimensional bi-flat $F$ in terms of a system of six first order ODEs that can be reduced to the full
 family of  P$_{VI}$ equation and we construct non-trivial examples of four dimensional tri-flat $F$ manifolds that are controlled by hypergeometric functions. 

In the second part of the paper we extend our analysis to include non-semisimple regular bi-flat and in general multi-flat $F$-manifolds. We show that in dimension three, regular non-semisimple bi-flat $F$-manifolds are locally parameterized by solutions of the full P$_{IV}$ and P$_{V}$ equations, according to the Jordan normal form of the endomorphism $L=E\circ$. As a consequence, combining this result with the local parametrization of $3$-dimensional semisimple bi-flat $F$-manifolds we have that confluences of  P$_{IV}$,  P$_{V}$  and P$_{VI}$  correspond to collisions of eigenvalues of $L$ preserving the regularity. Furthermore, we show that contrary to the semisimple situation, it is possible to construct regular non-semisimple multi-flat $F$-manifolds, with any number of compatible flat connections. This provides the first example of an $F$-manifold equipped with an infinite collection of non-trivial compatible flat structures.

\end{abstract}

\tableofcontents
\section{Introduction}
$F$-manifolds have been introduced in  \cite{HM} as a unifying geometric scheme that encompasses several areas of modern Mathematics, ranging from the theory of Frobenius manifolds  to special solutions of the oriented associativity equations (\cite{LoMa2}), from quantum $K$-theory (\cite{lee}) to differential-graded deformation theory (\cite {mer}).

An $F$-manifold $M$ is a smooth (or analytic) manifold equipped with a commutative and associative product $\circ : TM \times TM \rightarrow TM$ on sections of the tangent bundle $TM$, such that $\circ$ is $C(M)$-bilinear ($C(M)$ is the ring of smooth or analytic functions on $M$) and such that 
\beq\label{HM}P_{X\circ Y}(Z, W)=X\circ P_{Y}(Z,W)+Y\circ P_X(Z, W),\eeq
where $P_X(Z,W):=[X, Z\circ W]-[X,Z]\circ W-Z\circ[X, W].$
The condition \eqref{HM} is usually called the Hertling-Manin condition and it implies that the deviation of the structure $(TM, \circ, [\cdot, \cdot])$ from that of a Poisson algebra on $(TM, \circ)$ is not arbitrary. Usually $M$ is also required to be equipped with a distinguished vector  field $e$, called unity or identity, such that for every vector field $X$, $X\circ e=X$. 

Since the operation $\circ$ is $C(M)$-bilinear and commutative, it can be identified with a tensor field $c: S^2(TM)\rightarrow TM.$  Once $c$ is locally written in a coordinate system as $c^i_{jk}:=<c(\d_j, \d_k),dx^i>$, then the commutativity, the associativity and the Hertling-Manin condition \eqref{HM} translate respectively as
 \begin{eqnarray*}
&c^i_{jk}=c^i_{kj},\\
&c^i_{jl}c^l_{km}=c^i_{kl}c^l_{jm},\\
&c^s_{im}\d_s c^k_{jl}+c^k_{sl}\d_j c^s_{im}-c^s_{jl}\d_s c^k_{im}-c^k_{sm}\d_i c^s_{jl}-c^k_{si}\d_l c^s_{jm}-c^k_{js}\d_m c^s_{li}=0.
 \end{eqnarray*}
 
An $F$-manifold $(M, \circ, e)$ is called \emph{semisimple} if locally $(TM, \circ)$ is isomorphic to $C(M)^n$ (where $n$ is the dimension of the manifold $M$) with componentwise multiplication. This means that locally there exists a distinguished coordinate system such that, if $X$ and $Y$ are vector fields given in components as $X=(X^1, \dots, X^n),$ $Y=(Y^1, \dots, Y^n)$, then $(X\circ Y)^i=X^i Y^i$. This is equivalent to say that $c^i_{jk}=\delta^i_j \delta^i_k$ in this distinguished coordinate system (these are called {\em canonical coordinates} for $\circ$ whenever they exist). We will denote canonical coordinates with $\{u^1, \dots, u^n\}$. If $\circ$ is semisimple, then the identity vector field $e$ is given by $e=\sum_{i}\frac{\partial}{\partial u^i}$ in canonical coordinates for $\circ.$

Few years later, Manin introduced $F$-manifolds with compatible flat structure (\cite{manin}), which we call flat $F$-manifold for simplicity. In particular, he proved that many constructions related to Frobenius manifolds, such as Dubrovin's duality, do not require the presence of a (pseudo)-metric satisfying the condition $g(X\circ Y, Z)=g(X, Y\circ Z)$  for all vector fields $ X,Y,Z$ (such metrics 
 are said to be \emph{invariant}).  

\begin{defi}[\cite{manin}]\label{def111}A flat $F$-manifold $(M, \circ, \nabla, e)$ with identity is a manifold $M$ equipped with the following data: 
\begin{enumerate}
\item a commutative associate product $\circ : TM \times TM \rightarrow TM$ on sections of the tangent bundle $TM$, 
\item a distinguished vector field $e$ such that $X\circ e=X$ for every vector field $X$, 
\item a flat torsionless affine connection $\nabla,$
such that $\left(\nabla_X c\right)\left(Y,Z\right)=\left(\nabla_Y c\right)\left(X,Z\right)$ 
for all vector fields $X$, $Y$, and $Z$.
\item $\nabla e=0$ (flat identity).
\end{enumerate}
A semisimple flat $F$-manifold is defined analogously, with the requirement that the operation $\circ$ is semisimple. 
\end{defi}

Observe that in the Definition \ref{def111} there is no mention of the Hertling-Manin condition \eqref{HM} since the symmetry condition on $\nabla c$ forces \eqref{HM} to be automatically satisfied (see \cite{hert} for a proof). In any coordinate system, this condition reads $\nabla_i c^k_{lj}=\nabla_l c^k_{ij}.$ Let us mention also that the condition $\nabla e=0$ is the least important, and in many cases it is possible to modify the connection $\nabla$, preserving the other properties and in such a way to fulfill the condition $\nabla e=0$ even when it does not hold for the original connection (see for instance the example of $\vee$-systems below). 

The role played by flat $F$-manifolds in the study of integrable systems has been investigated in \cite{LPR,LP}. Further
  generalizations of these structures that suit very well the environment of integrable dispersionless PDEs  have been proposed in   \cite{ALimrn, AL, L2014}). In this paper,  following similar ideas, we introduce and study what we call {\em semisimple multi-flat $F$-manifolds}. They are a natural generalization of semisimple bi-flat $F$-manifolds (see \cite{AL, L2014}) and they are deeply related to the notion of eventual identities and duality introduced in \cite{manin}.

In order to define multi-flat $F$ manifolds we need to recall few facts about eventual identites:
\begin{defi}[\cite{manin}]
A vector field E on an $F$-manifold is called an \emph{eventual identity}, if it is invertible with respect to the product $\circ$, and if the bilinear product $*$ defined via
\beq\label{nm}
X *Y := X \circ Y \circ E^{-1},\qquad \text{ for all } X, Y \text{ vector fields}
\eeq
defines a new $F$-manifold structure on $M$. If $E$ satisfies the additional condition $[e,E]=e$ then it is called \emph{Euler vector field}.
\end{defi}
By definition, an eventual identity is the unity of the associated product $*$. A useful criterion to detect eventual identities is the following:
\begin{thm}\cite{DS}
An invertible vector field $E$ is an eventual identity for the $F$-manifold $(M, \circ, e)$ if and only if
\beq\label{DScond}
{\rm Lie}_E(\circ)(X,Y)=[e,E]\circ X\circ Y,\qquad\forall X,Y \text{ vector fields. }
\eeq
\end{thm}
In the semisimple case, it is actually easier to characterize eventual identities. We have indeed the following theorem.
\begin{thm}\label{theoremDSNijenhuis}\cite{ALimrn}
Let $(M, \circ, e)$ be a semisimple $F$-manifold and let $E$ be an invertible vector field and assume that the eigenvalues of the endomorphism of the tangent bundle $V=E\,\circ$ are distinct. Then condition \eqref{DScond}
is equivalent to the vanishing of the Nijenhuis torsion of $V$.
\end{thm}

In other words, in canonical coordinates for $\circ$ eventual identities are vector fields of the form 
$$E=\sum_{i=1}^n E^i(u^i)\f{\d}{\d u^i},$$ 
and the product $*$ has associated structure constants $c^{*i}_{jk}$ (again in canonical coordinates for $\circ$) given by :
\beq\label{dualp}
c^{*i}_{jk}=\frac{1}{E^i(u^i)}\delta^i_j\delta^i_k.
\eeq

We have now all the ingredients to define (semisimple) multi-flat $F$-manifolds. 
\begin{defi}\label{multiflatdefi}
Let $(M, \nabla, \circ, e)$ be a (semisimple) flat $F$-manifold with unity $e$. 
A \emph{multi-flat (semisimple)} $F$-manifold $(M,\nabla^{(l)},\circ, e, E,l=0...N-1)$ anchored at $(M, \nabla, \circ, e)$  is a manifold $M$ endowed with $N$ flat torsionless affine connections $\nabla^{(0)}:=\nabla,\, \nabla^{(1)},...,\nabla^{(N-1)}$, a commutative associative (semisimple) product $\circ$ on sections of the tangent bundle $TM$, an invertible vector field $E$ satisfying the following conditions:
\begin{enumerate}
\item $E$ is an Euler vector field (in the semisimple case we assume that the eigenvalues of $L:=E\circ$  are canonical coordinates for $\circ$).
%\item The endomorphism of the tangent bundle $L=E_{(1)}\circ$ has vanishing Nijenhuis torsion and its eigenvalues are canonical coordinates for $\circ$ (so that $E_{(1)}$ is an eventual identity) and $L^k=E_{(k)}\circ$. This means that in canonical coordinates for $\circ$, the products $\circ_{(l)}$, $l=1, \dots, N$ have associated tensor representatives $(c_{(l)})^i_{jk}=\frac{1}{E^i_{(l)}(u^i)}\delta^i_j\delta^i_k$ and $E^i_{(k)}(u^i)=(u^i)^k$.
\item Given $E_{(l)}:=E^{\circ l}=E\circ E\circ \dots \circ E$ $l$-times, $l=0, \dots, N-1$, (by definition, $E_{(0)}=e$, $E_{(1)}=E$), then we require $\nabla^{(l)}E_{(l)}=0.$
%$\nabla^{(l)} E_{(l)}=0$, where  $E_{(l)}:=E^{\circ l}=E\circ E\circ \dots \circ E$ $l$-times, $l=0, \dots, N-1$
 %(by definition, $E_{(0)}=e$, $E_{(1)}=E$).
\item Given $E_{(l)}$ and the related commutative, associative product $\circ_{(l)}$ (defined as $X\circ_{(l)}Y:=X\circ Y \circ E^{-1}_{(l)}, $ so that $\circ_{(0)}=\circ$ and $\circ_{(1)}=*$), we require that the connection $\nabla^{(l)}$ is compatible with $\circ_{(l)}$.  In other words we require that
\begin{equation}\label{scc}
\left(\nabla^{(l)}_X c_{(l)}\right)\left(Y,Z\right)=\left(\nabla^{(l)}_Y c_{(l)}\right)\left(X,Z\right),
\end{equation}
for all vector fields $X$, $Y$, and $Z$ for all $l=0, \dots N-1$.
  
\item The connections $\nabla^{(l)}$, $l=0, \dots, N-1$  are almost hydrodynamically equivalent (see \cite{AL}) i.e.
\beq\label{almostcomp}
(d_{\nabla^{(l)}}-d_{\nabla^{(l')}})(X\,\circ_{(l)})=0,
\eeq
for every vector fields $X$ and for every pair $l, l'$; 
 here $d_{\nabla^{(l)}}$ is the exterior covariant derivative constructed from the connection $\nabla^{(l)}$.  
\end{enumerate}
\end{defi}

\begin{rmk}
The last condition must be checked only for $l=0$. Indeed, due to the invertibility of the operator $E_{(l)}^{-1}\circ$  the condition \eqref{almostcomp} is equivalent to the condition
$$
(d_{\nabla^{(l)}}-d_{\nabla^{(l')}})(X\,\circ)=0,\qquad\forall X,
$$
which clearly follows from  the condition 
$$
(d_{\nabla}-d_{\nabla^{(l)}})(X\,\circ)=0,\qquad\forall X.
$$
\end{rmk}

\begin{rmk}
$M$ in Definition \ref{multiflatdefi} is  a real or complex $n$-manifold. In
the latter case $TM$ is intended as the holomorphic tangent bundle and
all the geometric data are supposed to be holomorphic. 
\end{rmk}
\begin{rmk}
The powers of the Euler vector fields are eventual identities. This follows from the fact that 
eventual identities form a subgroup of the group of invertible vector fields on an $F$-manifold \cite{DS}. The above definition can be easily generalized substituting the powers of the Euler vector field with general eventual identities.
\end{rmk}
\begin{rmk}
In the semisimple case the condition that $E_{(l)}:=E^{\circ l}$ implies that in canonical coordinates for $\circ$ the products $\circ_{(l)}$, $l=1, \dots N-1$ have associated tensor representatives $(c_{(l)})^i_{jk}=\frac{1}{E^i_{(l)}(u^i)}\delta^i_j\delta^i_k$ and $E^i_{(l)}(u^i)=(u^i)^l$. Furthermore the condition that the connections are almost hydrodynamically equivalent in canonical coordinates reduces to (see \cite{AL}):
\beq
\Gamma^{i}_{ij}=\Gamma^{(1)i}_{ij}=...=\Gamma^{(N-1)i}_{ij}.
\eeq
\end{rmk}

%\begin{rmk}
%Let us observe that in the very definition of multi-flat $F$ manifold it is built the fact that the 
%eventual identities $E_{(0)}=e, E_{(1)}=E,  \dots, E_{(N-1)}$ have the form $E^i_{(l)}=(u^i)^l$, $l=0, \dots, N-1$. This requirement is not too restrictive. For instance, it is easy to check that, under some genericity assumptions, all distributions of rank 2 generated by eventual identities are related to the standard one $\{E_{(0)}=e, E_{(1)}=E\}$ by a change of variable of the form $\tilde{u}^i=f^i(u^i)$. For distributions
% of rank 3 the analysis is much more involved. However considering only subalgebras  generated by the vector fields  $E^i_{(m)}=(u^i)^m$ (up to a shift in the index they
% coincide with the generators of the centerless Virasoro algebra) it is immediate to see that the generators of tridimensional subalgebras can be written as $ E^i_{(0)}=1, E^i_{(i)}=u^i, E^i_{(2)}=(u^i)^2$ in a suitable coordinate system (they are all isomorphic to $sl(2, \mathbb{C})$).
%\end{rmk}

In the first part of the paper we will study semisimple $F$-manifolds endowed with $N$ flat structures. In principle $N$ might be arbitrary, however we will see that the coexistence of more than $3$ flat structures is in general impossible. 
  The case of two structures has been studied in details in \cite{AL,L2014}. It turns out that tri-dimensional bi-flat $F$-manifolds are parametrized by solutions of Painlev\'e VI equation.
 In this paper we will find an alternative parametrization, in terms of the solutions of a system of $6$ ODEs admitting $5$ first integrals. 
 We will study in details also the case of tri-flat $F$-manifolds in the 3-component case. For more components, due to the appearance of some functional parameters the situation
 becomes more involved. We will find a class of solutions parametrized by hypergeometric functions.
% Moreover, we will show how in the study tri-flat $F$-manifolds the Lie algebra $sl(2, \mathbb{C})$ arises naturally

In the second part of the paper we will consider the non-semisimple case. First we will study regular non-semisimple bi-flat $F$-manifolds, leveraging on the recent results obtained in \cite{DH} unveiling a deep relation between regular bi-flat $F$-manifolds in dimension three on one side, and the full Painlev\'e equations P$_{VI}$, P$_{V}$ and P$_{IV}$ on the other. More precisely, regular bi-flat $F$-manifolds are characterized by the Jordan normal form of the operator $L=E\circ$. For three-dimensional manifolds, this gives rise to three cases, corresponding to $L_1, L_2$ and $L_3$ given by:
$$L_1:=\left( \begin{array}{ccc}
\lambda_1 & 0 & 0\\
0    & \lambda_2 & 0\\
0 & 0 & \lambda_3
\end{array}\right), \quad \quad L_2:=\left( \begin{array}{ccc}
\lambda_1 & 1 & 0\\
0    & \lambda_1 & 0\\
0 & 0 & \lambda_3
\end{array}\right), \quad \quad L_3:=\left( \begin{array}{ccc}
\lambda_1 & 1 & 0\\
0    & \lambda_1 & 1\\
0 & 0 & \lambda_1
\end{array}\right),$$
(here $\lambda_i$ with different indices are assumed to be distinct).
Regular bi-flat $F$-manifolds in dimension three whose endomophism $L$ has the form $L_1$ are actually semisimple and, as recalled above, are locally parameterized by solutions of the full  Painlev\'e VI. 

We will focus our attention on three-dimensional regular bi-flat $F$-manifolds whose operator $L$ has the form $L_2$ or $L_3$ and we will show that in the former case they are locally parameterized by solutions of the full P$_{V}$, while in the latter case they are locally parameterized by solutions of the full P$_{IV}.$

This highlight a striking parallelism between confluences of Painlev\' e equations and collision of eigenvalues of the endomorphism $L$ (preserving regularity), a fact which in our opinion deserves further investigation. It would be definitely interesting to extend this correspondence beyond the regular case. Unfortunately for the non-regular case there are no structural result similar to those developed in \cite{DH} at the moment. 

Let us remark that to the best of our knowledge, this is the first time in which other Painlev\' e trascendents, besides Painlev\' e VI appear in the analysis of geometric structures related to integrability or topological field theory. Our work provides a clear indication that the other Painlev\'e equations might be appear in the classification not only of non-semisimple bi-flat $F$-manifolds, besides the regular case treated here, but also in the analysis of non-semisimple Frobenius manifolds. 

We also point out that the approach championed in \cite{AL} and \cite{L2014} is based on the study of  a  generalized Darboux-Egorov system and cannot be applied to the semisimple case while the methodology developed here, in which the key role is played by a  geometric version of Tsarev's conditions  of integrability  paired with a  commutativity condition between the Lie derivative with respect to a set of eventual identities defining a subalgebra of the centerless Virasoro algebra and  the covariant derivative of the associated connections,  does not require the semisimplicity of the product.

Finally, in the second part of the paper we show the remarkable phenomenon that, while in the semisimple case there are in general obstructions to the existence of multi-flat $F$-manifolds, in the regular non-semisimple case it is possible to construct multi-flat $F$-manifolds with an {\em arbitrary} (countable) number of compatible flat connections  (all the  powers of the Euler vector field). This fact is in striking contrast to the semisimple situation, where the number of simultaneous compatible flat structures is severely limited. This is the first example of an $F$-manifold equipped with an infinite collection of non-trivial compatible flat structures.

The paper is structured as follows. In Section \ref{structuresec} we discuss the relations between geometric structures appearing in the study of $F$-manifolds and integrable dispersionless PDEs. We introduce Tsarev's conditions which will be essential to determine multi-flat $F$-structures once they are coupled with the necessary conditions determined in Section \ref{multiflatnesssec}. In Section \ref{examplesflatbiflatsec} we discuss some examples of flat and bi-flat $F$-manifolds. In particular, 
 we show that the theory of Lauricella structures recently developed in \cite{CHL, Looijenga} support non-trivial products in the sense of Manin. These structures are  related to the flat and bi-flat structures of the generalized $\epsilon$-system \cite{LP,AL,L2014}.

In Section \ref{multiflatnesssec}  we provide necessary conditions for the existence of multi-flat structures. In Section 5 we discuss the semisimple case proving that $N$-flat structures with $N>3$ can not exist in general. In Section \ref{biflatsec} we couple the necessary conditions for the existence of multi-flat structures found in the previous Section with Tsarev's conditions. This allows us to study in detail bi-flat $F$-manifolds in dimension $2$ and $3$, in particular we find that bi-flat $F$-manifold in dimension $3$ are parametrized by the solutions of a nonlinear non-autonomous system of first order quadratic ODEs, possessing $5$ independent  integral of motions. Moreover, we construct a one-parameter family of maps each of which associates a given solution of this system of ODEs to a solution of the Painlev{\'e} VI equation.

In Section \ref{triflatsec} we analyze tri-flat $F$-manifolds, construct a system of ODEs that parametrize them and find some special solutions of this system given by hypergeometric functions. 

In Section \ref{structuresec2}  we study non-semisimple regular bi-flat $F$-manifolds in dimension three according to the form of $L$. We provide a local model for these manifolds in the ``canonical coordinates" provided by \cite{DH} and show that the geometric data are controlled by two systems ODEs depending on the form of $L$. We give a detailed proof that these systems reduce in one case to the full P$_{V}$ and in the other to the full P$_{IV}.$ Although the reduction proof is completely elementary, it is highly non-trivial. 

In Section 9 we construct examples of non-semisimple regular tri-flat and multi-flat $F$-manifolds in dimension three (under the assumption that $L=E\circ$ has only one Jordan block). These examples show that the existence of multi-flat structures in the non-semisimple case is unexpectedly much more involved than in the semisimple case. In particular we show that in this situation it is possible to construct multi-flat $F$-manifolds with an arbitrary number of compatible flat structures. This happens essentially because once these $F$-manifolds are equipped with a quadri-flat structure, they are equipped automatically with infinitely many.

\section{Flat $F$-manifolds and Integrable dispersionless PDEs}\label{structuresec}
In this section, we survey the relationships between $F$-manifolds, flat $F$-manifolds and other geometric structures on one hand, and the theory of integrable dispersionless PDEs on the other. We also introduce Tsarev's conditions, which play a key role in determining multi-flat $F$-structures.

According to Tsarev's theory \cite{ts1,ts2}, integrable quasilinear systems of PDEs of the form
\beq\label{shs}
u^i_t=v^i(u)u^i_x,\qquad i=1,...,n
\eeq
are defined by  a set of functions $\Gamma^i_{ij}$ ($i\ne j$) satisfying the conditions (called {\em Tsarev's conditions})
\begin{eqnarray}
\label{rt1}
\d_j\Gamma^i_{ik}+\Gamma^i_{ij}\Gamma^i_{ik}-\Gamma^i_{ik}\Gamma^k_{kj}
-\Gamma^i_{ij}\Gamma^j_{jk}=0, \quad \mbox{if $i\ne k\ne j\ne i$}.
\end{eqnarray}
Once the conditions \eqref{rt1} are satisfied the solutions of the system
\beq\label{sym}
\d_j v^i=\Gamma^i_{ij}(v^j-v^i)
\eeq
define a set (depending on functional parameters) of commuting flows of the form \eqref{shs}. From \eqref{rt1} it follows that the solutions of
 \eqref{rt1} satisfy the conditions
\begin{equation}
\label{sh}
\partial_j\left(\frac{\partial_k v^i}{v^i-v^k}\right)=
\partial_k\left(\frac{\partial_j v^i}{v^i-v^j}\right)\hspace{1 
cm}\forall i\ne j\ne k\ne i,
\end{equation}
Conversely, given $v^i$ satisfying \eqref{sh} and using \eqref{sym} as definition of $\Gamma^i_{ij}$, the compatibility conditions \eqref{rt1}
 are automatically satisfied.
 
Quasilinear systems satisfying conditions \eqref{sh} are called \emph{semi-Hamiltonian} \cite{ts1,ts2} or \emph{rich} \cite{serre1,serre2}. Sevennec \cite{sevennec} later found
  a nice characterization of semi-Hamiltonian systems. He showed they coincide with diagonalizable systems
   of conservation laws.   

As the notation suggests, the functions $\Gamma^i_{ij}$ can be identified with (part of) the coefficients of a symmetric connection $\nabla$. The reconstruction of $\nabla$ can be done in essentially two non-equivalent ways. 

In the first case, we call the connection $\nabla$ a {\em Hamiltonian connection}. 
 In this case, $\nabla$ is the Levi-Civita connection of a diagonal metric $g$:
\beq\label{metric}
\d_j\ln{\sqrt{g_{ii}}}=\Gamma^i_{ij},\qquad j\ne i.
\eeq
Given a diagonal metric $g$ for which the functions $\Gamma^i_{ij}$ satisfy the above conditions,  all the remaining Christoffel symbols are uniquely defined through the classical Levi-Civita's formula.
 However, as it is easy to check,  the general solution
 of \eqref{metric}  depends on $n$ arbitary functions of a single variable: if $g_{ii}$ is a solution then $\varphi_i(u^i)g_{ii}$ is still a solution.  

The connections defined by \eqref{metric} have been introduced by Dubrovin and Novikov in \cite{DN}. We call them Hamiltonian connections since they are related to the Hamiltonian formalism. For instance, in the flat case (i.e. when $\nabla$ is flat), the differential operator
\beq
P^{ij}:=g^{ii}\delta^i_j\d_x-g^{il}\Gamma^j_{lk}u^k_x
\eeq  
defines a local Hamiltonian operators for the flows \eqref{shs} defined by the solutions of \eqref{metric}.

The  non-flat case is more involved: the Hamiltonian 
 operators are non-local and the non-local tail is related to the quadratic expansion of the Riemann tensor in terms of solutions of the system \eqref{sym}:
$$R^{ij}_{ij}=\sum_{\alpha}\epsilon_{\alpha}w^i_{\alpha}w^j_{\alpha}.$$
The existence of this quadratic expansion is a non-trivial property. It was conjectured by Ferapontov \cite{F} that all solutions of the system \eqref{metric} possess such a property.
 Ferapontov's conjecture has been checked for reductions of dKP and 2d Toda in \cite{GLR} and \cite{CLR}.

The other way to reconstruct a torsionless affine connection $\nabla$ having $\Gamma^i_{ij}$ as a subset of its Christoffel symbols in a distinguished coordinate system was devised in \cite{LP}. This leads to the notion of natural connections and $F$-manifold with compatible connection and flat unity \cite{LPR}. An {\em $F$-manifold with compatible connection and flat unity} is a semisimple $F$-manifold  $(M, \circ, e)$ equipped with a torsionless connection $\nabla$ (not necessarily flat) such that the following requirements hold
\begin{eqnarray*}
&&Z\circ R(W,Y)(X)+W \circ R(Y,Z)(X)+Y \circ R(Z,W)(X)=0,\\
&&\left(\nabla_X c\right)\left(Y,Z\right)=\left(\nabla_Y c\right)\left(X,Z\right),\\
&&\nabla e=0,
\end{eqnarray*}
where $R$ in the first condition above is the Riemann tensor and  $X, Y, Z, W$ are arbitrary vector fields.
Connections satisfying these conditions are called {\em natural connections}.
In the distinguished coordinate system given by the canonical coordinates of $\circ$, 
the first and second requirements imply Tsarev's condition \eqref{rt1} for $\Gamma^i_{ij}$ (\cite{LPR}), while the second and third one provide additional conditions that specify completely all the other Christoffel symbols. Indeed in canonical coordinates for $\circ$, given $\Gamma^i_{ij}$, the last two requirements above for $\nabla$ are equivalent to  
\begin{equation}\label{naturalc}
\begin{split}
\Gamma^{i}_{jk}&:=0,\qquad\forall i\ne j\ne k \ne i,\\
\Gamma^{i}_{jj}&:=-\Gamma^{i}_{ij},\qquad i\ne j,\\
\Gamma^{i}_{ii}&:=-\sum_{l\ne i}\Gamma^{i}_{li}.
\end{split}
\end{equation} 

Let us remark that the condition $\nabla e=0$ in the definition of natural connection $\nabla$ is the less important and can be dropped; as we have remarked in the Introduction it is not too restrictive, at least in some examples, since deforming the connection $\nabla$  with the product $\circ$ one can obtain $\nabla e=0$ for the deformed connection, while preserving the other two conditions.  

In this framework the quasilinear system \eqref{shs} can be written as
\beq\label{shs2}
u_t=X\circ u_x.
\eeq
 and the system \eqref{sym} reads
\beq\label{sym2}
c^i_{jl}\nabla_k X^l=c^i_{kl}\nabla_j X^l.
\eeq
In this setting the  characteristic velocities $v^i$ are thought as the components
 of the vector fields $X$ in canonical coordinates. 
Since the Riemann invariants are identified with the canonical
 coordinates, given a semi-Hamiltonian system, the associated natural connection is defined up to a reparameterization of the Riemann invariants, that is up to the choice of $n$ arbitrary functions of a single variable. 
 
Like in the case of Hamiltonian connections, the most interesting case is when the connection $\nabla$ is flat (we have discussed these manifolds in the Introduction, they are the flat $F$-manifolds introduced by Manin in \cite{manin}). In this case, a countable set of solutions of the system \eqref{sym} can be obtained  from a frame
 of flat vector fields $(X_{(1,0)},...,X_{(n,0)})$ via the following recursive relations (here $d_{\nabla}$ is the exterior covariant derivative)
\beq\label{rr}
d_{\nabla}X_{(p,\alpha+1)}=X_{(p,\alpha)}\circ.
\eeq
In flat coordinates the flows of the principal hierarchy are systems of conservation laws. Due to \eqref{rr}, the current associated to the ``time" $t_{(p,\alpha)}$ is given by the vector field $X_{(p,\alpha+1)}$.

In general the two ways we just described to reconstruct a torsionelss connection $\nabla$ starting from the functions $\Gamma^i_{ij}$ satisfying \eqref{rt1} are inequivalent: Hamiltonian connections are not natural connections and natural connections are not Hamiltonian. Indeed combining the conditions $\nabla g=0$ and $\nabla_i c^k_{lj}=\nabla_l c^k_{ij}$, one obtains
 $\d_j g_{ii}=\d_i g_{jj}$, which implies that in order to have a connection which is both Hamiltonian and natural, the metric must be potential in canonical coordinates (Egorov case). This is for instance the case of semisimple Frobenius manifolds.

In many examples (including Frobenius manifolds) besides the recursive relation \eqref{rr} there exists an additional one, which we called twisted Lenard-Magri chain (see \cite{ALimrn})
\beq\label{rr2}
d_{\nabla^{(1)}}(e\circ X_{(p,\alpha+1)})=d_{\nabla^{(2)}}(E\circ X_{(p,\alpha)}).
\eeq
It is based on the existence of an additional flat structure
 and on an eventual identity $E$. This leads  naturally to define the class of bi-flat $F$-manifolds that was extensively studied in \cite{AL,L2014}.

\begin{rmk}
In the semisimple case removing the condition $\nabla
e=0$ in the definition of natural connections one has the
freedom to choose the Christoffel symbols $\Gamma^i_{ii}$ \cite{LP}. The same freedom can be also described in terms of
the special family of connections \cite{DS2}
$$\tilde\nabla_X Y=\nabla_X Y+V\circ X\circ Y$$
This is a family of connections satisfying the symmetry
condition (1.5) (the product is not assumed to be
semisimple). Like in the semisimple case the condition
$\tilde\nabla e=0$ fixes uniquely the vector field $V$.
\end{rmk}
 
\section{Examples of flat and bi-flat $F$-manifolds}\label{examplesflatbiflatsec}
In this section we present some examples of flat and bi-flat $F$-manifolds. Despite their variety, these examples are all related to integrable systems. 

\subsection{$\vee$-systems}\label{veesystem}
$\vee$-systems were introduced by A. Veselov in \cite{Ve} to construct new solutions of generalized WDVV equations, starting from a special set of covectors. We want to point out that it is always possible to construct a flat $F$-manifold starting from a  $\vee$-system. 

First we recall the notion of $\vee$-systems (see \cite{Ve}).
Let $V$ be a finite dimensional vector space and let $V^*$ is dual.
Let $\mathcal{V}$ be a finite set of non-collinear covectors $\alpha\in V^*$ with the property that they span $V^*$. This means that the symmetric  bilinear form defined by $G^{\mathcal{V}}:=\sum_{\alpha\in  \mathcal{V}} \alpha\otimes \alpha$
is non-degenerate. The non-degeneracy of $G^{\mathcal{V}}$ is equivalent to require that the map $\phi_{\mathcal{V}}: V\rightarrow V^*$ defined by the 
formula 
$$(\phi_{\mathcal{V}}(u))(v):=G^{\mathcal{V}}(u,v), \quad u, v\in V,$$
is invertible. In this context, for each covector $\alpha$ it is possible to define the vector $\check \alpha\in V$ as
\begin{equation}\label{checkalpha}
\check{\alpha}:=\phi^{-1}_{\mathcal{V}}(\alpha), \quad \alpha\in V^*, 
\end{equation}
or, which is equivalent, as the unique vector in $V$ such that 
\begin{equation}\label{checkalpha2}\alpha=G^{\mathcal{V}}(\cdot, \check{\alpha}).\end{equation}

The finite spanning set $\mathcal{V}\subset V^*$   is called a $\vee$-system if for each two-dimensional plane $
\Pi\subset V^*$ one has
\begin{equation}\label{veeequation}
\sum_{\beta\in \Pi\cap \mathcal{V}}\beta(\check{\alpha})\check{\beta}=\lambda \check{\alpha},
\end{equation}
for each $\alpha \in \Pi\cap \mathcal{V}$ and for some $\lambda$, which may depend on $\Pi$ and $\alpha$. 

In this case, the (contravariant) metric is given by 
$$\check{G}=\sum_{\check{\alpha}\in\mathcal{V}}\check{\alpha}\otimes\check{\alpha},$$
while the product $\circ$ is defined by the following formula:
\beq\label{provsys}(X\circ Y)_u=\sum_{\alpha\in \mathcal{V}}\frac{\alpha(X) \alpha(Y)\, \check{\alpha}}{\alpha(u)}.\eeq
The product \eqref{provsys} is clearly commutative and the $\vee$-conditions guarantee that it is also associative. 
The flat connection $\nabla$ is the Levi-Civita connection associated to the standard flat metric obtained inverting $\check{G}$. It is a flat connection and it satisfies $\nabla_l c^i_{jk}=\nabla_j c^i_{lk}$. Indeed, in flat coordinates we have
$$\partial_l c^i_{jk}=-\sum_{\alpha \in \mathcal{V}}\frac{\alpha_j \alpha_k \alpha_l \check{\alpha}^i}{(\alpha(u))^2},$$
which is symmetric in $l,j,k$, so the condition holds. 
Therefore in order to have a flat $F$-manifold we need to check that $\nabla e=0$, where $e$ is the unit vector field of \eqref{provsys}.

It is immediate to see that the unity for the product \eqref{provsys} is given by the vector field $e_u:=(u^1,\dots, u^n)$ since for every vector field $X$
$$(X\circ e)_u:=\sum_{\alpha\in \mathcal{V}}\frac{\alpha(X) \alpha(u)\, \check{\alpha}}{\alpha(u)}=X_u,$$
due to the fact that $\sum_{\alpha\in \mathcal{V}}\alpha \otimes  \check{\alpha}$ is equal to the identity endomorphism.

The unity $e$ does not satisfy in general the condition $\nabla e=0$, but it is always possible to modify the Christoffel symbols of the Euclidean structure with the structure constants $c^i_{lp}$ so that for the new connection $\tilde \nabla$ one has $\tilde \nabla e=0$. 
Indeed,  for any vector field $V=v^i \partial_i$ we get
$$\tilde\nabla_V e=\left(v^i u^j \tilde\Gamma^k_{ij}+v^i \frac{\partial u^k}{\partial u^i}\right)\partial_k.$$
We have $\tilde\nabla_V e=0$ iff $v^i u^j\tilde\Gamma^k_{ij}+v^k=0$ for any choice of the vector field $V$.  This give the condition $u^j \tilde\Gamma^k_{ij}=-\delta^k_i$.  The last condition says that $e$ behaves like the unity for the product with structure constants  $-\tilde \Gamma^k_{ij}$, so it is natural to choose $\tilde \Gamma^k_{ij}=-c^k_{ij}$. 
 To get a flat $F$-manifold, we also need to check that $\tilde \nabla_l c^i_{jk}=\tilde\nabla_j c^i_{lk}$
 and that the modified connection $\tilde\nabla $ is still flat. Both properties follow from the associativity of the product and from the 
 condition $\nabla_l c^i_{jk}=\nabla_j c^i_{lk}$.
 
Let us point out that $\vee$-systems are also related to purely non-local Hamiltonian structures (see \cite{AL2014JMP}).

\subsection{Semisimple Frobenius manifolds}\label{frobsubsection}
As we mentioned before, Frobenius manifolds have a compatible flat structure which is the Levi-Civita connection of an invariant metric $g$.
 They possess also a second  flat metric, called intersection form that we denote with $\tilde{g}$. In canonical coordinates for a product $\circ$ compatible with $g$, the two metrics are related by the simple formula
$$\tilde{g}^{ii}=u^ig^{ii}, \forall i.$$
This implies that the Christoffel symbols $\Gamma^i_{ij}$ (with $i\ne j$) are the same. Moreover the Levi-Civita connection of the intesection form is compatible
 with the dual product $*$ whose structure constants are given by
$$c^{*i}_{jk}=\f{1}{u^i}\delta^i_j\delta^i_k.$$
In general the unity of the dual product is not flat with respect to the Levi-Civita connection of the intersection form. However it is always possible, modifying it in a suitable way (in particular modifying the Christoffel
 symbols $\Gamma^i_{ii}$) to get a second flat connection which satisfies also this further property. We will discuss later in more details this point. 

\subsection{Lauricella connections, $W(A_n)$ $\vee$-systems and Lauricella bi-flat $F$-manifolds}

An example of bi-flat $F$-manifold, which in general can not be recast in the framework of Frobenius manifold is provided by the generalized $\epsilon$-system. In this case, the Christoffel symbols that determine the connection are given by:
\beq\label{nablaepsilon}
\Gamma^{i}_{ij}=\f{\epsilon_j}{u^i-u^j}\qquad i\ne j,
\eeq
and in the coordinate system $\{u^1,\dots, u^n\}$ the structure constants of the product $\circ$ have the form $c^i_{jk}=\delta^i_j \delta^i_k.$ 
This case was treated in detail in \cite{LP} and \cite{L2014} and it
 is related to the Euler-Poisson-Darboux system
\beq\label{ddl}
dd_L k=dk\wedge da,
\eeq
where $L$ is an endomorphism of the tangent bundle given by $L={\rm diag}(u^1,....,u^n)$, 
$a$ is a function given by $a=\sum_{i=1}^n\epsilon_i u^i$ and $d_Lf(X)=(LX)(f)=df(LX)$, for every vector field $X$. 
Indeed one can write the solutions of the system \eqref{sym2} as $X^i=-\f{\d_i k}{\epsilon_i}$ where $k$ is a solution of \eqref{ddl}. The vector fields  $X^i_{(p,\alpha)}=-\f{\d_i k_{(p,\alpha)}}{\epsilon_i}$, which define the principal hierarchy correspond to special solutions of \eqref{ddl}: the flat vector fields  $X_{(p,0)}$ correspond to a set ($k_{(1,0)}=-a,k_{(2,0)},...,k_{(n,0)})$ of flat coordinates for the connection \eqref{nablaepsilon} with $\epsilon_i\to-\epsilon_i$ and, 
 up to inessential constant factors and a part from some resonant cases, the vector fields  $X_{(p,\alpha)}$ ($\alpha\ge 1$)  correspond to
 the solutions of \eqref{ddl} defined recursively  by 
 $dk_{(p,\alpha+1)}=d_Lk_{(p,\alpha)}-k_{(p,\alpha)}da$. For instance, it is easy to check that the vector field $X_{(1,1)}$ has components  $X^i_{(1,1)}=u^i-a$. The corresponding flow
$$u^i_{t_{(1,1)}}=\left(u^i-\sum_{k=1}^n\epsilon_k u^k\right)u^i_x,\qquad i=1,...,n,$$
is called \emph{the generalized $\epsilon$-system} \cite{Pavlovhydro}.

This example is related to the theory of Lauricella functions \cite{Lauricella} and Lauricella manifolds \cite{CHL,Looijenga}. Here the coordinates $u^1, \dots u^n$ are intended as complex coordinates. 

We begin by recalling the definition of Lauricella functions. Consider $n$ real numbers in the interval $(0,1)$, $(\epsilon_1, \dots, \epsilon_n):=\epsilon$, called the weight system $\epsilon$ and let $|\epsilon|:=\sum_{i=1}^n \epsilon_i$ be the total weight of $\epsilon.$ Let $\mathcal{H}:=\cup_{1\leq i<j\leq n}H_{ij} $ where $H_{ij}:=\{u\in \mathbb{C}^n | u^i=u^j\}.$ The value of the {\em Lauricella function} of weight $\epsilon$ at the point  $u:=(u^1, \dots, u^n)\in \mathbb{C}^{n}\setminus \mathcal{H}$ 
 is given by
$$\int_{\gamma_u}\eta_u=\int_{\gamma_u}(u^1-\zeta)^{-\epsilon_1}\dots (u^n-\zeta)^{-\epsilon_n}d\zeta.$$
Here  $\gamma_u$  is an oriented piecewise differentiable arc such that the end points of $\gamma_u$ lie in $\{u^1, \dots, u^n\}$  (but such that $\gamma_u$ does not meet this set elsewhere) and a determination of the multivalued differential $\eta_u$ is fixed (in general Lauricella functions are multivalued). Moreover the choice of the arc $\gamma_u$ and the choice of the determination of  $\eta_u$
 should depend continuously on $u$ (see \cite{Looijenga} for details).

To show that Lauricella functions provide (almost all) flat homogenous coordinates for the natural connection associated to the generalized $\epsilon$-system described above, we first recall the following 
\begin{prop}\label{lauricella}\cite{Looijenga}
Let $L^{\epsilon}_u$ be the complex vector space of germs of holomorphic Lauricella functions at $u\in \mathbb{C}^{n}\setminus \mathcal{H}$ with fixed weight system $\epsilon$. Then $\mathrm{dim}_{\mathbb{C}}(L^{\epsilon}_u)=n-1$ and $L^{\epsilon}_u$ contains the constant functions iff $|\epsilon|=1$. Moreover for any $f\in L^{\epsilon}_u$ the following hold
\begin{enumerate}
\item $e(f)=0$, where $e=\sum_{i=1}^n \frac{\partial}{\partial u^i}.$
\item $f$ is homogeneous of degree $(1-|\epsilon|)$.
\item $f$ satisfies the system of differential equations
\begin{equation}\label{lauricellaeq1}
\frac{\partial^2 f}{\partial u^i \partial u^j}=\frac{1}{u^i -u^j}\left(\epsilon_j \frac{\partial f}{\partial u^i}-\epsilon_i \frac{\partial f}{\partial u^j}\right), \quad 1\leq i<j\leq n.
\end{equation}
\end{enumerate}
\end{prop}

Notice that the above system \eqref{lauricellaeq1} coincides with the  Euler-Darboux-Poisson system \eqref{ddl} with $\epsilon_i\to-\epsilon_i$. Combining Proposition \ref{lauricella} with the results from Section 5.1 of \cite{L2014}, we obtain the following Corollary relating the generalized $\epsilon$-system with Lauricella functions:

\begin{cor}
Consider the generalized $\epsilon$-system in $n$-dimensions, $\epsilon=(\epsilon_1, \dots, \epsilon_n)$ and suppose that $0<\epsilon_i<1$ for all $i=1, \dots, n$ and that $|\epsilon|:=\sum_{i=1}^n \epsilon_i \neq 1$. Then any basis of Lauricella functions $\{f_l\}_{l=1}^{n-1}$ in $L^{\epsilon}_u$, $u\in \mathbb{C}^{n}\setminus\mathcal{H}$ gives rise to  $n-1$ of the $n$ flat coordinates of the natural connection associated to the generalized $\epsilon$-system.
\end{cor}

\emph{Proof}
Let $f$ be any element in a basis of Lauricella functions $\{f_l\}_{l=1}^{n-1}$ in $L^{\epsilon}_u$.
Introducing the notation $\theta_i=\frac{\partial f}{\partial u^i}$, equation \eqref{lauricellaeq1} can be written as ($\partial_i:=\frac{\partial }{\partial u^i}$): 
$$\partial_i \theta_j=\frac{1}{u^j-u^i}\left( \epsilon_i \theta_j-\epsilon_j\theta_i\right), \quad i=1, \dots n, \; i\neq j,$$
which is the first of the set of equations that characterize flat $1$-forms for the natural connection associated to the generalized $\epsilon$-system (see the first set of equations in formula $5.4$ in \cite{L2014}). Notice also the constraint $e(f)=0$ (the first point of Proposition \ref{lauricella}) immediately implies $\sum_{i=1}^n \partial_i \theta_j=0$ for all $j=1, \dots n$, which constitutes the other equation $5.4$ in \cite{L2014} characterizing flat $1$-forms for the natural connection. Therefore any basis of Lauricella functions gives rise to $n-1$ flat coordinates, provided these functions are not constant. This is the case since $|\epsilon|\neq 1$ by assumption and therefore by Proposition \ref{lauricella}  any basis of $L^{\epsilon}_u$ does not contain constant functions.  The natural connection has $n$ flat coordinates: the missing flat coordinate is the function $a=\sum_{i=1}^n\epsilon_i u^i$.
  
\begin{flushright}
$\Box$
\end{flushright}

Therefore, under suitable assumptions on the weights $\epsilon_i$, the Lauricella functions provide $n-1$ of the $n$ flat homogeneous coordinates for the generalized $\epsilon$-system. 

Finally we comment on the relation between the natural connection of the generalized $\epsilon$-system and the so called Lauricella connection.
We first recall the notion of Lauricella connection.
Consider the free smooth diagonal action $\psi: \mathbb{C}\times \mathbb{C}^n \rightarrow \mathbb{C}^n$ give by $\psi(\lambda, u)=(u^1+\lambda, \dots, u^n+\lambda)$. Call $V$ the quotient of $\mathbb{C}^n$ by this action and $\pi: \mathbb{C}^n \rightarrow V$ the corresponding quotient map. Denote with $e_1, \dots, e_n$ the standard basis of $\mathbb{C}^n$, which we identify with the global frame $\partial_1, \dots, \partial_n$ for its tangent bundle. Call $\mathcal{V}\mathbb{C}^n$ the line sub-bundle of $T\mathbb{C}^n$ given by $\mathrm{Ker}(\pi_*)$, this is just the vertical distribution and notice that it is spanned by $e=\sum_{i=1}^n \partial_i$.

Given now positive real numbers $\epsilon_1, \dots, \epsilon_n$, we define an inner product on $\mathbb{C}^n$ by $\langle e_i, e_j\rangle=\epsilon_i \delta_{ij}$. Using this inner product, it is possible to identify $V$ with the orthogonal complement of the main diagonal, i.e. with $Z_0:=\{(u^, \dots, u^n)| \sum_{i=1}^n \epsilon_i u^i=0\}$ and construct a global decomposition $T\mathbb{C}^n=\mathcal{V}\mathbb{C}^n\oplus \mathcal{C}$, where the sub-bundle $\mathcal{C}$ is orthogonal to $\mathcal{V}\mathbb{C}^n$ and it is spanned by the vector fields $\epsilon_j \partial_i-\epsilon_i \partial_j.$  All the integral leaves of $\mathcal{C}$ are just given by translations of the hyperplane $Z_0,$ namely they are $Z_c:=\{(u^1, \dots, u^n)| \sum_{i=1}^n \epsilon_i u^i=c\}$ as $c$ varies in $\mathbb{C}$.
 
 To each hyperplane $H_{ij}$ with $1\leq i<j\leq n$ in $\mathbb{C}^n$ there is associated a unique meromorphic differential with divisor $-H_{ij}$ and residue $1$ along $H_{ij}$, $\omega_{H_{ij}}:=\frac{d\phi_{H_{ij}}}{\phi_{H_{ij}}}$, where $\phi_{H_{ij}}$ is a linear equation for $H_{ij}.$ As $\phi_{H_{ij}}$ we can choose $u^i-u^j$. In this respect, Arnol'd has proved (see \cite{Arnold}) that the forms $\omega_{H_{ij}}$ are the generators of the cohomology ring of the colored braid groups (essentially the cohomology ring of the space of ordered subsets of $n$ different points of the plane $\mathbb{C}$).

Using the inner product introduced above, the orthogonal complement $H_{ij}^{\perp}$ of the hyperplane $H_{ij}$ is the line spanned by the vector $\epsilon_j e_i-\epsilon_i e_j$, since if $v=\sum_k v^k  e_k\in H_{ij}$, then $\langle v, 	\epsilon_j e_i-\epsilon_i e_j\rangle=\epsilon_j \epsilon_i(v^i-v^j)=0$. Consider the rank $1$ endomorphism $\rho_{H_{ij}}$ of $\mathbb{C}^n$ with kernel $H_{ij}$ and range given by $H_{ij}^{\perp}$. Any such endomorphism is self-adjoint with respect to the inner product $\langle \cdot, \cdot \rangle$ as it is immediate to check. We can fix a normalization for $\rho_{H_{ij}}$ imposing that $\epsilon_j e_i-\epsilon_i e_j$ is an eigenvector with eigenvalue $\epsilon_i+\epsilon_j$. This simply means that $\rho_{H_{ij}}$ has the form $u\mapsto (u^i-u^j)(\epsilon_j e_i-\epsilon_i e_j)$. Obviously we can also think of $\rho_{H_{ij}}$ as being vector fields valued, simply interpreting $\epsilon_j e_i-\epsilon_i e_j$ as $\epsilon_j \partial_i-\epsilon_i \partial_j$ which is what we do below discussing the connection $\nabla$. Moreover, one can view $\rho_{H_{ij}}$ as an endomorphism of the tangent bundle of $\mathbb{C}^n$ rather than an endomorphism of $\mathbb{C}^n$, in which case we write $\rho_{H_{ij}}=(du^i-du^j)\otimes (\epsilon_j \d_i-\epsilon_i \d_j).$ 

Observe also that $\rho_{H_{ij}}$ induces also an endomorphism $\rho^V_{H_{ij}}$ on $V$,  due to the fact that $\rho_{H_{ij}}$ is translation invariant and to the fact that its range lies in $Z_0$ (or seeing it as a vector fields valued map its range lies in $\mathcal{C}$). Similarly, since $\omega_{H_{ij}}$ is invariant under the action of $\psi$ extended to the tangent bundle,
  the corresponding form $\omega^V_{H_{ij}}$ on $V$  is well defined. 

 These details, although cumbersome, will be important to show that the Lauricella connection is a reduction of the natural connection of the generalized $\epsilon$-system in a sense detailed below.  

\begin{thm}\cite{CHL}\label{CHLthm1}
Let $\nabla^{0,V}$ be the standard, translation invariant flat connection on the tangent bundle of $V$. Then the connection 
 \begin{equation}\label{lauricellaconn}
\nabla^V:=\nabla^{0, V}-\sum_{1\leq i<j\leq n} \omega^V_{H_{ij}} \otimes\rho^V_{H_{ij}},
\end{equation}
 is called the Lauricella connection and it is flat. Furthermore if $0<\epsilon_i <1$ for all $i=1, \dots, n$, then the multivalued holomorphic Lauricella functions defined above are translation invariant, so they define multivalued holomorphic functions on $\tilde{V}$ (still called Lauricella functions) and their differentials are flat for the Lauricella connection. 
\end{thm}
Let us remark that Theorem \ref{CHLthm1} holds under slightly more general assumptions (see \cite{CHL} Section 2.3), but we recall it here in this form since we are interested to compare the Lauricella connection with the natural connection of the generalized $\epsilon$-system. 

First we have this straightforward characterization of the natural connection of the generalized $\epsilon$-system:
\begin{lemma}
Let $\nabla^0$ be the standard flat translation invariant connection of $\mathbb{C}^n$. Then 
the natural connection of the generalized $\epsilon$-system, intended as a connection on the
holomorphic tangent bundle, coincide with the
connection
$$\nabla:=\nabla^0-\sum_{1\leq m<l\leq n}\omega_{H_{ml}}\otimes \rho_{H_{ml}},$$
on $\mathbb{C}^n\setminus {\mathcal{H}}$. 
\end{lemma}
\emph{Proof} As above we identify the standard basis $\{e_1, \dots, e_n\}$ of $\mathbb{C}^n$ with
the global frame $\{\partial_1, \dots, \partial_n\}$ of its tangent bundle. Since $$\nabla_{\d_i} \d_j=\Gamma^k_{ij}\d_k=-\sum_{1\leq m<l\leq n}\omega_{H_{ml}}(\d_i)\rho_{H_{ml}}(\d_j),$$ it is immediately clear that $\Gamma^k_{ij}=0$ for $i\neq j\neq k\neq i,$
 due to the fact that the range of $\rho_{H_{ml}}$ is spanned by $\epsilon_l \d_m-\epsilon_m \d_l$ and that $\rho_{H_{ml}}(\d_j)=0$ for $m\neq j$, $l\neq j$. 
In general we have:
$$\nabla_{\d_i}\d_j=-\sum_{1\leq m<l\leq n}\frac{d(u^m-u^l)(\d_i)}{u^m-u^l} (du^m-du^l)(\d_j)(\epsilon_l \d_m -\epsilon_m \d_l).$$
For $i<j$ we obtain 
$$\nabla_{\d_i}\d_j=\frac{1}{u^i-u^j}\left( \epsilon_j \d_i-\epsilon_i \d_j\right),$$
so that $\Gamma^i_{ij}=\frac{\epsilon_j}{u^i-u^j}$, $\Gamma^j_{ji}=\frac{\epsilon_i}{u^j-u^i}$ and analogously for $i>j$. 
Finally 
$$\nabla_{\d_i}\d_i=\Gamma^k_{ii}\d_k=-\sum_{l>i}\frac{1}{u^i-u^l}(\epsilon_l \d_i-\epsilon_i \d_l)-\sum_{m<i}\frac{1}{u^m-u^i}(\epsilon_i \d_m -\epsilon_m \d_i)$$
$$=\sum_{l\neq i}\frac{\epsilon_i}{u^i-u^l}\d_l-\sum_{l\neq i}\frac{\epsilon_l}{u^i-u^l}\d_i,$$
from which we deduce that $\Gamma^i_{ii}=-\sum_{l\neq i}\frac{\epsilon_l}{u^i-u^l}=-\sum_{i\neq l}\Gamma^i_{il}$ and for $k\neq i$ that $\Gamma^k_{ii}=\frac{\epsilon_i}{u^i-u^k}=-\Gamma^k_{ki}.$ 
\begin{flushright}
$\Box$
\end{flushright}

%Given the projection map $\pi: \mathbb{C}^n \rightarrow V$, now we show that the natural connection of the generalized $\epsilon$-system is just the pull-back of the Lauricella connection on $V$ via $\pi$. 
The following Proposition clarifies the relation between the natural connection of the generalized $\epsilon$-system and the Lauricella connection:
\begin{prop}
Let $X, Y$ be vector fields on $V$ and denote with $X^{\mathcal{C}}$ and $Y^{\mathcal{C}}$ the unique vector fields on $\mathbb{C}^n$ such that $X^{\mathcal{C}}\subset \mathcal{C}$, $Y^{\mathcal{C}}\subset \mathcal{C}$ and such that $\pi_*(X^{\mathcal{C}}_u)=X_{\pi(u)}$ and $\pi_*(Y^{\mathcal{C}}_u)=Y_{\pi(u)}$. 
Then, with the notation introduced previously we have 
%\begin{equation}
%\nabla^V_XY=\pi_{*}\left(\nabla_{X^{\mathcal{C}}}Y^{\mathcal{C}}\right).
%\end{equation}
%Moreover, if one identifies $V$ with $Z_0$ or one of its translates, then we have 
$$\left(\nabla^V_XY\right)^{\mathcal{C}}=\nabla_{X^{\mathcal{C}}}Y^{\mathcal{C}}.$$
\end{prop}
\emph{Proof}  
Since $\nabla_{X^{\mathcal{C}}}Y^{\mathcal{C}}\subset \mathcal{C}$ we have to prove that 
\begin{equation}
\nabla^V_XY=\pi_{*}\left(\nabla_{X^{\mathcal{C}}}Y^{\mathcal{C}}\right).
\end{equation}
It is enough to prove the claim using constant vector fields, $X, Y$, in which case $X^{\mathcal{C}}$ and $Y^{\mathcal{C}}$ are also constant and the connections $\nabla^{0, V}$ and $\nabla^0$ do not play any role. By definition the function $\omega_{H_{ij}}(X^{\mathcal{C}})$ on $\mathbb{C}^n$  defines a function on $V$ which coincides with $\omega^V_{H_{ij}}(X)$ and $\rho^V_{H_{ij}}(Y)=\pi_*\rho_{H_{ij}}(Y^{\mathcal{C}})$. This implies
$$\omega^V_{H_{ij}}(X) \rho^V_{H_{ij}}(Y)=\pi_*\left(\omega_{H_{ij}}(X^{\mathcal{C}}) \rho_{H_{ij}}(Y^{\mathcal{C}})\right).$$

\begin{flushright}
$\Box$
\end{flushright}

Observe that in the case of $\vee$-systems one obtains a one-parameter family of flat connections in which the deformed Christoffel symbols are obtained via the product structure (see the discussion in Section \ref{veesystem} of this paper and \cite{AL2014JMP} for many more details). This leads to ask if also in the case of the Lauricella systems there is a product structure and if one can also obtains a one-parameter family of flat (Lauricella) connections. 

The answer to the latter question is positive and immediate. Indeed, due to the Proposition 2.3 in \cite{CHL}, in particular points (i) and (iv), the endomorphisms $\rho_{H_{ij}}$ can be rescaled by a common factor $\lambda \in \mathbb{C}^*$ without affecting the flatness of \eqref{lauricellaconn}. This means that $\nabla^{\lambda}:=\nabla^0-\lambda \sum_{1\leq i<j\leq n} \omega_{H_{ij}} \otimes\rho_{H_{ij}}$ is flat for any $\lambda$.

The answer to the former question is also positive, but we can actually provide two (in general) non-equivalent answers. 
Indeed, one way is to interpret the term $\sum_{1\leq m<l\leq n}\omega_{H_{ml}}\otimes \rho_{H_{ml}}$ in $\nabla$ as a deformation of $\nabla^0$ obtained using the structure constants of a non-trivial product. This leads to interpret this term as a $\vee$-system. To do this, recall that the metric $\langle \cdot, \cdot \rangle=\text{diag}(\epsilon^1, \dots, \epsilon^n)$  is diagonal in the coordinates $u^1, \dots, u^n$, and consider the vector fields $\check{\alpha}_{ij}:=\frac{1}{\sqrt{\epsilon_i \epsilon_j}}\left( \epsilon_j \partial_i-\epsilon_i \partial_j \right)=\sqrt{\epsilon_i \epsilon_j}\left(\frac{\partial_i}{\epsilon_i}-\frac{\partial_j}{\epsilon_j}\right).$
Now define forms $\alpha_{ij}:=\langle \check{\alpha}_{ij}, \cdot\rangle=\sqrt{\epsilon_i \epsilon_j}\left( du^i-du^j\right).$
With these definitions, it is immediate to check that the term that deforms $\nabla^0$ can be written as:
 $$\sum_{1\leq i<j\leq n}\omega_{H_{ij}}(X_u) \rho_{H_{ij}}(Y_u)=\sum_{1\leq i<j\leq n}\frac{\alpha_{ij}(X_u)\alpha_{ij}(Y_u)}{\alpha_{ij}(e_u)}\check{\alpha}_{ij},$$
where $X_u, Y_u\in T_u\mathbb{C}^n$ and $e_u=(u^1, \dots, u^n)$. 
Indeed, with these notations, $\omega_{H_{ij}, u}=\frac{\alpha_{ij}}{\alpha_{ij}(E_u)}$ and $\rho_{H_{ij}}=\alpha_{ij}\otimes \check{\alpha}_{ij}.$
The fact that the covectors $\alpha_{ij}$ form a $\vee$-system follows from the flatness of the connection
$$(\nabla_X Y)_u:=(\nabla^0_X Y)_u-\sum_{i,j} \frac{\alpha_{ij}(X_u) \alpha_{ij}(Y_u)\check{\alpha}_{ij}}{\alpha_{ij}(e_u)}.$$
This multiparametric family of $\vee$-systems appeared in \cite{CV}. Restricting it on the hypeplane $Z_0$ one gets a new family
 of $\vee$-system that, for $\epsilon_i=1$, correponds to the almost Frobenius structure for the Coxeter group $W(A_{n-1})$
 \cite{Ddual}.

The other possible product structure one can introduce is obtained by imposing that the standard basis of $\mathbb{C}^n$ is a basis of idempotents for a commutative associative product. This means that the standard basis for $\mathbb{C}^n$ provides canonical coordinates for a semisimple product. In a slightly different language this was observed for the first time in  \cite{LP} (see also \cite{L2014}). 

In other terms, depending on how one interprets the vector fields $\partial_i$, either as flat vector fields or as idempotents, one obtains two different examples of flat $F$-manifolds. Remarkably in the second case there is also a dual product  defined by the eventual identity $E=\sum_i u^i\frac{\d}{\d u^i}$ (see \cite{AL,L2014} for details). Due to the previous discussion we will call this structure on $\mathbb{C}^n\setminus {\mathcal{H}}$ the \emph{Lauricella bi-flat structure}.

\begin{rmk}
The generalized  $\epsilon$-system is a special example of a class of integrable quasilinear systems of PDEs associated to solutions
 of the Euler-Poisson-Darboux system \cite{Pavlovhydro,LM,L2006}. In \cite{KKS} it was observed that this class is related to Whitham $g$-phase Whitham equation by a sequence of Levy transformations generated by suitable Lauricella functions.
\end{rmk}

\begin{rmk} 
In the generalized $\epsilon$-system, the Christoffel symbols $\Gamma^{i}_{ij}$ depend only on the difference $u^i-u^j$. Under this assumption, the Tsarev's condition \eqref{rt1} reduces to the following algebraic system
\begin{eqnarray}
\label{rt1red}
\Gamma^i_{ij}\Gamma^i_{ik}=\Gamma^i_{ik}\Gamma^k_{kj}+\Gamma^i_{ij}\Gamma^j_{jk},\,\,\,\mbox{if $i\ne k\ne j\ne i$},
\end{eqnarray}
or
\begin{eqnarray}
\label{rt1redbis}
\f{\Gamma^k_{kj}}{\Gamma^i_{ij}}+\f{\Gamma^j_{jk}}{\Gamma^i_{ik}}=1\,\,\,\mbox{if $i\ne k\ne j\ne i$}.
\end{eqnarray}

%A new example fulfilling this condition is given  by
%\beq
%\label{eqgammaspecial}
%\Gamma^i_{ij}=\f{2\nu\epsilon_je^{\nu(u^i-u^j)}}{e^{\nu(u^i-u^j)}-e^{\nu(u^j-u^i)}}
%\eeq
%where $\nu$ and $\epsilon_j$ are arbitrary constants. It can be thought as  a deformation of the generalized $\epsilon$-system. Indeed, expanding at $\nu=0$ we get
%$$\Gamma^i_{ij}=\frac{\epsilon_j}{u^i-u^j}+\nu\epsilon+\f{1}{3}\epsilon(u^i-u^j)\nu^2+\mathcal{O}(\nu^3).$$
\end{rmk}

\section{Flatness conditions}\label{multiflatnesssec}
\subsection{Eventual identities and flatness conditions}
Given a {\it semisimple} $F$- manifold with an eventual identity $E$ we want to characterize \emph{flat} symmetric connections $\nabla$ compatible with the eventual identity $E$, i.e satisfying the following requirements
\begin{eqnarray*}
&&\left(\nabla_X c^*\right)\left(Y,Z\right)=\left(\nabla_Y c^*\right)\left(X,Z\right)\\
&&\nabla E=0,
\end{eqnarray*}
where $c^*$ is the $(1,2)$-tensor field associated to the dual product $*$.

Without loss of generality, we can analyze the situation in the canonical coordinates for the dual product $*$ induced by the eventual identity $E$; in this way  we can consider the case $E=e$. In this case, in canonical coordinates, the Christoffel symbols of the connection are uniquely specified once the coefficients
 $\Gamma^i_{ij}$ are given through the formula \eqref{naturalc}.
 
 It is possible to provide an intrinsic characterization of the flatness condition, which is given in Theorem \ref{flatnessintrinsic} below. Before stating this result and proving it, we elucidate a general fact:

\begin{lemma}\label{lemmaintrinsic2}
Let $M$ be a smooth manifold equipped with a $C^{\infty}(M)$-bilinear product $\circ$ on sections of its tangent bundle $\circ: TM\times TM \rightarrow TM$ and with a torsionless affine connection $\nabla$. Suppose $\circ$ is equipped with an unit vector field $e$ and that $\nabla$ and $\circ$ satisfy the following condition:
\beq\label{Rc+Rc+Rc2}
Z\circ R(W,Y)(X)+W \circ R(Y,Z)(X)+Y \circ R(Z,W)(X)=0, 
\eeq
where $R(X,Y):=\nabla_X\nabla_Y-\nabla_Y\nabla_X -\nabla_{[X,Y]}$ for all vector fields $X, Y, Z, W$. Then $\nabla$ is flat if and only if $R(e, W)=0$ for all vector fields $W$. 
\end{lemma}
\emph{Proof}
If $\nabla$ is flat, certainly $R(e, W)=0$ for all vector fields $W$. Conversely, suppose $R(e,W)=0$ for all vector fields. Then substituting $Z:=e$ in \eqref{Rc+Rc+Rc2} we get immediately 
$e\circ R(W, Y)(X)=0$, i.e. $R(W, Y)(X)=0$ for all vector fields $W, Y, X$, and we are done.\begin{flushright}
$\Box$
\end{flushright}

Observe that the condition \eqref{Rc+Rc+Rc2} appearing in the previous Lemma is exactly one of the conditions that define an $F$-manifold with compatible connection. However, there is no need for the product $\circ$ to be commutative, associative or semisimple, neither for the other conditions defining an $F$-manifold with compatible connection to be satisfied. 

\begin{thm}\label{flatnessintrinsic}
A semisimple $F$-manifold with compatible connection $\nabla$ (see Section 2) and flat unity $e$ is flat if and only if the operator ${\rm Lie}_e$ and the covariant derivative $\nabla$ satisfy the following condition:
\beq\label{flatness2}
{\rm Lie}_e (\nabla_X T)-\nabla_X ({\rm Lie}_e T)-\nabla_{[e, X]}T=0,
\eeq
for any vector field $X$ and for any tensor field $T$.
\end{thm}
\emph{Proof}
Since the unity $e$ is assumed to be a flat vector field, we have that ${\rm Lie}_e=\nabla_e$  and therefore the condition \eqref{flatness2} is equivalent to $R(e,X)(T)=0$. Now by Lemma \ref{lemmaintrinsic2} we know that $\nabla$ is flat if and only if $R(e,X)=0$ for all vector fields $X$.
\begin{flushright}
$\Box$
\end{flushright}
Using the fact that $X$ is an arbitrary vector field, we have the following Lemma

\begin{lemma}\label{flatness3lemma} Condition \eqref{flatness2} is equivalent to 
\beq\label{flatness3}{\rm Lie}_e(\nabla T)-\nabla ({\rm Lie}_eT)=0,\eeq
 for any tensor fied $T$. 
 \end{lemma} 
 \emph{Proof}
 Observe that we can write $\nabla_X T=(\nabla T)(X)=C(\nabla T\otimes X)$ for any vector field $X$, where $C$ is the contraction. 
 Therefore using the property that ${\rm Lie}_e$ commutes with contractions and it satisfies Leibniz rule with respect to the the tensor product we have $$[{\rm Lie}_e(\nabla T)-\nabla ({\rm Lie}_eT)](X)={\rm Lie}_e((\nabla T)(X))-\nabla T({\rm Lie}_e X)-\nabla_X({\rm Lie}_e T)=$$
 $$={\rm Lie}_e(\nabla_X T)-\nabla T([e, X])-\nabla_X({\rm Lie}_e T)={\rm Lie}_e(\nabla_X T)-\nabla_X({\rm Lie}_e T)-\nabla_{[e,X]}T.$$
 \begin{flushright}
$\Box$
\end{flushright}

 Observe that in the proof of Theorem \ref{flatnessintrinsic} no use has been made of two conditions, namely the semisimplicity of $\circ$ and the symmetry of $\nabla c$. The only hypotheses that were used are the presence of a flat identity $e$ for the product $\circ$ and condition \eqref{Rc+Rc+Rc2} for the torsionless connection $\nabla$.  

\begin{rmk}Let us observe that relation \eqref{flatness3} is reminiscent of the commutation relation between the Lie derivative with respect to $e$ and the differential $d_P$ associated to a Poisson structure $P$, when $P$ is part of an {\em exact} pencil of Poisson structures $Q-\lambda P$ (for more details see \cite{ALinher} and \cite{FL}).
\end{rmk}

\begin{rmk}
The flatness of $e$ and the condition \eqref{flatness3}, in general, do not imply the flatness of $\nabla$. Indeed the condition
 \eqref{flatness3} written for an arbitrary vector field $T$ reads
$$(\nabla_j\nabla_l e^i-R^i_{jkl}e^k)T^l=0.$$ 
\end{rmk}

\section{The semisimple case}

\subsection{Multi-flatness conditions in the semisimple case}
We apply now the flatness criterion discussed in the previous Section to study multi-flat structures in the semisimple case. As a consequence of the previous result we have the following

\begin{thm}\label{corollaryflatness}
A semisimple $F$-manifold with compatible connection $\nabla$ (see Section 2) and flat unity $e$ is flat if and only $e(\Gamma^i_{ij})=0$ for all $i\neq j$, where $\Gamma^i_{ij}$ are the Christoffel symbols of $\nabla$ in the canonical coordinates of $\circ$.
\end{thm}

\emph{Proof}
Under the current hypotheses, $\nabla$ is flat if and only if \eqref{flatness3} holds for an arbitrary tensor field $T$. However, notice that \eqref{flatness3} is automatically satisfied when $T$ is a function since covariant and Lie derivatives coincide on functions.
%$${\rm Lie}_e(\nabla_i T)-\nabla_i({\rm Lie}_eT)=e^l\d_l\d_i T+\d_l e^l \d_lT-\d_i(e^l\d_l T)=0.$$
 Moreover, the operators ${\rm Lie}_e$ and $\nabla_{i}$ commute with contractions and satisfy Leibniz rule with respect to tensor products. This easily implies that  \eqref{flatness3} holds for an arbitrary tensor fields $T$ if and only if it holds for an arbitrary {\em vector} field $T.$
Writing the right hand side of \eqref{flatness3} in canonical coordinates of $\circ$, for  $T$ an arbitrary vector field, we get
$$e\left(\d_j T^i+\Gamma^i_{jk}T^k\right)-\d_j(e(T^i))-\Gamma^i_{jk}e(T^k)=e(\Gamma^i_{jk})T^k,$$
since $e$ commutes with $\d_j$ in canonical coordinates. Therefore \eqref{flatness3} is fulfilled if and only if $e(\Gamma^i_{jk})=0$, due to the arbitrariness of $T$. On the other hand, for a natural connection in canonical coordinates one already has $\Gamma^i_{jk}=0$ $i\neq j\neq k\neq i$, while all the other non-vanishing components are expressed as linear combinations with constant coefficients of $\Gamma^i_{ij}$, $i\neq j$ (see formula \eqref{naturalc}).
%Since the unity $e$ is assumed to be a flat vector field, then by Lemma \ref{lemmaintrinsic1}, the condition \eqref{flatness2} is equivalent to $R(e,X)(T)=0$. Now by Lemma \ref{lemmaintrinsic2} we know that $\nabla$ is flat if and only if $R(e,X)=0$ for all vector fields $X$.
\begin{flushright}
$\Box$
\end{flushright}

It is clear from the proof of the previous Corollary, that the flatness of $\nabla$ is equivalent to $e(\Gamma^i_{jk})=0$ in canonical coordinates for $\circ$ {\em without} assuming the symmetry of $\nabla c$, so without using all the defining conditions for an $F$-manifold with compatible connection. That's because the symmetry of $\nabla c$ is the condition that forces $\Gamma^i_{jk}=0$ $i\neq j\neq k\neq i$ in canonical coordinates for $\circ$.

We have been stating results for $F$-manifolds with compatible connections and not under weaker assumptions, since  for our purposes we are interested in using the Tsarev's condition \eqref{rt1}. Now it turns out that Tsarev's condition is equivalent to $Z\circ R(W,Y)(X)+W \circ R(Y,Z)(X)+Y \circ R(Z,W)(X)=0$ {\em and} the symmetry of $\nabla c$, both expressed in the distinguished coordinates system given by canonical coordinates for $\circ.$
If follows also from the Corollary \ref{corollaryflatness} that, in canonical coordinates, all Christoffel symbols of $\nabla$ depend only on the differences $(u^i-u^j)$ of canonical coordinates.

Obviously, the flatness criterion provided by relation \eqref{flatness3} and its equivalent forms can be applied to the case of connections associated to general eventual identities.

It is easy to check that a symmetric connection $\nabla$
 compatible with the dual product defined by $E$ and satisfying the condition $\nabla E=0$ has Christoffel symbols (in canonical coordinates for $\circ$) of the form:
 \begin{equation}\label{dualgamma}
\begin{split}
\Gamma^{i}_{jk}&:=0,\qquad\forall i\ne j\ne k \ne i,\\
\Gamma^{i}_{jj}&:=-\f{E^i}{E^j}\Gamma^{i}_{ij},\qquad i\ne j,\\
\Gamma^{i}_{ii}&:=-\sum_{l\ne i}\f{E^l}{E^i}\Gamma^{i}_{li}-\f{\d_i E^i}{E^i}.
\end{split}
\end{equation}

Given an eventual identity $E$ with associated dual product $*$, it is useful to have relations characterizing the flatness of the connection given by \eqref{dualgamma} in the canonical coordinate for $\circ$. In this case $E$ does not reduce anymore to $e$. This characterization is provided by the following:

\begin{thm}\label{flatnessgeneralth}
Suppose that the functions $\Gamma_{ij}^i$ satisfy  Tsarev's conditions \eqref{rt1}, then in canonical coordinates for $\circ$, the symmetric connection \eqref{dualgamma} is flat if and only if
$$E(\Gamma^i_{ij})=-(\d_j E^j)\Gamma^i_{ij}, \qquad  i\ne j.$$
\end{thm}

\emph{Proof}. We use the invariant condition \eqref{flatness3}, expressed in canonical coordinates for $\circ$, where we choose as $T$ a vector field. In this case we get
\begin{eqnarray*}
[{\rm Lie}_E (\nabla T)]^i_j-[\nabla ({\rm Lie}_E T)]^i_j=0.
\end{eqnarray*}
Expanding this we get for $i\neq j$:
\begin{eqnarray*}
&&(E(\Gamma^i_{ij})+\Gamma^i_{ij}\d_j E^j)T^i+(E(\Gamma^i_{jj})-\Gamma^i_{jj}\d_i E^i+\Gamma^i_{jj}\d_j E^j+\Gamma^i_{jj}\d_jE^j)T^j=0,\\
\end{eqnarray*}
while for $i=j$ we obtain:
\begin{eqnarray*}
&&(E(\Gamma^i_{ii})+\d^2_iE^i +\Gamma^i_{ii}\d_iE^i)T^i+\sum_{l\ne i}(E(\Gamma^i_{il})+\Gamma^i_{il}\d_lE^l)T^l=0.
\end{eqnarray*}
Thus for the condition \eqref{flatness3} to be fulfilled we have that the following constraints have to be satisfied:
\begin{eqnarray*}
E(\Gamma^i_{ij})&=&-\Gamma^i_{ij}\d_j E^j,\\
E(\Gamma^i_{jj})&=&\Gamma^i_{jj}\d_i E^i-\Gamma^i_{jj}\d_j E^j-\Gamma^i_{jj}\d_jE^j,\\
E(\Gamma^i_{ii})&=&-\d^2_iE^i -\Gamma^i_{ii}\d_iE^i.
\end{eqnarray*}

The first condition is the statement of the Theorem. The second and third one follow using the first one, the defining relations of the natural connection and the obvious identities $E(E^i)=E^i\d_i E^i, \, E(\d_i E^i)=E^i\d_i^2 E^i$:
\begin{eqnarray*}
E(\Gamma^i_{jj})&=&E\left(-\f{E^i}{E^j}\Gamma^i_{ij}\right)=-\frac{E(E^i)}{E^j}\Gamma^i_{ij}+\frac{E^i}{E^j}\Gamma^i_{ij}\d_j E^j+\frac{E^i}{(E^j)^2}\Gamma^i_{ij}E(E^j)=\\
 &=&\Gamma^i_{jj}\d_i E^i-\Gamma^i_{jj}\d_j E^j-\Gamma^i_{jj}\d_jE^j,\\
E(\Gamma^i_{ii})&=&E\left(-\sum_{l\ne i}\f{E^l}{E^i}\Gamma^{i}_{il}-\f{\d_i E^i}{E^i}\right)=
     \sum_{l\neq i}\frac{E^l \Gamma^i_{il}E^i \d_i}{E^i}-\frac{E(\d_i E^i)}{E^i}+\frac{(\d_i E^i)^2}{E^i}=\\
& =&-\d^2_iE^i -\Gamma^i_{ii}\d_iE^i.
\end{eqnarray*}

\begin{flushright}
$\Box$
\end{flushright}

\begin{rmk}
In the case of semisimple Frobenius manifolds,  in canonical coordinates $\{u^1, \dots, u^n\}$ for $\circ$ we have that $E^i=u^i$ for all $i=1,\dots, n$, and $\tilde{g}^{ii}=u^ig^{ii}=u^i\d_i\varphi$. This implies that
\begin{eqnarray*}
\Gamma^i_{jk}&=&\tilde\Gamma^i_{jk}=0,\qquad\forall\; i\ne j\ne k\ne i,\\
\tilde\Gamma^i_{ji}&=&\tilde{g}^{ii}(\d_j\tilde{g}_{ii})=\Gamma^i_{ji},\qquad\forall\; i\ne j,\\
\tilde\Gamma^i_{jj}&=&\tilde{g}^{ii}(-\d_i\tilde{g}_{jj})=\f{u^i}{u^j}g^{ii}(-\d_ig_{jj})=\f{u^i}{u^j}g^{ii}(-\d_jg_{ii})
=-\f{u^i}{u^j}\Gamma^i_{ij}=-\f{u^i}{u^j}\tilde\Gamma^i_{ij},\qquad\forall\; i\ne j.\\
\end{eqnarray*}
Moreover, due to the homogeneity property, it is easy to check that $E(\Gamma^i_{ji})=-\Gamma^i_{ji}$. Due to the previous 
 theorem this means that the connection compatible with the dual product is flat. However this connection does not necessarily coincide
 with the Levi-Civita connection of the intersection form $\tilde{g}$ since, in general,
$$\tilde\Gamma^{i}_{ii}\ne-\sum_{l\ne i}\f{u^l}{u^i}\tilde\Gamma^{i}_{li}-\f{1}{u^i}.$$
\end{rmk}

\subsection{Non existence of semisimple $F$-manifolds with more than $3$ compatible connections}
We are going to apply Theorem \ref{flatnessgeneralth} to study the existence of multi-flat structures on $F$-manifolds.
Recall that, by definition, given an $N$-multi flat (semisimple) manifold, the $N$ connections $\nabla^{(l)}$, $l=0, \dots, N-1$ share the same Christoffel symbols $\Gamma^i_{ij}$, $i\neq j$ (say in the canonical coordinates for $\circ=\circ_{(0)}$), while the remaining ones are determined according to the formulas \eqref{dualgamma}, where $E$ is the corresponding eventual identity $E_{(l)}$.
Therefore, given the $E_{(l)}$, $l=0,\dots N-1$, it is possible to reconstruct $N$-multi-flat connections only if the system for $\Gamma^i_{ij}$ ($j$ is fixed):
\beq\label{orsysbis}
E_{(l)}(\Gamma^i_{ij})+(\d_j E^j_{(l)})\Gamma^i_{ij}=0, \quad l=0, \dots N-1
\eeq
admits non-trivial solutions $\Gamma^i_{ij}$ for all $i\neq j$. Indeed, \eqref{orsysbis} is just the flatness condition of Theorem \ref{flatnessgeneralth}.
It is possible to reduce the non-homogenous system \eqref{orsysbis} to a homogenous one. To do this we introduce a fictitious additional variable $u^{n+1}$ and assume that $\Gamma^i_{ij}$ is defined implicitly via $\phi(u^1, \dots , u^n, u^{n+1})=c$ where $c$ a constant and $u^{n+1}=\Gamma^i_{ij}(u^1, \dots, u^n)$. In this case the system \eqref{orsysbis} becomes
\beq\label{orsystris}
\hat{E}_{(l)}(\phi):=E_{(l)}(\phi)-(\d_j E^j_{(l)})u^{n+1}\d_{n+1}\phi=0, \quad l=0, \dots, N-1.
\eeq
In this way, determining $\phi$ can be interpreted as the problem of finding invariant functions for the distribution $\Delta$ generated by the vector fields $\{\hat{E}_{(l)}\}_{l=0, \dots, N-1}.$

Therefore we are interested in characterizing the integrable distributions generated by the extended vector fields $\hat{E}_{(l)}, l=,0, \dots N-1$, where by definition of multi-flat $F$-manifold the vector fields $E_{(l)}:=(u^1)^{l}\d_1+...+(u^n)^{l}\d_n$, $l=0, \dots, N-1.$

\begin{thm}\label{distri1}
Let $\Delta_{(i_1, \dots, i_k)}$ be the distribution spanned by the vector fields $\hat{E}_{(i_1)},\dots,  \hat{E}_{(i_k)}$ in the $n+1$-dimensional space with coordinates $(u^1, \dots, u^n, u^{n+1}).$  Then:
\begin{enumerate}
\item The distributions $\Delta_{(1,m)}$  with $m\in \mathbb{Z}\setminus\{1\}$ are integrable and these are the only integrable distribution of rank $2$ among $\Delta_{(i_1, i_2)}.$ 
\item  $\Delta_{(0,1,2)}$ is integrable.
\item  $\Delta_{(0,1,2,3)}$  is not integrable. Furthermore, at the points where $u^i\ne u^k$ ($i\ne k,i,k=1,...,n$)
 and $u^{n+1}\ne 0$ it is totally non-holonomic, that is the minimal integrable distribution $\bar{\Delta}$ containing $\Delta_{(0,1,2,3)}$ has dimension $n+1$.  
\item More in general $\Delta_{(i_1, \dots, i_k)}$, with $i_1<i_2<\dots<i_k$ is not integrable for $ 4\leq k \leq n$. 
\end{enumerate}
\end{thm}

\emph{Proof}. We have
\beq
[\hat{E}_{(l)},\hat{E}_{(m)}]^i=\left\{ \begin{array}{ll} (m-l)(u^i)^{l+m-1} & \textrm{if $i=1,...,n$} \\
-(m-l)(m+l-1)(u^j)^{m+l-2}u^{n+1}  & \textrm{if $i=n+1$}
\end{array}\right.
\eeq
that is
$$[\hat{E}_{(l)},\hat{E}_{(m)}]=(m-l)\hat{E}_{(m+l-1)}, \quad l\neq m,$$
$$[\hat{E}_{(l)}, \hat{E}_{(m)}]=0, \quad l=m.$$

Since $[\hat{E}_{(m)},\hat{E}_{(1)}]=\hat{E}_{(m)}$, the distribution  $\Delta_{(m,1)}$  is integrable. Moreover, any other distribution of rank $2$, $\Delta_{(i_1, i_2)}$ is not integrable since $[\hat{E}_{i_1}, \hat{E}_{i_2}]=(i_2-i_1)\hat{E}_{(i_1+i_2-1)},$  and $i_1+i_2-1=i_1$ or $i_1+i_2-1=i_2$ implies either $i_2=1$ or $i_1=1.$

Since  $\hat{E}_{(0)}$, $\hat{E}_{(1)}$ and $\hat{E}_{(2)}$ satisfy the commutation relations of $sl(2,\mathbb{C})$:  $[\hat{E}_{(0)},\hat{E}_{(1)}]=\hat{E}_{(0)},$  $[\hat{E}_{(0)},\hat{E}_{(2)}]=2\hat{E}_{(1)}$ and  $[\hat{E}_{(1)},\hat{E}_{(2)}]=\hat{E}_{(2)}$, we have that also the distribution  $\Delta_{(0,1,2)}$  is integrable.

With regard to the fourth point, consider $\Delta_{(i_1,\dots, i_k)}$, with $i_1<\dots<i_k$ and $4\leq k\leq n.$
If the two indices $i_{1}, i_{2}$ are both strictly negative or if $i_1<0$ and $i_2=0$, then $[\hat{E}_{(i_1)}, \hat{E}_{(i_2)}]\notin\Delta_{(i_1,\dots, i_k)}$, due to the commutation relations. Thus we can assume $i_1\geq 0$ and the indices $i_{k-1}, i_k$ strictly greater than $1$. Therefore again we have $[\hat{E}_{(i_{k-1})}, \hat{E}_{(i_k)}]=(i_k-i_{k-1})\hat{E}_{(i_k+i_{k-1}-1)}\notin \Delta_{(i_1,\dots, i_k)}$, since $i_k+i_{k-1}-1>i_k$.

Finally, it remains to prove the third point. By the fourth point, the distribution  $\Delta_{(0,1,2,3)}$ is not integrable. 

Before determining the minimal integrable distribution containing $\Delta_{(0,1,2,3)}$ we recall few definitions and a fundamental result. Given a collection of vector fields $\{\hat{E}_{(l)}\}_{l\in L}$ their Lie hull is the collection of all vector fields of the form $\{\hat{E}_{(l)}, [\hat{E}_{(l)}, $
$\hat{E}_{(m)}],  [\hat{E}_{(n)},[\hat{E}_{(l)}, \hat{E}_{(m)}]], \dots\}$ generated by the iterated Lie brackets. 
The minimal integrable distribution containing $\Delta_{(i_1, \dots, i_k)}$ is the minimal integrable distribution containing the Lie hull of the vector fields $\{\hat{E}_{(i_1)}, \dots \hat{E}_{(i_k)}\}.$
The distribution $\Delta_{(i_1, \dots, i_k)}$ (or equivalently the associated collection of vector fields) is called bracket generating if its Lie hull spans the whole tangent bundle in an open set. In this case the minimal integrable distribution containing $\Delta_{(i_1, \dots, i_k)}$ has integral leaf equal to the entire $n+1$-dimensional space (this is the Chow-Rashevsky Theorem, see \cite{C,R}). 

We apply this result to compute the minimal integrable distribution containing $\Delta_{(0,1,2,3)}$ and we show that it is the $n+1$-dimensional space. 

In order to compute the minimal integrable distribution containing $\Delta_{(0,1,2,3)}$, we consider the sub-bundle of the tangent bundle spanned by $\hat{E}_{(0)},\hat{E}_{(1)},\hat{E}_{(2)},\hat{E}_{(3)},\hat{E}_{(m)}=\frac{1}{m-2}[\hat{E}_{(2)},\hat{E}_{(m-1)}],\,m=4,5,...,n$. To show that its rank is $n+1$, it is sufficient to show that the determinant of the $A$ matrix does not vanish on an open set, where 
$$A:=\begin{pmatrix}
1 & \dots & 1  & 0\\
u^1 & \dots & u^n & -u^{n+1} \\
\vdots & \ddots & \vdots &\vdots\\
(u^1)^{n} & \dots & (u^n)^{n} & -n(u^j)^{n-1}u^{n+1}
\end{pmatrix}.$$
The matrix $A$ can be written as
$$A=
\begin{pmatrix}
1 & \dots & 1 & -u^{n+1}\f{\d}{\d u^{n+1}}_{|u^{n+1}=u^j}1\\
u^1 & \dots & u^n &-u^{n+1}\f{\d}{\d u^{n+1}}_{|u^{n+1}=u^j}(u^{n+1}) \\
\vdots & \ddots & \vdots &\vdots\\
(u^1)^{n} & \dots & (u^n)^{n} &-u^{n+1}\f{\d}{\d u^{n+1}}_{|u^{n+1}=u^j}(u^{n+1})^{n}
\end{pmatrix}.
$$

Expanding the determinant of $A$ along the last column, we get
$$\det(A)=\sum_{k=0}^{n} A_k\left(-u^{n+1}\f{\d}{\d u^{n+1}}_{|u^{n+1}=u^j}(u^{n+1})^{k}\right),$$
where $A_k$ are the corresponding minors. Since the $A_k$'s do not depend on $u^{n+1}$ we can factor the derivative operator in front of the expansion and get:
$$\det(A)=-u^{n+1}\f{\d}{\d u^{n+1}}_{|u^{n+1}=u^j}\left(\sum_{k=0}^{n} A_k(u^{n+1})^{k} \right)=-u^{n+1}\f{\d}{\d u^{n+1}}_{|u^{n+1}=u^j}\det(V_{0,\dots, n}),$$
where $V_{0,\dots, n}$ is the Vandermonde matrix. By the form of the Vandermonde determinant, it is clear that $\det(A)\neq 0$ in the open subset $\Omega:=\{(u^1, \dots, u^n , u^{n+1})\; | \;u^i\ne u^k (i\ne k,i,k=1,...,n),  u^{n+1}\ne 0\}.$

\begin{flushright}
$\Box$
\end{flushright}

\begin{rmk}
Notice that the extended vector fields $Z_{(l)}:=\hat{E}_{(l+1)}$ satisfy the commutation relation
$$[Z_{(l)},Z_{(m)}]=[\hat{E}_{(l+1)},\hat{E}_{(m+1)}]=(m-l)\hat{E}_{(m+l+1)}=(m-l)Z_{(m+l)},$$
of the centerless Virasoro algebra. 
\end{rmk}

Theorem \ref{distri1} shows that in general multi-flat $F$-structures with more than three distinct products can not exist. Indeed, if a distribution $\Delta_{(i_1, \dots, i_k)}$  is totally non-holonomic, then the only solutions of the system \eqref{orsystris} are given by functions $\phi$ that are constant everywhere and therefore they give rise to trivial Christoffel symbols $\Gamma^i_{ij}$. Below we point out that instead multi-Hamiltonian structures are allowed also in the case $N>3$.

\subsection{Multi-flat structures vs multi-Hamiltonian structures}\label{multiflatvsHamiltonian}

Unlike the multi-flat case we have been analyzing, it is possible to have multi-Hamiltonian structures encompassing more than three structures. An example is given by
the following $n+1$ metrics introduced in \cite{FP}:
\beq
g_{ii}^{(\alpha)}=\f{\prod_{k\ne i}(u^k-u^i)}{(u^i)^{\alpha}},\qquad\alpha=0,...,n.
\eeq
They are flat and thus their inverses define $n+1$ Hamiltonian structures of hydrodynamic type that turns out to be compatible among each other. The corresponding Levi-Civita connections are defined by
\begin{equation}
\begin{split}
\Gamma^i_{ij}&=\d_j\ln{\sqrt{g^{(\alpha)}_{ii}}}=-\f{1}{2}\f{1}{u^i-u^j},\\
\Gamma^{i}_{jk}&=0,\qquad\forall i\ne j\ne k \ne i,\\
\Gamma^{i}_{jj}&=-\f{1}{2}\f{\d_ig^{(\alpha)}_{jj}}{g^{(\alpha)}_{ii}},\\
\Gamma^{i}_{ii}&=\d_i\ln{\sqrt{g^{(\alpha)}_{ii}}}.
\end{split}
\end{equation}
Observe that the Christoffel symbols $\Gamma^i_{ij}$ coincide with the Christoffel symbols $\Gamma^i_{ij}$ for the $\epsilon$-system in the case $\epsilon=-\f{1}{2}$.  It is also immediate to check directly that
 \begin{eqnarray*}
E_{(0)}(\Gamma^{i}_{ij})&=&[\d_1+\d_2+\dots+\d_n]\Gamma^{i}_{ij}=0,\\
E_{(1)}(\Gamma^{i}_{ij})&=&[u^1\d_1+u^2\d_2+\dots+u^n\d_n]\Gamma^{i}_{ij}=- \Gamma^{i}_{ij},
\end{eqnarray*}
while
$$E_{(k)}(\Gamma^{i}_{ij})=[(u^1)^{k}\d_1+(u^2)^k\d_2+\dots+(u^n)^{k}\d_n]\Gamma^{i}_{ij}\ne-k(u^j)^{k-1} \Gamma^{i}_{ij}$$
for $k\ge 2$. This means that among the natural connections  
\begin{equation}
\begin{split}
\Gamma^i_{ij}&=-\f{1}{2}\f{1}{u^i-u^j},\\
\Gamma^{i}_{jk}&=0, \qquad\forall i\ne j\ne k \ne i,\\
\Gamma^{i}_{jj}&=\f{1}{2}\f{E_{(k)}^i}{E_{(k)}^j}\f{1}{u^i-u^j}, \qquad i\ne j,\\
\Gamma^{i}_{ii}&=\f{1}{2}\sum_{l\ne i}\f{E_{(k)}^l}{E_{(k)}^i}\f{1}{u^i-u^l}-\f{\d_i E_{(k)}^i}{E_{(k)}^i},
\end{split}
\end{equation}
 only those compatible with the eventual identities $E_{(0)}$ and $E_{(1)}$ are flat. 

\subsection{A comment on more general eventual identities}
Here we drop the assumption $ E^k_{(m)}=(u^k)^m$ and we consider the case of two vector fields  $\hat{E}_{(l)}, \hat{E}_{(m)}$ and see when the distribution they span is integrable. We will see that we obtain again the same results we proved  assuming $E^k_{(m)}=(u^k)^m.$ 
To this end we compute
\begin{eqnarray}\label{comgen1}
&&[\hat{E}_{(l)}, \hat{E}_{(m)}]^k=E^k_{(l)}\d_k E^k_{(m)}-E^k_{(m)}\d_k E^k_{(l)}, \quad k=1, \dots n\\\label{comgen2}
&&[\hat{E}_{(l)}, \hat{E}_{(m)}]^{n+1}=-\left(E^j_{(l)}\d^2_j E^j_{(m)}-E^j_{(m)}\d^2_j E^j_{(l)} \right) u^{n+1}.
\end{eqnarray}
In the open set where $E^i_{(l)}(u^i)\neq 0$, $i=1, \dots n$, we can introduce a change of variables of the form $\frac{d\tilde u^i}{d u^i}=\frac{1}{E^i_{(l)}(u^i)}$, $i=1, \dots n$, so that in the new coordinates $\tilde u^i$ $E_{(l)}$ becomes ${\tilde E}^i_{(l)}=\frac{d\tilde u^i}{d u^i}E^i_{(l)}=1$ for each $i=1,\dots n$. Since the vector field commutators above do not depend on the coordinate system, we can use them also in the coordinate system $\tilde u^i$ (anyway we drop the notation $\tilde E_{(l)}$ and so on to simplify the notation).
So without loss of generality on a an open set, we can assume $E_{(l)}=e=(1, \dots, 1)$ while $E_{(m)}=E=(E^1(u^1),\dots,  E^n(u^n))$ is arbitrary. 
Substituting this information in the computation of vector field commutators above, we get
$[\hat{E}_{(0)}, \hat{E}_{(m)}]^i=\d_i E^i$, $i=1, \dots n$ and $[\hat{E}_{(0)}, \hat{E}_{(m)}]^{n+1}=-\d_j^2E^ju^{n+1}.$ Imposing that the distribution $\Delta$ spanned by $\hat{E}_{(0)}, \hat{E}_{(m)}$ is integrable, i.e. $[\hat{E}_{(0)}, \hat{E}_{(m)}]=\alpha \hat{E}_{(0)}+\beta \hat{E}_{(m)}$ for some functions $\alpha(u^1, \dots, u^n, u^{n+1}), \beta(u^1, \dots, u^n, u^{n+1})$ we get:
\begin{eqnarray}\label{biflatsystem}
&&\d_i E^i=\alpha +\beta E^i, \quad i=1, \dots, n\\ 
\label{biflatsystem2}
&&\d_j^2 E^j=\beta\d_j E^j, \quad \text{ for fixed }j. 
\end{eqnarray}
\begin{lemma}
Suppose that $\d_j E^j\ne 0$, then the functions $\alpha$ and $\beta$ in the system above are necessarily constants. 
\end{lemma}
\emph{Proof}
%The general solution of  \eqref{biflatsystem2} for arbitrary function $\beta$ is given by 
%$$E^j_{(m)}=c_1+c_2\int^{u^j}\exp\left(\int^t \beta(u^1, \dots,u^{j-1}, s, u^{j+1}, \dots, u^{n+1})\, ds\right)\, dt.$$
%But imposing $\d_k E^j_{(m)}=0$ for $k\neq j,$ $k=1, \dots n+1$ we get that 
%$$c_2\int^{u^j}\exp\left(\int^t \beta(u^1, \dots,u^{j-1}, s, u^{j+1}, \dots, u^{n+1})\, ds\right)\left(\int^t\frac{\d \beta}{\d %u^k}\, ds\right) \, dt=0,$$
From \eqref{biflatsystem2} it follows immediately that $\beta=\beta(u^j)$. Now we turn to \eqref{biflatsystem} with $i=j$ so we have
$$\d_j E^j=\alpha(u^1, \dots u^{n+1}) +\beta(u^j) E^j.$$
Taking derivative with respect to $u^k$, $k\neq j$ we get that $\alpha=\alpha(u^j)$ only. 
%The general solution is given by 
%$$E^j=\left(\int^{u^j}\alpha(u^1, \dots, u^{j-1}, t, u^{j+1}, \dots, u^{n+1})\exp\left(-\int^t %\beta(s)\, ds \right)\, dt+c_1\right)\exp\left(\int^{u^j}\beta(s)\, ds \right).$$
%Now imposing that $\d_k E^j=0$ for $k\neq j$, $k=1, \dots, n+1$ we get that 
%$$\left(\int^{u^j}\frac{\d\alpha}{\d u^k}\exp\left(-\int^t \beta(s)\, ds \right)\, dt\right)\exp\left(\int^{u^j}\beta(s)\, ds \right)=0,$$
%and thus $\alpha=\alpha(u^j)$ only. 
Finally we look at \eqref{biflatsystem} with $i\neq j$. For this we have
$$ \d_i E^i(u^i)=\alpha(u^j) +\beta(u^j) E^i(u^i).$$
If $\beta=0$, then $\alpha(u_j)$ has to a be a constant $k$ and this gives $E^i=ku^i,$
 for all $i=1, \dots n$. If $\beta\neq 0$, then taking again derivative with respect to $u^i$ we get
$$\d_i^2 E^i(u^i)=\beta(u^j)\d_i E^i(u^i).$$
Assuming $\d_i E^i(u^i)\neq 0$ this gives that $\beta$ has to be a constant $k$. Going back to $\d_i E^i(u^i)=\alpha(u^j) +k E^i(u^i)$ and taking derivative with respect to $u^j$ this shows that $\alpha$ is also a constant. 
\begin{flushright}
$\Box$
\end{flushright}

Now we classify eventual identities $\{e, E\}$ related to bi-flat structures. 
\begin{thm}
Consider the bi-flat structures $\{e, E\}$. Then $E$ is either given by $\alpha (u^1, \dots, u^n)$ with $\alpha$ non-zero constant or $E=(c_1 e^{\beta u^1}-\frac{\alpha}{\beta}, \dots, c_n e^{\beta u^n}-\frac{\alpha}{\beta}).$
\end{thm}
\emph{Proof}
Since $\alpha$ and $\beta$ are constants, then equation \eqref{biflatsystem2} is just a consequence of equation \eqref{biflatsystem} for $i=j$. Therefore it is enough to classify all the solutions of \eqref{biflatsystem}. 
If $\beta=0$ and $\alpha\neq 0$, then $E^i=\alpha u^i$, $i=1, \dots n$, while 
if $\beta\neq 0$ then $E^i=c_i e^{\beta u^i}-\frac{\alpha}{\beta}$, $i=1, \dots, n$ (where the constants $c_1^i$ in general might be chosen to depend on $i$). 

\begin{flushright}
$\Box$
\end{flushright}

Without loss of generality we can assume $\alpha=1$ in the first case and $\alpha=0$ in the second case (the distribution does not change). Moreover assuming $c_i \ne 0,\,\, \forall i$ we can reduce the second case to the first one with a simple change of coordinates:
 $\tilde{u}^i=-\frac{1}{\beta c_i}e^{-\beta u^i}$.

\section{Bi-flat $F$-manifolds}\label{biflatsec}
Bi-flat $F$-manifolds were introduced in \cite{AL} and further studied and classified in \cite{L2014}. In this Section we present the classification in dimension two and three, using  Tsarev's conditions instead of a generalized Darboux-Egorov system. Due to the results of the previous section semisimple bi-flat $F$-manifolds are parametrized by the solutions of the system
\begin{eqnarray}
\label{BF1}
&&\d_k\Gamma^i_{ij}=-\Gamma^i_{ij}\Gamma^i_{ik}+\Gamma^i_{ij}\Gamma^j_{jk}
+\Gamma^i_{ik}\Gamma^k_{kj}, \quad i\ne k\ne j\ne i,\\
\label{BF2}
&&E_{(0)}(\Gamma^i_{ij})=0,\qquad i\ne j\\
\label{BF3}
&&E_{(1)}(\Gamma^i_{ij})=-\Gamma^i_{ij},\qquad i\ne  j
\end{eqnarray}
where $E_{(0)}=\sum_{i=1}^n\d_i$ and  $E_{(1)}=\sum_{i=1}^nu^i\d_i$. It is possible to prove (see Appendix 1 for details) that the above system is compatible and thus its general solution depends
 on $n(n-1)$ arbitrary constants.

\subsection{Two dimensional bi-flat $F$-manifolds}

For $n=2$ Tsarev's conditions \eqref{BF1} are empty. The general solution of the remaining conditions \eqref{BF2} and \eqref{BF3} depends on two arbitrary constants $\epsilon_1$ and $\epsilon_2$. It coincides with the  two-component generalized $\epsilon$-system (see \cite{L2014}):
$$\Gamma^{i}_{ij}=\f{\epsilon_j}{u^i-u^j},\qquad i\ne j.$$ 

\subsection{Three-dimensional bi-flat $F$-manifolds }

Tridimensional  bi-flat $F$-manifolds are parametrized by solutions of Painlev\'e VI  equation \cite{AL,L2014}. This result has been obtained reducing a generalized version of the Darboux-Egorov system 
 for the rotation coefficients $\beta_{ij}$ to a sistem of ODEs equivalent to the sigma form of Painlev\'e VI. Given a solution of  Painlev\'e VI, the natural connection is defined as
\beq\label{gammabeta}
\Gamma^i_{ij}=\f{H_j}{H_i}\beta_{ij},
\eeq
where $\beta_{ij}$ is the corresponding solution of the generalized  Darboux-Egorov system and the function $H_i$ are the  Lam\'e coefficients satisfying the further conditions
 $e(H_i)=0$ and $E(H_i)=d_iH_i$, (see \cite{AL,L2014} for details).

In this Section, we follow a different approach, based on the study of the sytem (\ref{BF1},\ref{BF2},\ref{BF3}). 
In particular we show that this system is equivalent to a system of six first order ODEs admitting  $4$ independent first integrals.
  Moreover we provide an explicit relation between the solutions of this system and the solutions of the generic Painlev\'e VI equation.  
 The value of the $4$ parameters of the  Painlev\'e VI equation is related to the value of the first integrals of the system.

 As a first step we have to solve the system 
\begin{eqnarray*}
E_{(0)}(\Gamma^{i}_{ij})&=&[\d_1+\d_2+\d_3]\Gamma^{i}_{ij}=0,\\
E_{(1)}(\Gamma^{i}_{ij})&=&[u^1\d_1+u^2\d_2+u^3\d_3]\Gamma^{i}_{ij}=- \Gamma^{i}_{ij},
\end{eqnarray*}
the solutions of which are given by 
\begin{eqnarray*}
&&\Gamma^1_{12}=\f{F_{12}\left(\f{u^2-u^3}{u^1-u^2}\right)}{u^1-u^2},\qquad\Gamma^1_{13}=\f{F_{13}\left(\f{u^2-u^3}{u^1-u^2}\right)}{u^1-u^3},\qquad\Gamma^2_{21}=\f{F_{21}\left(\f{u^2-u^3}{u^1-u^2}\right)}{u^2-u^1},\\
&&\Gamma^2_{23}=\f{F_{23}\left(\f{u^2-u^3}{u^1-u^2}\right)}{u^2-u^3},\qquad\Gamma^3_{31}=\f{F_{31}\left(\f{u^2-u^3}{u^1-u^2}\right)}{u^3-u^1},\qquad\Gamma^3_{32}=\f{F_{32}\left(\f{u^2-u^3}{u^1-u^2}\right)}{u^3-u^2}.
\end{eqnarray*}
where $F_{ij}$, $i\neq j$ are arbitrary smooth functions. 
Imposing Tsarev's conditions and introducing the auxiliary variable $z=\f{u^2-u^3}{u^1-u^2},$ we obtain the system
\begin{equation}\label{mainsys}
\begin{split}
\f{dF_{12}}{dz}&=-\f{(F_{12}(z)F_{23}(z)-F_{12}(z)F_{13}(z))z-F_{12}(z)F_{23}(z)+F_{32}(z)F_{13}(z)}{z(z-1)},\\
\f{dF_{21}}{dz}&=\f{(F_{21}(z)F_{23}(z)-F_{21}(z)F_{13}(z))z+F_{23}(z)F_{31}(z)-F_{23}(z)F_{21}(z)}{z(z-1)},\\
\f{dF_{13}}{dz}&=\f{(F_{12}(z)F_{23}(z)-F_{12}(z)F_{13}(z))z-F_{12}(z)F_{23}(z)+F_{32}(z)F_{13}(z)}{z(z-1)},\\
\f{dF_{31}}{dz}&=-\f{(-F_{31}(z)F_{12}(z)+F_{21}(z)F_{32}(z))z+F_{31}(z)F_{32}(z)-F_{21}(z)F_{32}(z)}{z(z-1)},\\
\f{dF_{23}}{dz}&=-\f{(F_{21}(z)F_{23}(z)-F_{21}(z)F_{13}(z))z+F_{23}(z)F_{31}(z)-F_{23}(z)F_{21}(z)}{z(z-1)},\\
\f{dF_{32}}{dz}&=\f{(-F_{31}(z)F_{12}(z)+F_{21}(z)F_{32}(z))z+F_{31}(z)F_{32}(z)-F_{21}(z)F_{32}(z)}{z(z-1)}
\end{split}
\end{equation}
or, in the more compact form
\begin{equation}\label{mainsys2}
\begin{split}
\f{d\ln F_{12}}{dz}&=-\f{1}{z}F_{23}(z)+\f{1}{z-1}F_{13}(z)-\f{1}{z(z-1)}\f{F_{13}(z)F_{32}(z)F_{21}(z)}{F_{12}(z)F_{21}(z)},\\
\f{d\ln F_{21}}{dz}&=\f{1}{z}F_{23}(z)-\f{1}{z-1}F_{13}(z)+\f{1}{z(z-1)}\f{F_{12}(z)F_{23}(z)F_{31}(z)}{F_{12}(z)F_{21}(z)},\\
\f{d\ln F_{13}}{dz}&=-\f{1}{z-1}F_{12}(z)+\f{1}{z(z-1)}F_{32}(z)+\f{1}{z}\f{F_{12}(z)F_{23}(z)F_{21}(z)}{F_{13}(z)F_{31}(z)},\\
\f{d\ln F_{31}}{dz}&=\f{1}{z-1}F_{12}(z)-\f{1}{z(z-1)}F_{32}(z)-\f{1}{z}\f{F_{21}(z)F_{13}(z)F_{32}(z)}{F_{13}(z)F_{31}(z)},\\
\f{d\ln F_{23}}{dz}&=-\f{1}{z}F_{21}(z)-\f{1}{z(z-1)}F_{31}(z)+\f{1}{z-1}\f{F_{32}(z)F_{13}(z)F_{21}(z)}{F_{23}(z)F_{32}(z)},\\
\f{d\ln F_{32}}{dz}&=\f{1}{z}F_{21}(z)+\f{1}{z(z-1)}F_{31}(z)-\f{1}{z-1}\f{F_{12}(z)F_{23}(z)F_{31}(z)}{F_{23}(z)F_{32}(z)}.\\
\end{split}
\end{equation}
It is straightforward to check that the above system admits three linear first integrals 
\begin{eqnarray}
\label{I1}
I_1&=&F_{12}+F_{13},\\
\label{I2}
I_2&=&F_{23}+F_{21},\\
\label{I3}
I_3&=&F_{31}+F_{32},
\end{eqnarray}
and one quadratic first integral
\beq\label{I4}
I_4=F_{31}F_{13}+F_{12}F_{21}+F_{23}F_{32}.
\eeq
We consider also the cubic first integral
\beq\label{I5}
I_5=-I_3I_4+I_1I_2I_3=F_{21}F_{13}F_{32}+F_{12}F_{23}F_{31}+(I_2-I_3)F_{13}F_{31}+(I_1-I_3)F_{23}F_{32},
\eeq
where $I_1,I_2,I_3$ are given by \eqref{I1}, \eqref{I2} and  \eqref{I3} respectively. 

On the affine subspace $S$ defined by $I_1=d_1,I_2=d_2,I_3=d_3$ we can reduce the original system of six first order ODEs to a system of three first order ODEs in the variables $F_{12}(z),F_{23}(z)$ and $F_{31}(z)$:
\begin{eqnarray*}
\f{dF_{12}}{dz}&=&-\f{(F_{12}F_{23}-F_{12}(d_1-F_{12}))z-F_{12}F_{23}+(d_3-F_{31})(d_1-F_{12})}{z(z-1)},\\
\f{dF_{31}}{dz}&=&-\f{(-F_{31}F_{12}+(d_2-F_{23})(d_3-F_{31}))z+F_{31}(d_3-F_{31})-(d_2-F_{23})(d_3-F_{31})}{z(z-1)},\\
\f{dF_{23}}{dz}&=&-\f{((d_2-F_{23})F_{23}-(d_2-F_{23})(d_1-F_{12}))z+F_{23}F_{31}-F_{23}(d_2-F_{23})}{z(z-1)}.
\end{eqnarray*}
On the subspace $S$ the functions $I_4$ and $I_5$ become dependent
$$(I_4)_{|S}=F_{31}(d_1-F_{12})+F_{12}(d_2-F_{23})+F_{23}(d_3-F_{31}),\,(I_5)_{|S}=-d_3(I_4)_{|S}+d_1d_2d_3$$
Using this first integral we can further reduce the above system to a system of two non-autonomous first order ODEs. To prove 
 the relation between the system \eqref{mainsys} and  the Painlev\'e VI transcendents we observe that from \eqref{gammabeta} it follows immediately
\beq\label{mainID}
\f{F_{ij}\left(\f{u^2-u^3}{u^1-u^2}\right)}{u^i-u^j}\f{F_{ji}\left(\f{u^2-u^3}{u^1-u^2}\right)}{u^j-u^i}=\Gamma^i_{ij}\Gamma^j_{ji}=\beta_{ij}\beta_{ji}=\f{\tilde{F}_{ij}\left(\f{u^2-u^3}{u^1-u^2}\right)}{u^i-u^j}\f{\tilde{F}_{ji}\left(\f{u^2-u^3}{u^1-u^2}\right)}{u^j-u^i}.
\eeq
The last identity is due to the fact the rotation coefficients satisfy the conditions
$$E_{(0)}\beta_{ij}=0,\qquad E_{(1)}\beta_{ij}=-\beta_{ij}.$$
The remaining condition
$$\d_k\beta_{ij}=\beta_{ik}\beta_{kj}$$
becomes a system of ODEs for the unknown functions $\tilde{F}_{ij}(z)$. This system of ODEs admits a quadratic and a cubic
 first integrals that are very similar to the first integrals $I_4$ and $I_5$ of the system \eqref{mainsys}. Up to a sign the only difference is the fact that in the present case the quantities  $I_1,I_2,I_3$ appearing in $I_5$ are not constant but first integrals while in \cite{L2014} they coincide with the degree of homogeneity $d_1,d_2,d_3$ of the Lam\'e coefficients. Taking into account the results of \cite{L2014} and the identity \eqref{mainID} it is clear that a function $f(z)$ satisfying the condition $f'=F_{12}F_{21}$ must be  a solution of Painlev\'e VI equation. Actually, it turns out that the correspondence between solutions of the system \eqref{mainsys} and solutions of the Painlev\'e VI equation is given in terms of purely {\it algebraic} operations, as it is highlighted by the following Theorem (see also the Appendix 2):

\begin{thm}\label{thmreductionsigma}
Let $(F_{12}(z),F_{21}(z),F_{13}(z),F_{31}(z),F_{23}(z),F_{32}(z))$ be a  solution of the system \eqref{mainsys}, then the function $f(z)=F_{23}F_{32}+zF_{12}F_{21}-\f{q_1}{2}$ is a solution 
 of the equation 
 \begin{equation}\label{PVImod}
\begin{split}
[z(z-1)f'']^2=&[q_2-(d_2-d_3)g_2-(d_1-d_3)g_1]^2-4f'g_1g_2,
\end{split}
\end{equation}
where $g_1=f-zf'+\f{q_1}{2}$ and $ g_2=(z-1)f'-f+\f{q_1}{2}$ and the parameters $d_1,d_2,d_3,q_1,q_2$ coincide with the values
 of the first integrals  $I_1,I_2,I_3,I_4,I_5$ on the given solution of \eqref{mainsys}. Furthermore, equation \eqref{PVImod}
  can be reduced to the sigma form of the generic Painlev\'e VI equation. 
%Conversely, given a solution $f(z)$ of the equation \eqref{PVImod}  one can reconstruct algebraically a  solution $\{F_{ij}(z)\}$ of \eqref{mainsys}.
\end{thm}
\emph{Proof} Let $(F_{12}(z),F_{21}(z),F_{13}(z),F_{31}(z),F_{23}(z),F_{32}(z))$ be a  solution of the system \eqref{mainsys} and $d_1,d_2,d_3,q_1,q_2$ the corresponding values of the first integrals $I_1,I_2,I_3,I_4,I_5$. 
In analogy with \cite{AL,L2014} we introduce the function $f(z)=F_{23}F_{32}+zF_{12}F_{21}-\f{q_1}{2}$ satisfying  $f':=F_{12}F_{21}$.  Indeed
\begin{eqnarray*}
\f{d}{dz}\left(F_{12}F_{21}\right)&=&\f{F_{23}F_{31}F_{12}-F_{13}F_{32}F_{21}}{z(z-1)},\\
\f{d}{dz}\left(F_{23}F_{32}\right)&=&-\f{F_{23}F_{31}F_{12}-F_{13}F_{32}F_{21}}{z-1},\\
\f{d}{dz}\left(F_{13}F_{31}\right)&=&\f{F_{23}F_{31}F_{12}-F_{13}F_{32}F_{21}}{z}.
\end{eqnarray*}
Summarizing we have
\begin{eqnarray*}
F_{12}F_{21}&=&f',\\
F_{23}F_{32}&=&g_1:=f-zf'+\f{q_1}{2}.
\end{eqnarray*} 
Taking into account that
$$F_{31}F_{13}+F_{12}F_{21}+F_{23}F_{32}=q_1,$$
we obtain
$$F_{31}F_{13}=g_2:=(z-1)f'-f+\f{q_1}{2}.$$
Using these relations we get
\begin{equation*}
\begin{split}
[z(z-1)f'']^2=&[F_{23}F_{31}F_{12}-F_{13}F_{32}F_{21}]^2=\\
&[q_2-(d_2-d_3)F_{13}F_{31}-(d_1-d_3)F_{23}F_{32}]^2-4F_{23}F_{31}F_{12}F_{13}F_{32}F_{21}=\\
&[q_2-(d_2-d_3)g_2-(d_1-d_3)g_1]^2-4f'g_1g_2.
\end{split}
\end{equation*}
Up to an inessential sign the above equation coincides with the equation $(4.3)$ appearing in \cite{L2014} and, as a consequence,  it is equivalent to the sigma form of the generic Painlev\'e VI equation
 (see \cite{L2014} for details).

\begin{flushright}
$\Box$
\end{flushright}

 This proves that each solution of \eqref{mainsys} determines a specific Painlev\'e VI equation (namely it fixes its parameters) and it identifies a unique solution of the corresponding Painlev\'e VI equation itself. Moreover the correspondence is clearly algebraic. The converse statement is also true and it will be proved in the Appendix 2. 

\section{Tri-flat $F$-manifolds}\label{triflatsec}
In this Section we provide a complete classification of tri-flat $F$-manifolds in dimension two and a partial classification in dimension three. We first briefly discuss the relation between tri-flat $F$-manifolds and tri-Hamiltonian Frobenius manifolds, introduced and studied in \cite{Ro}.

\subsection{Tri-flat $F$ manifolds and the augmented Darboux-Egorov system}
\emph{Tri-Hamiltonian} Frobenius manifolds exist only in even dimensions \cite{Ro}, while we will see that general tri-flat structures exist also in odd dimensions.
 Notice that  (semisimple) Frobenius manifolds are special examples of bi-flat $F$-manifolds as was pointed out in Section \ref{frobsubsection}, but tri-Hamiltonian Frobenius manifolds, in general, do not constitute a special subclass of tri-flat $F$-manifolds. To see this, we proceed as follows.
 First of all,  tri-Hamiltonian Frobenius manifolds are related to solutions of the following system (see \cite{Ro})
$$E_{(0)}(\beta_{ij})=0,\qquad E_{(1)}(\beta_{ij})=-\beta_{ij},\qquad E_{(2)}(\beta_{ij})=-(u^i+u^j)\beta_{ij}.$$
and the last equation in general is not compatible with $E_{(2)}(\Gamma^{i}_{ij})=-2u^j \Gamma^{i}_{ij}$.

 We have the following theorem that elucidate the relationship between tri-flat $F$-manifolds and the augmented Darboux-Egorov system for the rotation coefficients $\beta_{ij}$ and for the Lam\'e coefficients $H_i$. 
\begin{thm}
Let $\beta_{ij},\,i\ne j$ be a solution of the system
\begin{eqnarray}
\label{ED1}
&&\d_k\beta_{ij}=\beta_{ik}\beta_{kj},\qquad k\ne i\ne j\ne k\\
\label{ED2}
&&E_{(0)}(\beta_{ij})=0,\\
\label{ED3}
&&E_{(1)}(\beta_{ij})=-\beta_{ij},\\
\label{ED4}
&&E_{(2)}(\beta_{ij})=[2d_iu^i-2(d_j+1) u^j+c_i-c_j]\beta_{ij},
\end{eqnarray}
(where $d_1,...,d_n$, $c_1,...,c_n$ are constants) and let $(H_1,...,H_n)$ be a solution  of the system 
\begin{eqnarray}
\label{L1}
&&\d_j H_i=\beta_{ij}H_j,\qquad i\ne j\\
\label{L2}
&&E_{(0)}(H_i)=0,\\
\label{L3}
&&E_{(1)}(H_i)=d_iH_i,\\
\label{L4}
&&E_{(2)}(H_i)=(2d_iu^i+c_i)H_i.
\end{eqnarray} 
Then 
\begin{itemize}
\item the connection $\nabla^{(0)}$ defined by
\begin{equation}\label{naturalc1}
\begin{split}
\Gamma^{(0)i}_{jk}&:=0,\qquad\forall i\ne j\ne k \ne i,\\
\Gamma^{(0)i}_{jj}&:=-\Gamma^{(0)i}_{ij},\qquad i\ne j,\\
\Gamma^{(0)i}_{ij}&:=\f{H_j}{H_i}\beta_{ij},\qquad i\ne j,\\
\Gamma^{(0)i}_{ii}&:=-\sum_{l\ne i}\Gamma^{(0)i}_{li},
\end{split}
\end{equation} 
\item the connection $\nabla^{(1)}$ defined by
\begin{equation}\label{dualnabla}
\begin{split}
\Gamma^{(1)i}_{jk}&:=0,\qquad\forall i\ne j\ne k \ne i,\\
\Gamma^{(1)i}_{jj}&:=-\f{u^i}{u^j}\Gamma^{(1)i}_{ij},\qquad i\ne j,\\
\Gamma^{(1)i}_{ij}&:=\f{H_j}{H_i}\beta_{ij},\qquad i\ne j,\\
\Gamma^{(1)i}_{ii}&:=-\sum_{l\ne i}\f{u^l}{u^i}\Gamma^{(1)i}_{li}-\f{1}{u^i},
\end{split}
\end{equation}
\item the connection $\nabla^{(2)}$ defined by
\begin{equation}\label{dualnablaplus}
\begin{split}
\Gamma^{(2)i}_{jk}&:=0,\qquad\forall i\ne j\ne k \ne i,\\
\Gamma^{(2)i}_{jj}&:=-\f{(u^i)^2}{(u^j)^2}\Gamma^{(2)i}_{ij},\qquad i\ne j,\\
\Gamma^{(2)i}_{ij}&:=\f{H_j}{H_i}\beta_{ij},\qquad i\ne j,\\
\Gamma^{(2)i}_{ii}&:=-\sum_{l\ne i}\f{(u^l)^2}{(u^i)^2}\Gamma^{(2)i}_{li}-\f{2}{u^i},
\end{split}
\end{equation} 
\item the vector fields $E_{(0)},E_{(1)}$, $E_{(2)}$ and the corresponding products $\circ_{(0)}, \circ_{(1)}$. $\circ_{(2)}$,  
\end{itemize}
define a semisimple tri-flat $F$-manifold. Moreover any  semisimple tri-flat $F$-manifold can be obtained in this way.
\end{thm}

\noindent
\emph{Proof}. Given a solution of the system (\ref{ED1},\ref{ED2},\ref{ED3},\ref{ED4},\ref{L1},\ref{L2},\ref{L3},\ref{L4})
 to prove that the above formulas define a semisimple tri-flat $F$-manifold is an elementary straightforward computation. The converse statement was partially proved in \cite{L2014} (the part involving $E_{(0)}$ and $E_{(1)}$). The part involving   
 $E_{(2)}$ can be proved in a similar way. 
 
 First we observe that $E_{(2)}(\Gamma^{(k)i}_{ij})=-2u^j \Gamma^{(k)i}_{ij}$ for $k=0,1,2$, for $i\neq j$ because we are starting from a tri-flat $F$-manifold. Since $\Gamma^{(k)i}_{ij}=\beta_{ij}\frac{H_j}{H_i}$, we can rewrite $\Gamma^{(k)i}_{ij}=\d_j \ln(H_i)$ if $\d_j H_i=\beta_{ij}H_j$. Now we obtain
 $$\d_j\left(E_{(2)}(\ln{H_i})\right)=E_{(2)}\left(\d_j\ln{H_i}\right)+2u^j\d_j\ln{H_i}=E_{(2)}(\Gamma^{(k)i}_{ij})+2u^j \Gamma^{(k)i}_{ij}=0,\qquad\forall j\ne i.$$
 This is equivalent to $\frac{\d_ j E_{(2)}(H_i)}{E_{(2)}(H_i)}=\frac{\d_ j H^i}{H^i}$ which has  solution $E_{(2)}(H_i)=f_i(u^i)H_i.$
%\begin{itemize}
%\item First we observe that
%$$\d_j\left(E_{(2)}(\ln{H_i})\right)=E_{(2)}\left(\d_j\ln{H_i}\right)+2u^j\d_j\ln{H_i}=0,\qquad\forall j\ne i.$$
%This means that $E_{(2)}(H_i)=f_i(u^i)H_i$.  
%\item To determine the coefficients $f_i$ we compute their derivative
To check that equation \eqref{L4} is satisfied, we need to compute the coefficients $f_i(u^i)$. 
To do so we determine their derivative
\begin{eqnarray*}
\d_i f_i&=&\d_i\left(\f{E_{(2)}(H_i)}{H_i}\right)=\f{E_{(2)}(\d_i H_i)+2u^i\d_i H_i}{H_i}-\f{E_{(2)}(H_i)\d_i H_i}{H_i^2}=\\
&&\f{E_{(2)}\left(-\sum_{l\ne i}\d_l H_i\right)-2u^i\sum_{l\ne i}\d_l H_i+f^i\sum_{l\ne i}\d_l H_i}{H_i},\\
\end{eqnarray*}
where we have used equation \eqref{L2} which is equivalent to $\d_i H_i=-\sum_{l\neq i}\d_l H_i$ (we are allowed to use it since by the results of \cite{L2014} we already know the converse for $E_{(0)}$ and $E_{(1)}$). Using $E_{(2)}(-\d_l H_i)=2u^l \d_l H_i-\d_l E_{(2)}(H_i)$ and the fact that $\d_l E_{(2)}(H_i)=f^i\d_l H_i$ the last expression becomes
\begin{eqnarray*}
&&\f{2\sum_{l\ne i}u^l\d_l H_i-2u^i\sum_{l\ne i}\d_l H_i}{H_i}=\\
&&\f{2\sum_{l=1}^nu^l\d_l H_i-2u^i\sum_{l=1}^n\d_l H_i}{H_i}=2d_i,\\
\end{eqnarray*}
by equations \eqref{L2} and \eqref{L3}.
This means that the Lam\'e coefficients $H_i$ satisfy the condition
\beq\label{LC}
E_{(2)}(H_i)=(2d_iu^i+c_i)H_i,
\eeq
where $d_i=\f{E_{(1)}(H_i)}{H_i}$ and $c_i$ are constants.
%\end{itemize}

\begin{flushright}
$\Box$
\end{flushright} 

Comparing \eqref{ED4} with Romano's condition (see \cite{Ro})
$$E_{(2)}(\beta_{ij})=-(u^i+u^j)\beta_{ij},$$
we observe that they coincide iff $d_i=d_j=-\f{1}{2}$ and $c_i=c_j$.

\subsection{Three-dimensional tri-flat $F$-manifolds }

Let us consider the case corresponding to the subalgebra generated by $Z_{(-1)},Z_{(0)},Z_{(1)}$ (or, which is equivalent, the subalgebra generated by $\hat{E}_{(0)}, \hat{E}_{(1)}, \hat{E}_{(2)}$).

First of all we have to solve the systems (for $j=1,2,3$)
\begin{eqnarray*}
E_{(0)}(\Gamma^{i}_{ij})&=&[\d_1+\d_2+\d_3]\Gamma^{i}_{ij}=0,\\
E_{(1)}(\Gamma^{i}_{ij})&=&[u^1\d_1+u^2\d_2+u^3\d_3]\Gamma^{i}_{ij}=- \Gamma^{i}_{ij},\\
E_{(2)}(\Gamma^{i}_{ij})&=&[(u^1)^{2}\d_1+(u^2)^2\d_2+(u^3)^{2}\d_3]\Gamma^{i}_{ij}=-2u^j \Gamma^{i}_{ij}.
\end{eqnarray*}

The general solution is given by
\begin{eqnarray*}
&&\Gamma^1_{12}=\frac{C_{12}(u^3-u^1)}{(u^2-u^1)(u^2-u^3)},\,\Gamma^1_{13}=\frac{C_{13}(u^1-u^2)}{(u^3-u^1)(u^3-u^2)},\,\Gamma^2_{21}=\f{C_{21}(u^2-u^3)}{(u^1-u^3)(u^1-u^2)},\\
&&\Gamma^2_{23}=\frac{C_{23}(u^1-u^2)}{(u^3-u^1)(u^3-u^2)},\,\Gamma^3_{31}=\frac{C_{31}(u^2-u^3)}{(u^1-u^3)(u^1-u^2)},\,\Gamma^3_{32}=\f{C_{32}(u^3-u^1)}{(u^2-u^1)(u^2-u^3)},
\end{eqnarray*}
where $C_{12},C_{21},C_{13},C_{31},C_{23},C_{32}$ are arbitrary constants. Imposing Tsarev's condition we obtain immediately the following constraints
$$C_{13}=-C_{12},\qquad C_{23}=-C_{21},\qquad C_{32}=-C_{31},\qquad C_{12}+C_{23}+C_{31}=1.$$

\begin{rmk}
In some cases it might be convenient to work in canonical coordinates for the dual product rectifying the Euler vector field. In this case the generators of the
 $sl(2, \mathbb{C})$ algebra have the exponential form
\beq\label{distri2eq}\hat{E}_{(l)}:=e^{lu^1}\d_1+\dots e^{lu^n}\d_n-le^{lu^j}u_{n+1}\d_{n+1},\quad  l=-1,0,1.\eeq
All the formulas obtained in this paper can be immediately rephrased in this dual framework. For instance, the Christoffel symbols  defining three dimensional  tri-flat $F$-manifolds  have the following form
 
\begin{eqnarray*}
&&\Gamma^1_{12}=\f{C_{12}(e^{u^3}-e^{u^1})e^{u^2}}{(e^{u^2}-e^{u^3})(e^{u^2}-e^{u^1})},\,
\Gamma^1_{13}=-\f{C_{13}(e^{u^1}-e^{u^2})e^{u^3}}{(e^{u^3}-e^{u^1})(e^{u^3}-e^{u^2})},\,\\
&&\Gamma^2_{21}=-\f{C_{21}(e^{u^2}-e^{u^3})e^{u^1}}{(e^{u^1}-e^{u^3})(e^{u^1}-e^{u^2})},\,\,\Gamma^2_{23}=\f{C_{23}(e^{u^1}-e^{u^2})e^{u^3}}{(e^{u^3}-e^{u^1})(e^{u^3}-e^{u^2})},\\
&&\Gamma^3_{31}=\f{C_{31}(e^{u^2}-e^{u^3})e^{u^1}}{(e^{u^1}-e^{u^3})(e^{u^1}-e^{u^2})},\,\Gamma^3_{32}=-\f{C_{32}(e^{u^3}-e^{u^1})e^{u^2}}{(e^{u^2}-e^{u^3})(e^{u^2}-e^{u^1})},
\end{eqnarray*}
with
$$C_{13}=-C_{12},\qquad C_{21}=-C_{23},\qquad C_{32}=-C_{31},\qquad C_{12}+C_{23}+C_{31}=1.$$
\end{rmk}

\subsection{Four-dimensional tri-flat $F$-manifolds}

The study of tri-flat $F$-manifolds in higher dimensions in much more complicated due to the appearance of functional parameters.
 For instance four dimensional tri-flat $F$-manifolds are related to the solutions of
 the system (with $j=1,2,3,4$)
 \begin{eqnarray*}
E_{(0)}(\Gamma^{i}_{ij})&=&[\d_1+\d_2+\d_3+\d_4]\Gamma^{i}_{ij}=0,\\
E_{(1)}(\Gamma^{i}_{ij})&=&[u^1\d_1+u^2\d_2+u^3\d_3+u^4\d_4]\Gamma^{i}_{ij}=- \Gamma^{i}_{ij},\\
E_{(2)}(\Gamma^{i}_{ij})&=&[(u^1)^{2}\d_1+(u^2)^2\d_2+(u^3)^{2}\d_3+(u^4)^{2}\d_4]\Gamma^{i}_{ij}=-2u^j \Gamma^{i}_{ij}.
\end{eqnarray*}
satisfying the Tsarev's condition \eqref{rt1}. After the first step we obtain
\begin{eqnarray*}
\Gamma^i_{i1}&=& F_{i1}\left(\f{(u^1-u^2)(u^3-u^4)}{(u^2-u^3)(u^1-u^4)}\right)\f{u^3-u^2}{(u^1-u^3)(u^1-u^2)},\quad i=2,3,4,\\
\Gamma^i_{i2}&=& F_{i2}\left(\f{(u^1-u^2)(u^3-u^4)}{(u^2-u^3)(u^1-u^4)}\right)\f{u^3-u^1}{(u^2-u^3)(u^2-u^1)},\quad i=1,3,4,\\
\Gamma^i_{i3}&=& F_{i3}\left(\f{(u^1-u^2)(u^3-u^4)}{(u^2-u^3)(u^1-u^4)}\right)\f{u^2-u^1}{(u^3-u^1)(u^3-u^2)},\quad i=1,2,4,\\
\Gamma^i_{i4}&=& F_{i4}\left(\f{(u^1-u^2)(u^3-u^4)}{(u^2-u^3)(u^1-u^4)}\right)\f{u^1-u^3}{(u^4-u^1)(u^4-u^3)},\quad i=1,2,3.
\end{eqnarray*}
The second step seems very difficult. We have to solve a system of 24 equations (Tsarev's conditions) for the 12 unknown functions $F_{ij}$.
 This system can be written as a system of ODEs (\emph{two for each unknown function}) in the variable $z=\f{(u^1-u^2)(u^3-u^4)}{(u^2-u^3)(u^1-u^4)}$ for the unknown functions $F_{ij}(z):$
%\begin{footnotesize}
\begin{eqnarray*}
&&\f{dF_{12}}{dz}=-\f{-F_{12}F_{13}+F_{12}F_{23}+F_{32}F_{13}+F_{12}}{z-1}=-\f{-F_{42}F_{14}+F_{12}F_{14}-F_{12}F_{24}}{z},\\
&&\f{dF_{13}}{dz}=\f{F_{12}F_{23}-F_{12}F_{13}+F_{32}F_{13}-F_{13}}{z}=\f{-F_{14}F_{13}+F_{14}F_{43}+F_{34}F_{13}}{z},\\
&&\f{dF_{14}}{dz}=-\f{-F_{42}F_{14}+F_{12}F_{14}-F_{12}F_{24}}{z}=-\f{(F_{34}F_{13}+F_{14}F_{43}-F_{14}F_{13})z+F_{14}}{z(z-1)},\\
&&\f{dF_{21}}{dz}=-\f{F_{23}F_{21}-F_{13}F_{21}-F_{23}F_{31}+F_{21}}{z-1}=-\f{-F_{24}F_{21}+F_{24}F_{41}+F_{14}F_{21}}{z},\\
&&\f{dF_{23}}{dz}=-\f{-F_{13}F_{21}+F_{23}F_{21}-F_{23}F_{31}-F_{23}}{(z-1)z}=\f{F_{23}F_{34}-F_{23}F_{24}+F_{43}F_{24}}{z},\\
&&\f{dF_{24}}{dz}=\f{F_{14}F_{21}-F_{24}F_{21}+F_{24}F_{41}-F_{24}z}{(z-1)z}=-\f{z(F_{34}F_{23}-F_{24}F_{23}+F_{24}F_{43})+F_{24}}{(z-1)z},\\
&&\f{dF_{31}}{dz}=-\f{-F_{31}F_{14}+F_{31}F_{34}-F_{41}F_{34}}{z}=\f{F_{31}F_{12}+F_{21}F_{32}-F_{31}F_{32}+F_{31}}{z},\\
&&\f{dF_{32}}{dz}=\f{F_{31}F_{12}+F_{21}F_{32}-F_{31}F_{32}-F_{32}}{(z-1)z}=\f{F_{34}F_{42}-F_{34}F_{32}+F_{24}F_{32}}{z},\\
&&\f{dF_{34}}{dz}=-\f{F_{31}F_{34}-F_{41}F_{34}-F_{31}F_{14}+F_{34}z}{(z-1)z}=\f{F_{34}F_{42}-F_{34}F_{32}+F_{24}F_{32}}{z},\\
&&\f{dF_{41}}{dz}=\f{F_{41}F_{12}+F_{21}F_{42}-F_{41}F_{42}+F_{41}}{z}=-\f{F_{31}F_{43}+F_{41}F_{13}-F_{41}F_{43}-F_{41}}{z-1},\\
&&\f{dF_{42}}{dz}=\f{F_{41}F_{12}+F_{21}F_{42}-F_{41}F_{42}-F_{42}}{(z-1)z}=-\f{F_{42}F_{23}-F_{42}F_{43}+F_{32}F_{43}+F_{42}}{z-1},\\
&&\f{dF_{43}}{dz}=\f{F_{31}F_{43}-F_{41}F_{43}+F_{41}F_{13}+F_{43}}{(z-1)z}=\f{F_{42}F_{23}-F_{42}F_{43}+F_{32}F_{43}-F_{43}}{z}.
\end{eqnarray*}
%\end{footnotesize}
Comparing the right and sides of the above equations we obtain some constraints on the functions $F_{ij}$. We have the following relations
\begin{eqnarray*}
(z-1)\f{dF_{12}(z)}{dz}+F_{12}(z)&=&-z\f{dF_{13}(z)}{dz}-F_{13}(z),\\
\f{dF_{12}(z)}{dz}&=&\f{dF_{14}(z)}{dz},\\
-(z-1)\f{dF_{14}(z)}{dz}-\f{F_{14}(z)}{z}&=&z\f{dF_{13}(z)}{dz},\\
z(z-1)\f{dF_{23}(z)}{dz}-F_{23}(z)&=&-(z-1)\f{dF_{21}(z)}{dz}+F_{21}(z),\\
z\f{dF_{21}(z)}{dz}&=&z(z-1)\f{dF_{24}(z)}{dz}+zF_{24}(z),\\
-(z-1)\f{dF_{24}(z)}{dz}-\f{F_{24}}{z}&=&z\f{dF_{23}(z)}{dz},\\
z(z-1)\f{dF_{34}(z)}{dz}+zF_{34}(z)&=&z\f{dF_{31}(z)}{dz},\\
z(z-1)\f{dF_{32}(z)}{dz}+F_{32}(z)&=&z\f{dF_{31}(z)}{dz}-F_{31}(z),\\
\f{dF_{32}(z)}{dz}&=&\f{dF_{34}(z)}{dz},\\
z(z-1)\f{dF_{42}(z)}{dz}+F_{42}(z)&=&z\f{dF_{41}(z)}{dz}-F_{41}(z),\\
z(z-1)\f{dF_{43}(z)}{dz}-F_{43}(z)&=&-(z-1)\f{dF_{41}(z)}{dz}+F_{41}(z),\\
z\f{dF_{43}(z)}{dz}+F_{43}(z)&=&-(z-1)\f{dF_{42}(z)}{dz}-F_{42}(z),\\
\end{eqnarray*}
which imply
\begin{eqnarray*}
F_{14}(z)-F_{12}(z)&=&C_1,\\
zF_{13}(z)+(z-1)F_{12}(z)&=&C_1,\\
F_{32}(z)-F_{34}(z)&=&C_2,\\
(z-1) F_{34}(z)-F_{31}(z)&=&C_2,\\
-zF_{43}(z)-(z-1)F_{42}(z)&=&C_3,\\
\f{F_{41}(z)}{z}-\f{(z-1)}{z}F_{42}(z)&=&C_3,\\
\f{zF_{23}(z)}{z-1}+\f{F_{21}(z)}{z-1}&=&C_7,\\
(z-1)F_{24}(z)-F_{21}(z)&=&C_7.
\end{eqnarray*}
Since for each unknown we have two equations, we have still to impose that such equations coincide. In general 
 this seems a very complicate task. However, assuming $C_1=0$ we obtain the following additional constraints
\begin{eqnarray*}
C_7 &=& C_2+C_3-2,\\
F_{42}(z)&=&\f{(1 -C_3)z+F_{34}(z) (z-1)-C_2}{z-1},\\ 
F_{21}(z)&=&F_{34}(z) (z-1)+1-C_2,\\
F_{34}(z)&=&C_3+F_{12}(z)-1.
\end{eqnarray*}
After this, all the equations of the original system reduce to the first order equation
\beq\label{finalEQ}
\f{dF_{12}(z)}{dz}=-\f{F_{12}(z)[(F_{12}(z)+C_3-1)(1-z)+C_2]}{z(z-1)}
\eeq
whose general solution is given by
\beq
F_{12}(z) =\f{C_9z^{C_2}(z-1)^{-C_2}}{C_8C_9z^{C_9}+{\rm hypergeom}([C_2, C_9], [1+C_9], \f{1}{z})}
\eeq
where $C_9=1-C_3$ and $C_8$ is an additional integration constant.

\section{Non-semisimple regular bi-flat $F$ manifolds in dimension three and Painlev\'e equations}\label{structuresec2}

We consider now non-semisimple multi-flat $F$ manifolds. 

%Let us recall their  definition.  
%\begin{defi}\label{multiflatdefiBIS}
%Let $(M, \nabla, \circ, e)$ be a semisimple flat $F$-manifold with unity $e$. 
%A \emph{multi-flat} $F$-manifold $(M,\nabla^{(l)},\circ, e, E,l=0...N-1)$  is a manifold $M$ endowed with $N$ flat torsionless affine connections $\nabla^{(0)}:=\nabla,\, \nabla^{(1)},...,\nabla^{(N-1)}$, a commutative associative product $\circ$ on sections of the tangent bundle $TM$, an invertible vector field $E$ satisfying the following conditions:
%\begin{enumerate}
%\item $E$ is an Euler vector field.
%\item Given $E_{(l)}:=E^{\circ l}=E\circ E\circ \dots \circ E$ $l$-times, $l=0, \dots, N-1$,  we require $\nabla^{(l)}E_{(l)}=0.$
%\item Given $E_{(l)}$ and the related commutative, associative product $\circ_{(l)}$  we require that
%\begin{equation}\label{scc}
%\left(\nabla^{(l)}_X c_{(l)}\right)\left(Y,Z\right)=\left(\nabla^{(l)}_Y c_{(l)}\right)\left(X,Z\right),
%\end{equation}
%for all vector fields $X$, $Y$, and $Z$ for all $l=0, \dots N-1$.  
%\item The connections $\nabla^{(l)}$, $l=0, \dots, N-1$  are almost hydrodynamically equivalent.
%\end{enumerate}
%\end{defi}

%\begin{rmk}\label{rmk2}
%We want to point out that the requirement that $E$ is an eventual identity for $\circ$, implies that the endomorphism $L:=E \circ $ has vanishing Nijenhuis torsion (see \cite{ALimrn}, Theorem 3.4). However, the vanishing of the Nijenhuis torsion for $L$ does not in general imply that $E$ is an eventual identity for $\circ$, unless $\circ$ is semisimple. 
%\end{rmk}

According to the results of the Section 4, the flatness of $\nabla_{(l)}$ is equivalent to the following pair of conditions: 
\begin{itemize}
\item $[{\rm Lie}_{E_{(l)}}, \nabla_{(l)}](T)=0, $ for any tensor field $T$.
\item For every vector field $X, Y, Z, W$ we have 
$$Z \circ_{(l)} R_{(l)}(W, Y )(X) +W \circ_{(l)} R_{(l)}(Y,Z)(X) + Y \circ_{(l)} R_{(l)}(Z,W)(X) = 0,$$
where $R_{(l)}$ is the Riemann operator associated to the torsionless connection $\nabla_{(l)}$.  
\end{itemize}

In this Section we are interested in $F$-manifolds that are not semisimple, but that satisfy still a regularity condition. In order to deal with the non-semisimple regular case we will use a result of David and Hertling (see \cite{DH}) about the existence
 of local  ``canonical coordinates" for non-semisimple regular $F$-manifolds with an Euler vector field. Let us summarize the main results of their work which are relevant for our situation.

%\begin{defi}
%A Euler vector field on an $F$-manifold $(M, \circ, e)$ is a vector field such that 
%$$\rm{Lie}_E(\circ)(X,Y)=dX\circ Y,$$
%where $d$ is a constant called the weight of the Euler vector field.
%\end{defi}
%In the case that interests us $d=1$. 
\begin{defi}[\cite{DH}]
An $F$-manifold $(M, \circ, e, E)$ where $E$ is an Euler vector field  is called \emph{regular} if for each $p\in M$ the endomorphism $L_p:=E_p \circ : T_pM \rightarrow T_pM$ has exactly one Jordan block for each distinct eigenvalue.
\end{defi}

Here is the result from \cite{DH} which is relevant for our analysis:
\begin{thm}[\cite{DH}]\label{DavidHertlingth}
Let $(M, \circ, e, E)$ be a regular $F$-manifold of dimension greater or equal to $2$ with an Euler vector field $E$ of weight one. Furthermore assume that locally around a point $p\in M$, the operator $L$ has only one eigenvalue. Then
there exists locally around $p$ a distinguished system of coordinates $\{u^1, \dots, u^n\}$ ( a sort of ``generalized canonical coordinates" for $\circ$) such that 
\begin{eqnarray} \label{canonical1}
e&=&\partial_{u^1},\\ \label{canonical2}
c^k_{ij}&=&\delta^k_{i+j-1},\\ \label{canonical3}
E&=&u^1\partial_{u^1}+\dots+u^n\partial_{u^n}.
\end{eqnarray}
\end{thm}

%The $F$-manifolds considered in \cite{DH} are of the form $(M, \circ, e, E)$ where $E$ is an Euler vector field of weight one and where $\rm{dim}(M)\geq 2$. Furthermore the $F$-manifold $M$ is assumed to be \emph{regular}, which means that for each $p\in M$ the endomorphism $L_p:=E_p \circ : T_pM \rightarrow T_pM$ has exactly one Jordan block for each distinct eigenvalue. 

%Under the assumption that $M$ is regular and that $L_p$ has only one eigenvalue for all $p\in M$, it was proved in \cite{DH} that there exists locally around $p$ a distinguished system of coordinates $\{u^1, \dots, u^n\}$ ( a sort of ``generalized canonical coordinates" for $\circ$) such that 
%\begin{eqnarray} %\label{canonical1}
%e&=&\partial_{u^1},\\ %\label{canonical2}
%c^k_{ij}&=&\delta^k_{i+j-1},\\ \label{canonical3}
%E&=&u^1\partial_{u^1}+\dots+u^n\partial_{u^n}.
%\end{eqnarray}
(Here we have performed a shift of the variables $u^1$ and $u^2$ compared to the coordinate system identified in \cite{DH} to obtain simpler formulas. In particular the operator $L_p$ has Jordan normal form with one Jordan block and all eigenvalues equal to $a$ at the point $p$ with coordinates $u^1=a, u^2=1, u^3=\dots=u^n=0$.)

%It is immediate to check that for the data $(M, \circ, e, E)$ as in Theorem \ref{DavidHertlingth}  $E$ is an eventual identity and therefore it gives rise to a dual product $*$, as indicated in the following:

%\begin{lemma}
%The Euler vector field $E$ in \eqref{canonical3} is indeed an eventual identity according to Definition \ref{eventualdefi}. 
%\end{lemma}
%\proof
%This is a straightforward computation left to the reader.
%First we need to prove that $E$ is invertible, namely that there exists a vector %field $F$ such that $F\circ E=e$. 
%This amounts to solve the system of $n$ algebraic equations:
%$$(u^1+a)F^j \delta^k_j+(u^2+1)F^j \delta^k_{j+1}+u^3F^j \delta^k_{j+2}+%\dots+ u^n \delta^k_{j+n-1}F^j =\delta^k_1, \quad k=1, \dots n.$$
%This system is clearly lower triangular with nonzero diagonal elements, as it can %be observed writing down few of the equations, so that the solution always %exists and it is unique. For instance, in the specific case $n=3$ we find that 
%$$F=\left(\frac{1}{u^1+a}, -\frac{u^2+1}{(u^1+a)^2}, -\frac{au^3+u^1u^3-%(u^2)^2-2u^2-1}{(u^1+a)^3} \right).$$
%Since $E$ is invertible, it is an eventual identity if and only if it satisfies the %David-Strachan criterion (see \cite{DS}), namely 
%$$\rm{Lie}_E(\circ)(X, Y)=[e, E]\circ X \circ Y.$$
%Since $E$ is an Euler vector field of weight one $\rm{Lie}_E(\circ)(X, Y)=X\circ Y$. On the other hand a straightforward computation in canonical coordinates shows that $[e, E]=e$ and therefore the David-Strachan criterion is fulfilled. 
\endproof

Let us point out that if the endomorphisms $L_p:=E_p\circ$ consist of different Jordan blocks with distinct eigenvalues, then the results of \cite{DH} can be readily extended using Hertling's Decomposition Lemma (see \cite{H}). However, in the case in which there are multiple Jordan blocks with the same eigenvalues no results are available to the best of our knowledge. 
In the three dimensional case, assuming regularity, one has only three possibilities. One is the semisimple case (in which $L$ has the form $L_1$), one is the case with one Jordan block and all eigenvalues equal (this corresponds to $L=L_3$) and this is the situation we analyze in detail in the first part of the next section. In the third case (corresponding to $L=L_2$) there is a non-trivial $2\times 2$ Jordan block with one eigenvalue and a second distinct eigenvalue. This last case is analyzed in detail in the second part of the next section

\subsection{The case of one single eigenvalue and one Jordan block}

In this section, we use canonical coordinates for a regular nonsemisimple bi-flat $F$-manifold in dimension three to show that locally these structures are parameterized by solutions of a three-parameter second order ODE  that contains the full Painlev\'e IV for a special choice of one of these parameters. 
%In this section, we construct nonsemisimple bi-flat $F$-manifolds, in dimension three under the assumption of regularity. We show that they are essentially parameterized by solutions of of a three-parameters second order ODE that contains for a special choice of one of the parameter the full Painlev\'e IV equation. 

\begin{thm}\label{thm1}
Let $(M, \nabla_1, \nabla_2, \circ, *, e, E)$ be a regular bi-flat $F$-manifold in dimension three such that $L_p$ has three equal eigenvalues. Then there exist local coordinates $\{u^1, u^2, u^3\}$ such that 
\begin{enumerate}
\item $e, E, \circ$ are given by \eqref{canonical1}, \eqref{canonical2}, \eqref{canonical3}.
\item The Christoffel symbols $\Gamma_{jk}^{(1)i}$ for $\nabla_1$ are given by:
$$\Gamma_{23}^{(1)1}=\Gamma_{32}^{(1)1}=\Gamma_{33}^{(1)2}=\frac{F_1\left(\frac{u^3}{u^2} \right)}{u^2}, \;
\Gamma_{32}^{(1)3}=\Gamma_{23}^{(1)3}=\frac{F_2\left(\frac{u^3}{u^2} \right)}{u^2},\; \Gamma_{32}^{(1)2}=\Gamma_{23}^{(1)2}=\frac{F_3\left(\frac{u^3}{u^2} \right)}{u^2},$$
$$\Gamma_{22}^{(1)1}=\frac{F_4\left(\frac{u^3}{u^2} \right)}{u^2}, \;\Gamma_{22}^{(1)2}=\frac{F_5\left(\frac{u^3}{u^2} \right)}{u^2}, \; 
\Gamma_{22}^{(1)3}=\frac{F_6\left(\frac{u^3}{u^2} \right)}{u^2}, \; 
\Gamma_{33}^{(1)3}=\frac{F_3\left(\frac{u^3}{u^2} \right)-F_4\left(\frac{u^3}{u^2} \right)}{u^2},$$
 where the functions $F_1, \dots, F_6$ satisfy the system
\begin{eqnarray}
\label{F1}
\frac{d F_1}{dz}&=&0,\\
\label{F2}
\frac{d F_2}{dz}&=&2F_4F_3z+2F_2F_1z-2F_5F_1z+F_6F_1-F_2F_3+F_4-F_3,\\
\label{F3}
\frac{d F_3}{dz}&=&-F_4F_3-F_2F_1+F_5F_1-F_1,\\
\label{F4}
\frac{d F_4}{dz}&=&F_4F_3+F_2F_1-F_5F_1-F_1,\\
\label{F5}
\frac{d F_5}{dz}&=&F_4F_3z+F_2F_1z-F_5F_1z-F_6F_1+F_2F_3+F_1z-F_3,\\
\label{F6}
\frac{d F_6}{dz}&=&-2F_4F_3z^2-2F_2F_1z^2+2F_5F_1z^2-F_6F_1z+F_2F_3z+\\
\nonumber
&& F_4F_6-F_4z+F_2^2-F_2F_5+F_3z-F_2.
\end{eqnarray}
in the variable $z=\frac{u^3}{u^2}$ while the other symbols are identically zero.
\item The Christoffel symbols $\Gamma_{jk}^{(2)i}$ for $\nabla_2$ are uniquely determined by the Christoffel symbols of $\nabla_1$ via the following formulas:
\begin{eqnarray*}
\Gamma_{11}^{(2)1}&=&\frac{1}{(u^1)^3}\left[- (u^1)^2+\Gamma_{22}^{(1)1} u^1 (u^2)^2+ \Gamma_{32}^{(1)1}(2u^1 u^2 u^3- (u^2)^3)\right],\\
\Gamma_{11}^{(2)2}&=&\frac{1}{(u^1)^3}\left[\Gamma_{22}^{(1)2}u^1 (u^2)^2+ \Gamma_{32}^{(1)2} (2u^1 u^2 u^3-(u^2)^3)+u^2 u^1+\Gamma_{32}^{(1)1}( u^1 (u^3)^2-(u^2)^2 u^3)\right],\\
\Gamma_{11}^{(2)3}&=&\frac{1}{(u^1)^3}\left[\Gamma_{22}^{(1)3} u^1 (u^2)^2+\Gamma_{32}^{(1)3}(2 u^1 u^2 u^3-(u^2)^3)
+\Gamma_{32}^{(1)2}(u^1 (u^3)^2-(u^2)^2u^3)+ u^1 u^3\right.\\
&&\left. -(u^2)^2+\Gamma_{22}^{(1)1} ((u^2)^2 u^3-(u^3)^2u^1)\right],\\
\Gamma_{21}^{(2)1}&=&-\frac{1}{(u^1)^2}\left[\Gamma_{22}^{(1)1}u^1 u^2+\Gamma_{32}^{(1)1}( u^1 u^3-(u^2)^2)\right],\\
\end{eqnarray*}
\begin{eqnarray*}
\Gamma_{21}^{(2)2}&=&-\frac{1}{(u^1)^2}\left[\Gamma_{22}^{(1)2} u^1 u^2+\Gamma_{32}^{(1)2}( u^1 u^3-(u^2)^2)+ u^1-\Gamma_{32}^{(1)1} u^2 u^3\right],\\
\Gamma_{21}^{(2)3}&=&-\frac{1}{(u^1)^2}\left[\Gamma_{22}^{(1)3} u^1 u^2+\Gamma_{32}^{(1)3}( u^1 u^3-(u^2)^2 )+(\Gamma_{22}^{(1)1}-\Gamma_{32}^{(1)2}) u^2 u^3- u^2\right],\\
\Gamma_{31}^{(2)1}& =& -\frac{u^2}{u^1}\Gamma_{32}^{(1)1},\\
\Gamma_{31}^{(2)2}&=&-\frac{1}{u^1}\left[\Gamma_{32}^{(1)2} u^2+\Gamma_{32}^{(1)1} u^3\right],\\
\Gamma_{31}^{(2)3}&=& \frac{1}{u^1}\left[-\Gamma_{32}^{(1)3}u^2+(\Gamma_{22}^{(1)1}-\Gamma_{32}^{(1)2})u^3-1\right],\\
%\end{eqnarray*}
%\begin{eqnarray*}
\Gamma_{22}^{(2)1} &=& -\frac{\Gamma_{32}^{(1)1}u^2-\Gamma_{22}^{(1)1}u^1}{u^1},\\
\Gamma_{22}^{(2)2} &=& \frac{\Gamma_{22}^{(1)2}u^1-\Gamma_{32}^{(1)1}u^3-\Gamma_{32}^{(1)2}u^2}{u^1},\\
\Gamma_{22}^{(2)3}&=&\frac{\Gamma_{22}^{(1)3}u^1-\Gamma_{32}^{(1)3}u^2+u^3\Gamma_{22}^{(1)1}-\Gamma_{32}^{(1)2}u^3-1}{u^1},\\
\Gamma_{22}^{(2)1} &=& \Gamma_{32}^{(1)1}, \; \;
  \Gamma_{23}^{(2)2}= \Gamma_{32}^{(1)2},\; \;
  \Gamma_{23}^{(2)3}= \Gamma_{32}^{(1)3},\\
   \Gamma_{32}^{(2)1} &=& \Gamma_{32}^{(1)1},\; \;
  \Gamma_{32}^{(2)2}=\Gamma_{32}^{(1)2},\; \;
  \Gamma_{32}^{(2)3} = \Gamma_{32}^{(1)3}, \\
  \Gamma_{33}^{(2)1}& =& 0,\; \;
  \Gamma_{33}^{(2)2} = \Gamma_{32}^{(1)1},\; \; \Gamma_{33}^{(2)3}=  -\Gamma_{22}^{(1)1}+\Gamma_{32}^{(1)2}.
\end{eqnarray*}
\item The dual product $*$ is obtained via formula \eqref{nm} using $\circ$ and $E$.
\end{enumerate}
\end{thm}

\n
\emph{ Proof}. Due to David-Hertling result there exist local coordinates such that $e, E, \circ$ are given by \eqref{canonical1}, \eqref{canonical2}, \eqref{canonical3}. To determine the Christoffel symbols $\Gamma_{ij}^{(1)k}$ for the torsionless connection $\nabla_1$ in these coordinates we start imposing the following conditions: 
\begin{itemize}
\item compatibility with $\circ$:
$$\Gamma_{ml}^{(1)i}c_{jk}^m-\Gamma_{lk}^{(1)m}c_{jm}^i-\Gamma_{mj}^{(1)i}c_{lk}^m+\Gamma_{jk}^{(1)m}c_{lm}^i=0, \; 1\leq l,j,k \leq 3,$$
\item symmetry of the connection:
$$\Gamma_{ij}^{(1)k}=\Gamma_{ji}^{(1)k},$$
\item flatness of unity:
$$\nabla e=0 \iff \Gamma_{1j}^{(1)i}=0.$$
\end{itemize}
This provides a system of algebraic equations for $\Gamma^{(1)k}_{ij}$. These symbols are in general functions of $u^1, u^2, u^3$. 

Imposing the commutativity of $\nabla_1$ and $\rm{Lie}_e$, coming from the flatness of $\nabla_1$ (see Remark \label{rmk1}) we obtain that  the symbols $\Gamma^{(1)k}_{i,j}$ do not depend on $u^1$. 

Now we link the Christoffel symbols $\nabla_{(2)}$ to the Christofell symbols of $\nabla_{(1)}$ imposing that the two connections are almost hydrodynamically equivalent, i.e. 
$d_{\nabla_{(1)}}\left( X \circ\right)=d_{\nabla_{(2)}}\left( X \circ \right),$ we obtain the following constraints on $\Gamma_{ij}^{(2)k}$. 

 $$\Gamma_{2 2}^{(2) 1} = \Gamma_{22}^{(1)1}(u^2, u^3)+\Gamma_{31}^{(2) 1}(u^2, u^3)$$
  $$\Gamma_{2 2}^{(2) 2} = \Gamma_{22}^{(1)2}(u^2, u^3)+\Gamma_{31}^{(2) 2}(u^2,u^3),$$
  $$ \Gamma_{22}^{(2)3} = \Gamma_{2 2}^{(1) 3}(u^2, u^3)+\Gamma_{31}^{(2) 3}(u^2,u^3), $$
  $$\Gamma_{22}^{(2)1} = \Gamma_{32}^{(1)1}(u^2, u^3), $$
  $$\Gamma_{23}^{(2)2}= \Gamma_{32}^{(1)2}(u^2, u^3),$$
  $$\Gamma_{23}^{(2)3}= \Gamma_{32}^{(1)3}(u^2, u^3),$$
  $$ \Gamma_{32}^{(2)1} = \Gamma_{32}^{(1)1}(u^2, u^3),$$
  $$\Gamma_{32}^{(2)2}=\Gamma_{32}^{(1)2}(u^2, u^3),$$
  $$\Gamma_{32}^{(2)3} = \Gamma_{32}^{(1)3}(u^2, u^3), $$
  $$\Gamma_{33}^{(2)1}=0,$$
  $$\Gamma_{33}^{(2)2} = \Gamma_{32}^{(1)1}(u^2, u^3),$$
  $$\Gamma_{33}^{(2)3}=  -\Gamma_{22}^{(1)1}(u^2, u^3)+\Gamma_{32}^{(1)2}(u^2, u^3)$$

Imposing that $\nabla_{(2)}E=0$ and using the constraints obtained so far, we are able to express uniquely all the Christoffel symbols of $\nabla_{(2)}$ in terms of the Christoffel symbols of $\nabla_{(1)}$. 
We get indeed the further constraints:
\begin{eqnarray*}
\Gamma_{11}^{(2)1}&=&\frac{1}{(u^1)^3}\left[- (u^1)^2+\Gamma_{22}^{(1)1} u^1 (u^2)^2+ \Gamma_{32}^{(1)1}(2u^1 u^2 u^3- (u^2)^3)\right],\\
\Gamma_{11}^{(2)2}&=&\frac{1}{(u^1)^3}\left[\Gamma_{22}^{(1)2}u^1 (u^2)^2+ \Gamma_{32}^{(1)2} (2u^1 u^2 u^3-(u^2)^3)+u^2 u^1+\Gamma_{32}^{(1)1}( u^1 (u^3)^2-(u^2)^2 u^3)\right],\\
\Gamma_{11}^{(2)3}&=&\frac{1}{(u^1)^3}\left[\Gamma_{22}^{(1)3} u^1 (u^2)^2+\Gamma_{32}^{(1)3}(2 u^1 u^2 u^3-(u^2)^3)
+\Gamma_{32}^{(1)2}(u^1 (u^3)^2-(u^2)^2u^3)+ u^1 u^3\right.\\
&&\left. -(u^2)^2+\Gamma_{22}^{(1)1} ((u^2)^2 u^3-(u^3)^2u^1)\right],\\
\Gamma_{21}^{(2)1}&=&-\frac{1}{(u^1)^2}\left[\Gamma_{22}^{(1)1}u^1 u^2+\Gamma_{32}^{(1)1}( u^1 u^3-(u^2)^2)\right],\\
\Gamma_{21}^{(2)2}&=&-\frac{1}{(u^1)^2}\left[\Gamma_{22}^{(1)2} u^1 u^2+\Gamma_{32}^{(1)2}(u^1 u^3-(u^2)^2)+ u^1-\Gamma_{32}^{(1)1} u^2 u^3\right],\\
\end{eqnarray*}
\begin{eqnarray*}
\Gamma_{21}^{(2)3}&=&-\frac{1}{(u^1)^2}\left[\Gamma_{22}^{(1)3} u^1 u^2+\Gamma_{32}^{(1)3}( u^1 u^3-(u^2)^2 )+(\Gamma_{22}^{(1)1}-\Gamma_{32}^{(1)2}) u^2 u^3- u^2\right],\\
\Gamma_{31}^{(2)1}& =& -\frac{u^2}{u^1}\Gamma_{32}^{(1)1},\\
\Gamma_{31}^{(2)2}&=&-\frac{1}{u^1}\left[\Gamma_{32}^{(1)2} u^2+\Gamma_{32}^{(1)1} u^3\right],\\
\Gamma_{31}^{(2)3}&=& \frac{1}{u^1}\left[-\Gamma_{32}^{(1)3}u^2+(\Gamma_{22}^{(1)1}-\Gamma_{32}^{(1)2})u^3-1\right].
\end{eqnarray*}

Now we use the expression of the Euler vector field in the canonical coordinates and impose the commutativity of $\nabla_{(2)}$ with $\rm{Lie}_E$, coming from the flatness of $\nabla_{(2)}$. Let $T$ be a general vector field, then we impose 
$$\rm{Lie}_E\nabla_{(2)j} T^i-\nabla_{(2)j} (\rm{Lie}_E T^i)=0,$$
that is 
$$E(\partial_j T^i+\Gamma^{(2)i}_{jk}T^k)-(\nabla_{(2)j} T^l)\partial_l E^i+(\nabla_{(2)l} T^i)\partial_j E^l-\partial_j(\rm{Lie}_ET^i)-\Gamma^i_{jk}\rm{Lie}_ET^k=0.$$
Expanding further we obtain the following system of PDEs for $\Gamma^{(2)i}_{jk}(u^2, u^3)$:
$$E^m\partial_m \Gamma^{(2)i}_{jk}-\Gamma^{(2)m}_{jk}\partial_m E^i+\Gamma^{(2)i}_{mk}\partial_j E^m+\Gamma^{(2)i}_{jm}\partial_k E^m+\partial_j\partial_kE^i=0.$$
Since $\Gamma^{(2)i}_{jk}$ are expressed uniquely in terms of $\Gamma^{(1)i}_{jk}$, the previous system of PDEs reduces to a system for the unknown $\Gamma^{(1)i}_{jk}$. 

In particular, we observe that for $[j,k,i]=[3,2,2]$ we get the PDE:
$$ u^2(\partial_{u^2}\Gamma_{32}^{(1)2})+\partial_{u^3}\Gamma_{32}^{(1)2}u^3+\Gamma_{32}^{(1)2}=0,
$$
for $[j,k,i]=[2,3,3]$ we obtain the PDE:
$$ u^2(\partial_{u^2}\Gamma_{32}^{(1)3})+\partial_{u^3}\Gamma_{32}^{(1)3}u^3+\Gamma_{32}^{(1)3}=0,
$$
and finally for $[j,k,i]=[3,1,1]$ we get the PDE:
$$ 
u^2(\partial_{u^2}\Gamma_{32}^{(1)1})+\partial_{u^3}\Gamma_{32}^{(1)1}u^3+\Gamma_{32}^{(1)1}=0.
$$
The general solutions of these PDEs can be obtained directly with the method of characteristics yielding 
$$\Gamma_{32}^{(1)1}=F_1\left(\frac{u^3}{u^2} \right)\frac{1}{u^2}, \;
\Gamma_{32}^{(1)3}=F_2\left(\frac{u^3}{u^2} \right)\frac{1}{u^2},\; \Gamma_{32}^{(1)3}=F_3\left(\frac{u^3}{u^2} \right)\frac{1}{u^2}.$$
Substituting these solutions in the remaining equations, we obtain for $[j,k,i]=[2,2,2]$, for $[j,k,i]=[2,2,1]$ and for $[j,k,i]=[2,2,3]$ identical PDEs for $\Gamma_{22}^{(1)2}$, $\Gamma_{22}^{(1)1}$ and $\Gamma_{22}^{(1)3}$.
These yield the solutions:
$$\Gamma_{22}^{(1)1}=F_4\left(\frac{u^3}{u^2} \right)\frac{1}{u^2}, \;\Gamma_{22}^{(1)2}=F_5\left(\frac{u^3}{u^2} \right)\frac{1}{u^2}, \; 
\Gamma_{22}^{(1)3}=F_6\left(\frac{u^3}{u^2} \right)\frac{1}{u^2}.$$

%Combining this system of PDEs with the algebraic constraints obtained above we get the following general solution:
%$$\Gamma_{1j}^k=0, \; 1\leq j,k\leq 3,\;
%\Gamma_{22}^1=F_5\left(\frac{u^3}{u^2+1} \right)\frac{1}{u^2+1},$$
%$$\Gamma_{22}^2=F_1\left(\frac{u^3}{u^2+1} \right)\frac{1}{u^2+1}$$
%$$\Gamma_{22}^3=F_2\left(\frac{u^3}{u^2+1} \right)\frac{1}{u^2+1}$$
%$$\Gamma_{23}^1=F_3\left(\frac{u^3}{u^2+1} \right)\frac{1}{u^2+1}$$
%$$\Gamma_{23}^2=F_6\left(\frac{u^3}{u^2+1} \right)\frac{1}{u^2+1}$$
%$$\Gamma_{23}^3=F_4\left(\frac{u^3}{u^2+1} \right)\frac{1}{u^2+1}$$
%$$\Gamma_{32}^1=F_3\left(\frac{u^3}{u^2+1} \right)\frac{1}{u^2+1}$$
%$$\Gamma_{32}^3=F_4\left(\frac{u^3}{u^2+1} \right)\frac{1}{u^2+1}$$
%$$\Gamma_{33}^1=0$$
%$$\Gamma_{33}^2=F_3\left(\frac{u^3}{u^2+1} \right)\frac{1}{u^2+1}$$
%$$\Gamma_{33}^3=-F_5\left(\frac{u^3}{u^2+1} \right)\frac{1}{u^2+1}+F_6\left(\frac{u^3}%{u^2+1} \right)\frac{1}{u^2+1}$$

Imposing the zero curvature conditions for $\nabla_{(1)}$, we obtain the system of equations (\ref{F1},\,\ref{F2},\,\ref{F3},\,\ref{F4},\,\ref{F5},\,\ref{F6}) for the unknown functions $F_i$ in the variable $z=\frac{u^3}{u^2}$.

To conclude we observe that it is easy to check by straightforward computations that the remaining conditions (namely the flatness of $\nabla_2$ and the compatibility of $\nabla_2$ with $*$) are automatically satisfied once the functions $F_i$ are chosen among the solutions of the system (\ref{F1},\,\ref{F2},\,\ref{F3},\,\ref{F4},\,\ref{F5},\,\ref{F6}).
\begin{flushright}
$\Box$
\end{flushright}

The system (\ref{F1},\,\ref{F2},\,\ref{F3},\,\ref{F4},\,\ref{F5},\,\ref{F6}) reduces to the full Painlev\' e IV family of equations. 

\begin{thm}
Regular bi-flat $F$-manifolds in dimension three such that $L_p$ has three equal eigenvalues and one Jordan block are locally parameterized by solutions of the full Painlev\' e IV equation. 
\end{thm}

\emph{ Proof}. 
It is straightforward to check that the system of ordinary differential equations given by (\ref{F1},\,\ref{F2},\,\ref{F3},\,\ref{F4},\,\ref{F5},\,\ref{F6}) admits the following integrals of motion:
$$I_1=F_1,\;  I_2=2F_1z+F_3+F_4, \; I_3=-2F_3z+F_4z-F_2-F_5.$$
Using these first integrals, the system can be reduced to a system of three ODEs given by:
\begin{eqnarray*}
\frac{d F_4}{dz}&=&4I_1^2z^2-2I_1I_2z+F_4I_1z+I_2F_4-I_3I_1-F_4^2-2I_1F_5-I_1,\\
\frac{d F_5}{dz}&=&-4I_1^2z^3+6I_2I_1z^2-9F_4I_1z^2-2I_2^2z+6I_2F_4z+I_3I_1z-4F_4^2z-I_2I_3\\
&&-I_2F_5+I_3F_4+F_4F_5-I_1F_6+3I_1z-I_2+F_4,\\
\frac{d F_6}{dz}&=& -4I_2I_1z^3+12F_4I_1z^3+2I_2^2z^2-9I_2F_4z^2-4I_3I_1z^2+8F_4^2z^2\\
&&-6I_1F_5z^2+3I_2I_3z+5I_2F_5z-5I_3F_4z-8F_4F_5z-I_1F_6z-6I_1z^2\\
&&+3I_2z+I_3^2+3I_3F_5+F_6F_4-5F_4z+2F_5^2+I_3+F_5.
\end{eqnarray*}

We further reduce this system to a second order ODE in the following way. First we express $F_5(z)$ in terms of $F_4(z)$ and its first derivative using the first equation, obtaining (here and thereafter we assume $I_1\neq 0$):
$$F_5=\frac{1}{2I_1}\left(4I_1^2z^2-2I_2I_1z+F_4I_1z+I_2F_4-I_3I_1-F_4^2-\frac{dF_4}{dz}-I_1\right).$$
We substitute this in the second equation and solve for $F_6$:
\begin{eqnarray*}
F_6& =& -\frac{1}{2I_1^2} \left(8I_1^3z^3-8I_1^2I_2z^2+14I_1^2F_4z^2+2I_1I_2^2z-9I_1F_4I_2z-2I_1^2I_3z+7I_1F_4^2z+F_4I_2^2\right.\\
&&\left.+I_1I_2I_3-2F_4^2I_2-I_1F_4I_3+\frac{dF_4}{dz}zI_1+F_4^3+2I_1^2z-I_1I_2-F_4\frac{dF_4}{dz}-\frac{d^2F_4}{dz^2}\right). 
\end{eqnarray*}
Substituting these expressions for $F_5$ and $F_6$ in terms of $F_4$ and its derivatives in the last ODE of the system above, we obtain a third order nonlinear ODE for $F_4$. Multiplying it by $(-I_1z+I_2-F_4)$ and by $2I_1^2$, it is possible to recognize it that it is a total derivative with respect to $z$ of an expression involving the second derivative of $F_4$. Integrating this expression one obtains the nonlinear second order ODE:
\begin{eqnarray*}
0&=&8I_2^2I_1^2z^2-I_3I_1^3z^2-10I_2I_1^3z^3+13I_1^3z^3F_4-2I_2^3I_1z-I_3F_4^2I_1+8F_4^3I_1z-2I_2zI_1^2\\
&&-2I_1F_4I_2+2F_4zI_1^2+\frac{7}{2}I_2^2F_4^2+\frac{31}{2}F_4^2I_1^2z^2+2I_2I_3I_1^2z-23I_1^2I_2z^2F_4
+11I_1I_2^2zF_4\\
&&-17I_2F_4^2I_1z-2I_1^2I_3zF_4+2I_1I_2I_3F_4+\frac{3}{2}F_4^4+\frac{1}{2}\left(\frac{dF_4}{dz}\right)^2+C_1\\
&&+I_1\frac{dF_4}{dz}+I_1^3z^2+I_1F_4^2+(-I_1z+I_2-F_4)\frac{d^2F_4}{dz^2}+4I_1^4z^4-I_2^3F_4-4I_2F_4^3,
\end{eqnarray*}
where $C_1$ is the constant of integration. 
Now we show that this ODE can be reduced a three-parameters ODE that contains the full Painlev\'e IV equation for a special value of one of the parameters. 

First we do a change of variables of the form 
$F_4(z)=f(z)-I_1z+I_2$ in order to obtain a term of the form $f(z)\frac{d^2 f}{dz^2}$ which is the term that appears in Painlev\'e IV. 
Doing this we obtain the following ODE:
\begin{eqnarray*}
0&=&\frac{3}{2}f(z)^4+\left(2zI_1+2I_2\right)f(z)^3+\left(-I_3I_1+\frac{1}{2}I_1^2z^2+I_1zI_2+I_1+\frac{1}{2}I_2^2\right)f(z)^2\\
&&-f(z)\frac{d^2f(z)}{dz^2}+I_3I_1I_2^2-\frac{1}{2}I_1^2-I_1I_2^2+\frac{1}{2}\left(\frac{d^2f(z)}{dz^2}\right)+C1.
\end{eqnarray*}
Since in Painlev\'e IV the coefficient in front of $f(z)^3$ is of $4z$, we introduce the affine transformation $z=w-\frac{I_2}{I_1}$ (assuming $I_1\neq 0$) and we call $g(w)=f\left(w-\frac{I_2}{I_1}\right)$. In this way, the previous ODE becomes:
\begin{eqnarray*}
g(w)\frac{d^2 g(w)}{dw^2}&=&\frac{3}{2}g(w)^4+2I_1wg(w)^3+\left(-I_3I_1+I_1+\frac{1}{2}I_1^2w^2\right)g(w)^2\\
&&+\frac{1}{2}\left(\frac{dg(w)}{dw}\right)^2+I_3I_1I_2^2-I_1I_2^2-\frac{1}{2}I_1^2+C_1.
\end{eqnarray*}

The previous ODE depends on the following combination of constants: $I_1$, $I_1-I_3I_1$ and on $I_3I_1I_2^2-I_1I_2^2-\frac{1}{2}I_1^2+C_1$. So calling $A=I_1, B=I_1-I_3I_1, C=I_3I_1I_2^2-I_1I_2^2-\frac{1}{2}I_1^2+C_1$ we can rewrite it as
\begin{eqnarray*}
g(w)\frac{d^2 g(w)}{dw^2}&=&\frac{3}{2}g(w)^4+2Awg(w)^3+\left(B+\frac{1}{2}A^2w^2\right)g(w)^2\\
&&+\frac{1}{2}\left(\frac{dg(w)}{dw}\right)^2+C.
\end{eqnarray*}
The solutions of this ODE parameterize locally bi-flat regular $F$-manifolds in dimension three.
Now we prove that this equation is equivalent to the full Painlev\'e IV family, using a suitable double scaling of the independent and dependent variables.   

We introduce the variable $t=\alpha w$ and rescale $g$ to $\beta g$. Moreover, we introduce the notation $y(t)=g\left(\frac{t}{\alpha}\right)=g(w)$. Then 
$$\frac{d g(w)}{dw}=\frac{d y(t)}{dt}\frac{dt}{dw}=\frac{dy(t)}{dt}\alpha.$$
Performing both transformations the previous ODE gets rescaled to 
\begin{eqnarray*}
\beta^2\alpha^2 y(t)\frac{d^2 y(t)}{dt^2}&=&\frac{3}{2}\beta^4 y(t)^4+2\frac{A}{\alpha}t\beta^3 y(t)^3+\left(B+\frac{1}{2}\frac{A^2}{\alpha^2}t^2\right)\beta^2y(t)^2\\
&&+\frac{1}{2}\beta^2\alpha^2\left(\frac{dy(t)}{dt}\right)^2+C.
\end{eqnarray*}
To reduce this to a constant multiple $\gamma$ of the full Painlev\' e IV family, we look for a nontrivial solution of the following algebraic system:
$$\beta^2\alpha^2 =\gamma, \;\; \beta^4=\gamma, \; \; \frac{A}{\alpha}\beta^3=2\gamma, \; \; \frac{1}{2}\frac{A^2}{\alpha^2}\beta^2=2\gamma.$$
A nontrivial solution is given by $\alpha=\beta=\sqrt{\frac{A}{2}}, \; \gamma=\frac{A^2}{4}.$ With this choice the ODE becomes:
\begin{eqnarray*}
\frac{A^2}{4} y(t)\frac{d^2 y(t)}{dt^2}&=&\frac{3}{2}\frac{A^2}{4} y(t)^4+A^2t y(t)^3+\left(B\frac{A}{2}+\frac{1}{2}A^2t^2\right)y(t)^2\\
&&+\frac{1}{2}\frac{A^2}{4}\left(\frac{dy(t)}{dt}\right)^2+C.
\end{eqnarray*}
If $A=I_1\neq 0$, then dividing both sides by $\frac{A^2}{4}$ we obtain 
\begin{eqnarray*}
 y(t)\frac{d^2 y(t)}{dt^2}&=&\frac{3}{2} y(t)^4+4t y(t)^3+\left(2\frac{B}{A}+2t^2\right)y(t)^2\\
&&+\frac{1}{2}\left(\frac{dy(t)}{dt}\right)^2+\frac{CA^2}{4}
\end{eqnarray*}
Introducing the constants $c=\frac{CA^2}{4}$ and $b=-\frac{B}{A}$ we obtain 
\begin{eqnarray*}
 y(t)\frac{d^2 y(t)}{dt^2}&=&\frac{3}{2} y(t)^4+4t y(t)^3+2\left(t^2-b\right)y(t)^2\\
&&+\frac{1}{2}\left(\frac{dy(t)}{dt}\right)^2+c,
\end{eqnarray*}
which is indeed the full Painlev\' e IV family.

\begin{flushright}
$\Box$
\end{flushright}

\begin{rmk}
In the proof of the previous Theorem we have assumed that $I_1\neq 0$, hence the genericity statement. If $I_1=0$ then the system  (\ref{F1},\,\ref{F2},\,\ref{F3},\,\ref{F4},\,\ref{F5},\,\ref{F6}) reduces to a system of ODEs that can be integrated explicitly. In particular, using the integrals of motion $I_1=0$, $I_2$ and $I_3$, the system obtained by reduction and 
involving only $F_4$, $F_5$ and $F_6$ is lower triangular. 
\end{rmk}

\subsection{The case of two distinct eigenvalues and two Jordan blocks}

In this subsection we analyze the case in which the operator $L_p$ has two distinct eigevanlues, one eigenvalue with algebraic multiplicity two (and nontrivial $2\times2$ Jordan block), while the other eigenvalue is simple. In this case, we use Hertling's Decomposition Lemma (Thereom 2.11   from \cite{H}) to obtain the following result. 

\begin{thm}
Let $(M, \nabla_1, \nabla_2, \circ, *, e, E)$ be  non-semisimple regular bi-flat $F$-manifold in dimension three such that $L_p$ has exactly two distinct eigenvalues and two Jordan blocks. Then there exist local coordinates $\{u^1, u^2, u^3\}$ such that 
\begin{enumerate}
\item $e, E, \circ $ are given by 
\begin{eqnarray}
e&=&\partial_{u^1}+\partial_{u^3}\\
E&=&u^1\partial_{u^1}+u^2\partial_{u^2}+u^3\partial_{u^3}\\
c^i_{jk}&=&\delta^k_{i+j-1} \;\text{ if }\; 1\leq i,j,k\leq 2\\
c^3_{33}&=&1\\
c^i_{jk}&=& 0 \; \;\text{ in all other cases }
\end{eqnarray} 
\item The Christoffel symbols $\Gamma_{jk}^{(1)i}$ for $\nabla_1$ are given by:
$$\Gamma_{13}^{(1)3}=\frac{F_4\left(\frac{u^3-u^1}{u^2}\right)}{u^2},\; \;
\Gamma_{22}^{(1)1}= \frac{F_3\left(\frac{u^3-u^1}{u^2}\right)}{u^2},\;\; 
\Gamma_{22}^{(1)2}=\frac{F_6\left(\frac{u^3-u^1}{u^2}\right)}{u^2},$$
$$
\Gamma_{23}^{(1)3}=\frac{F_1\left(\frac{u^3-u^1}{u^2}\right)}{u^2},\;\;
\Gamma_{31}^{(1)1}=\frac{F_2\left(\frac{u^3-u^1}{u^2}\right)}{u^2},\;\;
\Gamma_{31}^{(1)2}=\frac{F_5\left(\frac{u^3-u^1}{u^2}\right)}{u^2},\;\;,$$
$$\Gamma_{11}^{(1)1}=-\Gamma_{31}^{(1)1},\;\; 
\Gamma_{11}^{(1)2}=-\Gamma_{31}^{(1)2},\;\;
\Gamma_{11}^{(1)3}=-\Gamma_{13}^{(1)3},\;\;  
 \Gamma_{12}^{(1)2}=-\Gamma_{31}^{(1)1},\;\;      
  \Gamma_{12}^{(1)3}=-\Gamma_{23}^{(1)3},\;\; $$
$$   \Gamma_{21}^{(1)2}=-\Gamma_{31}^{(1)1},\;\; 
 \Gamma_{21}^{(1)3}=-\Gamma_{23}^{(1)3},\;\; \Gamma_{23}^{(1)2}=\Gamma_{31}^{(1)1},\;\;$$
 $$ \Gamma_{33}^{(1)1}=-\Gamma_{31}^{(1)1},\;\; \Gamma_{33}^{(1)2}=-\Gamma_{31}^{(1)2},\;\; \Gamma_{33}^{(1)3}=-\Gamma_{13}^{(1)3},$$
 where the functions $F_1, \dots F_6$ satisfy the system 
\begin{eqnarray}
\label{F1bis}
\frac{dF_1}{dz}& =& -\frac{F_3F_4-F_1^2+F_1F_6+F_1}{z},\\
\label{F2bis}
\frac{d F_2}{dz}&=& \frac{F_3F_5-F_2F_1-F_2}{z},\\
\label{F3bis}
\frac{d F_3}{dz}&=&0,\\
\label{F4bis}
\frac{d F_4}{dz}&=&-\frac{F_3F_4-F_1^2+F_1F_6+F_4z+F_1}{z^2},\\
\label{F5bis}
\frac{d F_5}{dz}&=&\frac{-F_5F_1 z+F_5F_6z+F_2F_4z+F_3 F_5-F_5z-F_2F_1-F_2}{z^2},\\
\label{F6bis}
\frac{d F_6}{dz}&=&-2F_3F_5+2F_2F_1.
\end{eqnarray}
 in the variable $z=\frac{u^3-u^1}{u^2}$ while the other symbols not obtainable from the  above list using the symmetry of the connection are identically zero.         
\item The Christoffel symbols $\Gamma_{jk}^{(2)i}$ for $\nabla_2$ are uniquely determined by the Christoffel symbols of $\nabla_1$ via the following formulas:
$$
\Gamma_{11}^{(2)1} = \frac{\Gamma_{22}^{(1)1}(u^2)^2-\Gamma_{31}^{(1)1}u^3u^1-u^1}{(u^1)^2},\; \;    
\Gamma_{11}^{(2)2}  = \frac{\Gamma_{31}^{(1)1}u^3u^2-\Gamma_{31}^{(1)2}u^3u^1+\Gamma_{22}^{(1)2}(u^2)^2+u^2}{(u^1)^2},$$
$$
\Gamma_{11}^{(2)3}= \frac{\Gamma_{23}^{(1)3}u^2u^3-\Gamma_{13}^{(1)3}u^1u^3}{(u^1)^2},\; \; 
\Gamma_{12}^{(2)1}= -\frac{\Gamma_{22}^{(1)1}u^2}{u^1},\;\;
\Gamma_{12}^{(2)2}= -\frac{\Gamma_{31}^{(1)1}u^3+\Gamma_{22}^{(1)2}u^2+1}{u^1},$$
$$
\Gamma_{12}^{(2)3}= -\frac{\Gamma_{23}^{(1)3}u^3}{u^1},\;\;
\Gamma_{13}^{(2)1}= \Gamma_{31}^{(1)1},\;\;
\Gamma_{13}^{(2)2}= \Gamma_{31}^{(1)2},\;\;
\Gamma_{13}^{(2)3}= \Gamma_{13}^{(1)3},$$
%\Gamma_{21}^{(2)1}&=& _\frac{F_5}{u^1}\\
%\Gamma_{21}^{(2)2}&=& _F6(-(u[1]-u[3])*(1/u[2]))*(1/u[2])
$$
\Gamma_{22}^{(2)1}=\Gamma_{22}^{(1)1},\;\;
\Gamma_{22}^{(2)2}=\Gamma_{22}^{(1)2},\;\;
\Gamma_{23}^{(2)2}= \Gamma_{31}^{(1)1},\;\;
\Gamma_{23}^{(2)3}=\Gamma_{23}^{(1)3},\;\; 
\Gamma_{33}^{(2)1}=-\frac{\Gamma_{31}^{(1)1} u^1}{u^3},$$
$$
\Gamma_{33}^{(2)2}= -\frac{\Gamma_{31}^{(1)2}u^1+\Gamma_{32}^{(1)2}u^2}{u^3},\;\;
\Gamma_{33}^{(2)3}= -\frac{\Gamma_{23}^{(1)3}u^2+\Gamma_{13}^{(1)3}u^1+1}{u^3},
$$     
while the other symbols not obtainable from the  above list using the symmetry of the connection vanish identically. 
\item The dual product $*$ is obtained via formula \eqref{nm} using $\circ$ and $E$.
\end{enumerate}
\end{thm}
\emph{ Proof}
The first point of the Theorem is a direct consequence of the results of \cite{DH} and of 
Hertling's Decomposition Lemma (Thereom 2.11  from \cite{H}) .
Imposing that $\nabla_{(1)}$ is torsionless, that it is compatible with $\circ$, and that it satisfies $\nabla_{(1)}e=0$, we obtain the following constraints on $\Gamma_{ij}^{(1)k}$:
$$
\Gamma_{11}^{(1)1} = -\Gamma_{31}^{(1)1},\;\; \Gamma_{11}^{(1)2} = -\Gamma_{31}^{(1)2},\; \; \Gamma_{11}^{(1)3}= -\Gamma_{13}^{(1)3},\;\; \Gamma_{12}^{(1)1} = 0,\;\; \Gamma_{12}^{(1)2} = -\Gamma_{31}^{(1)1},$$
$$\Gamma_{33}^{(1)1} = -\Gamma_{31}^{(1)1}, \;\;  \Gamma_{33}^{(1)2} = -\Gamma_{31}^{(1)2},\;\; \Gamma_{33}^{(1)3} = -\Gamma_{13}^{(1)3},
$$
$$ \Gamma_{12}^{(1)3} = -\Gamma_{23}^{(1)3},  \; \; \Gamma_{21}^{(1)1} = 0,\;\;  \Gamma_{21}^{(1)3}= -\Gamma_{23}^{(1)3}, $$
$$ \Gamma_{22}^{(1)3}= 0,\;\; \Gamma_{23}^{(1)1} = 0,\;\; \Gamma_{23}^{(1)2} = \Gamma_{31}^{(1)1},$$  
together with the trivial constraints $\Gamma_{ij}^{(1)k}=\Gamma_{ji}^{(1)k}$. 

Then we impose a series of constraints on the dual connection $\nabla_{(2)}$ that are sufficient to determine uniquely the Christoffel symbols $\Gamma_{ij}^{(2)k}$ in terms of the Christoffel symbols $\Gamma_{ij}^{(1)k}$. These constraints are  the requirement that $\nabla_{(2)}$ is almost hydrodynamically equivalent to $\nabla_{(1)}$,  i.e.
$(d_{\nabla_1}-d_{\nabla_2})(X\,\circ)=0$ and $\nabla_{(2)}E=0$. These constraints give the formulas in the third point, that express $\Gamma_{ij}^{(2)k}$ in terms of $\Gamma_{ij}^{(1)k}$. 

Once we have expressed the Christoffel symbols of $\nabla_{(2)}$ in terms of the Christoffel symbols of $\nabla_{(1)}$, we obtain a system of PDEs in $\Gamma_{ij}^{(1)k}$, imposing that commutativity of $\nabla_{(2)}$ with ${\rm Lie}_E$ and the commutativity of ${\rm Lie}_e$ with $\nabla_{(1)}$. The latter system in particular implies that $\Gamma_{ij}^{(1)k}(u^1, u^2, u^3)$ can be expressed as functions of two variables, as $\Gamma_{ij}^{(1)k}(u^2, u^3-u^1)$. 
Following a procedure similar to process described in the proof of Theorem \ref{thm1}, we can solve the two systems and we find that (here $z=\frac{u^3-u^1}{u^2}$):
$$\Gamma_{13}^{(1)3} = \frac{F_4(z)}{u^2},\;\; \Gamma_{22}^{(1)1} = \frac{F_3(z)}{u^2}, \;\;\Gamma_{22}^{(1)2} =  \frac{F_6(z)}{u^2}, $$
$$\Gamma_{23}^{(1)3} =  \frac{F_1(z)}{u^2},\;\; \Gamma_{31}^{(1)1} =  \frac{F_2(z)}{u^2},\;\;\Gamma_{31}^{(1)2}=  \frac{F_5(z)}{u^2}, \;\;\Gamma_{32}^{(1)2} =  \frac{F_7(z)}{u^2},$$
for arbitrary smooth functions $F_i(z)$. 
At this point, we impose that $\nabla_{(2)}$ is a torsionless connection and this gives the only constraint $\Gamma_{31}^{(1)1}=\Gamma_{32}^{(1)2}$ or equivalently $F_2(z)=F_7(z)$. 

Imposing the zero curvature conditions for $\nabla_{(1)}$, we obtain the system of equations (\ref{F1bis},\ref{F2bis},\ref{F3bis},\ref{F4bis},\ref{F5bis},\ref{F6bis}).

To conclude we observe that it is easy to check by straightforward computations that the remaining conditions (namely the flatness of $\nabla_2$ and the compatibility of $\nabla_2$ with $*$) are automatically satisfied once the functions $F_i$ are chosen among the solutions of the system (\ref{F1bis}, \ref{F2bis}, \ref{F3bis}, \ref{F4bis}, \ref{F5bis}, \ref{F6bis}).

\begin{flushright}
$\Box$
\end{flushright}

Now we prove that the system (\ref{F1bis}, \,\ref{F2bis}, \,\ref{F3bis}, \,\ref{F4bis}, \,\ref{F5bis}, \,\ref{F6bis}) can be reduced to Painlev\' e V equation. 
\begin{thm}
Regular bi-flat $F$-manifolds in dimension three such that $L_p$ has two distinct eigenvalues and two Jordan blocks are  locally parameterized by solutions of the full Painlev\' e V equation. 
\end{thm}
\emph{Proof}
Since $\frac{d F_3}{dz}=0$  we set $F_3=I_1$. It is straightforward to check via direct computation that the system above has two additional integrals of motion, namely $F_1-F_4z=I_2$ and $F_6+2F_2=I_3$, where we have indicated with $I_2$ and $I_3$ the corresponding values of the integrals of motion. Using these three integrals of motion in the system above we reduce it to the following three ODEs:
\begin{equation}
\label{eqaux1}
 -F_4F_2z+I_1F_5-I_2F_2-\frac{dF_2}{dz}z-F_2=0, 
 \end{equation}
 \begin{equation}
 \label{eqaux2}
  -F_4F_5z-2F_2F_5z-I_2F_5+I_3F_5+F_4F_2-\frac{dF_5}{dz}z-F_5+\frac{dF_2}{dz}=0,
  \end{equation}
  \begin{equation}
  \label{eqaux3}
  F_4^2z^2+2F_4F_2z^2+2I_2F_4z+2I_2F_2z-I_3F_4z-F_4I_1+I_2^2-I_2I_3-\left(\frac{dF_4}{dz}z+F_4\right)z-F_4z-I_2=0.
  \end{equation}
We solve for $F_5$ in \eqref{eqaux1} and we substitute in \eqref{eqaux2} to obtain a second order ODE in $F_2$ and $F_4$, call it $\alpha$. We solve for $F_2$ in the third equation \eqref{eqaux3} and we substitute in $\alpha$ thus obtaining a complicate nonlinear \emph{third order} ODE involving only $F_4$, given by (we have renamed $F_4$ with $F$):
$$
-z^8\left(\frac{dF}{dz}\right)^3+z\left(4Fz^6+7I_2z^5\right)\left(\frac{dF}{dz}\right)^2+z\left(2Fz^7+2z^6I_2\right)\left(\frac{dF}{dz}\right)\left(\frac{d^2F}{dz^2}\right)+$$
$$z(3F^4z^7+12I_2F^3z^6-2I_3F^3z^6-2I_1F^3z^5+18I_2^2F^2z^5-6I_2I_3F^2z^5-2F^3z^6$$
$$-6I_1I_2F^2z^4+12I_2^3Fz^4-6I_2^2I_3Fz^4-6I_2F^2z^5-6I_1I_2^2Fz^3+3I_2^4z^3-2I_2^3I_3z^3$$
$$-6I_2^2Fz^4-3F^2z^5-2I_1I_2^3z^2-2I_2^3z^3-10I_2Fz^4
+I_1^2I_2^2z-10I_2^2z^3)\left(\frac{dF}{dz}\right)$$
$$+z\left(-5F^2z^6-12I_2Fz^5-7I_2^2z^4\right)\left(\frac{d^2F}{dz^2}\right)+z\left(-F^2z^7-2I_2Fz^6-I_2^2z^5\right)\left(\frac{d^3F}{dz^3}\right)$$
$$+4F^5z^7+z(17I_2z^5-3I_3z^5-I_1z^4-3z^5)F^4$$
$$+z\left(28I_2^2z^4-10I_2I_3z^4-2I_1I_2z^3-I_1I_3z^3-10I_2z^4-I_1^2z^2-I_1z^3\right)F^3$$
$$+z\left(22I_2^3z^3-12I_2^2I_3z^3-3I_1I_2I_3z^2-12I_2^2z^3-3I_1^2I_2z-3I_1I_2z^2-I_2z^3\right)F^2$$
$$+z\left(8I_2^4z^2-6I_2^3I_3z^2+2I_1I_2^3z-3I_1I_2^2I_3z-6I_2^3z^2-I_1^2I_2^2-3I_1I_2^2z-2I_2^2z^2\right)F$$
$$+z\left(I_2^5z-I_2^4I_3z+I_1I_2^4-I_1I_2^3I_3-I_2^4z-I_1I_2^3\right)=0$$

This ODE can be reduced to a second order ODE since it admits a nontrivial integrating factor $\mu$ given by 
$$\mu=\frac{-Fz^2-I_2z+I_1}{z^3(Fz+I_2)^3}.$$
The resulting second order nonlinear ODE is given by:
$$
\left(-F^3z^8-3I_2^2z^7+I_1F^2z^6-3I_2^2Fz^6+2I_1I_2Fz^5-I_2^3z^5+z^4I_1I_2^2\right)\left(\frac{d^2F}{dz^2}\right)$$
$$+\left(F^2z^8+I_2^2z^6+2I_2Fz^7-\frac{1}{2}I_1Fz^6-\frac{1}{2}I_2z^5I_1\right)\left(\frac{dF}{dz}\right)^2+$$
$$+\left(-z^7F^3-5z^6I_2F^2+3z^5I_1F^2-7z^5I_2^2I_2^2F+7z^4I_1I_2F-3I_2^3z^4+4z^3I_1I_2^2\right)\left(\frac{dF}{dz}\right)+$$
$$+F^6z^8+\left(-z^7-I_3z^7+6I_2z^7-\frac{5}{2}I_1z^6\right)F^5+$$
$$+\left(-5I_2I_3z^6-\frac{25}{2}I_2z^5I_1+2I_1I_3z^5+15I_2^2z^6-5I_2z^6+2I_1^2z^4+2I_1z^5\right)F^4$$
$$+\left(6I_1I_2I_3z^4-I_1^2I_3z^3-10I_2^2I_3z^5-\frac{45}{2}z^4I_1I_2^2+8I_1^2I_2z^3+\right.$$
$$\left.+6I_1I_2 z^4+Cz^4-I_2z^5-\frac{1}{2}I_1^3z^2-I_1^2z^3+20I_2^3z^5-10I_2^2z^5\right)F^3$$
$$ +\left(6I_1I_2^2I_3z^3-3I_1^2I_2I_3z^2-I_1I_2z^3+6z^3I_1I_2^2-\frac{3}{2}I_1^3I_2z-3I_1^2I_2 z^2+\right.$$
$$\left.+3I_2Cz^3-10I_2^3I_3z^4-\frac{35}{2}I_1I_2^3z^3+11I_1^2I_2^2z^2+15I_2^4z^4-10I_2^3z^4-2 I_2^2z^4\right)F^2+$$
$$+\left(2I_1I_2^3I_3z^2-3I_1^2I_2^2I_3z-5I_1I_2^4z^2+6I_1^2I_2^3z+2I_1I_2^3z^2-3I_1^2I_2^2 z-\frac{5}{2}I_1I_2^2z^2\right.$$
$$\left.+3I_2^2Cz^2-5I_2^4I_3z^3-I_2^3z^3-I_1^3I_2^2+6I_2^5z^3-5I_2^4 z^3\right)F$$
$$+I_2^6z^2+I_1^2I_2^4-I_2^5z^2-I_1^2I_2^3-I_2^5I_3z^2-I_1^2I_2^3I_3+I_2^3Cz-\frac{3}{2}I_1I_2^3z=0.$$
Now we show that this equation can be reduced to Painlev\' e V through a series of nonlinear transformations. 

First we consider the transformation $F\mapsto I_1\frac{F}{z^2}-\frac{I_2}{z}$, in this way the second order ODE above becomes (we have multiplied it by $z^3$):
$$\frac{I_1^3F(-2I_1F^2z^4+2I_1z^4F)}{2z}\left(\frac{d^2F}{dz^2}\right)
+\frac{I_1^3F(2I_1z^4F-I_1z^4)}{2z}\left(\frac{dF}{dz}\right)^2$$
$$+\frac{I_1^3F(-2I_1F^2z^3+2I_1Fz^3)}{2z}\left(\frac{dF}{dz}\right)$$
 $$ +\frac{I_1^3F}{2z}(2I_1^3F^5-2I_1^2I_3F^4z-5I_1^3F^4+4I_1^2I_3F^3z-2I_1^2F^4z+5I_1I_2^2F^2z^2-4I_1I_2I_3F^2z^2+z^2I_1I_2^2$$
 $$+4I_1^3F^3-2I_1^2I_3F^2z+4I_1^2F^3z-2I_1I_2^2Fz^2-4I_1I_2F^2z^2-I_1^3F^2-2I_1^2F^2z-4I_1F^2z^2+2F^2Cz^2) = 0.$$
Then we introduce the transformation $F\mapsto \frac{1}{1-F^{-1}}$ and the constant $\alpha=4I_1^4I_2^2-4I_1^4I_2I_3-4I_1^4I_2-4I_1^4+2I_1^3C$ and express $C$ in terms of $\alpha$ and in terms of the other constants. In this way the equation becomes (after factoring out common factors):
$$(2I_1^4F^2z^2-2I_1^4z^2F)\left(\frac{d^2F}{dz^2}\right)
+(-3I_1^4z^2F+I_1^4z^2)\left(\frac{dF}{dz}\right)^2+$$
$$+(2I_1^4zF^2-2I_1^4zF)\left(\frac{dF}{dz}\right)
+\alpha F^5-3F^4\alpha$$
$$+\left(-2I_1^5\frac{I_3}{z}+\frac{I_1^6}{z^2}+I_1^4I_2^2+3\alpha-2\frac{I_1^5}{z}\right)F^3+\left(2\frac{I_1^5I_3}{z}+\frac{I_1^6}{z^2}-3I_1^4I_2^2-\alpha+2\frac{I_1^5}{z}\right)F^2$$
$$+3I_1^4FI_2^2-I_1^4I_2^2=0.$$

Finally we introduce a transformation of the independent variable, setting $z=\frac{1}{s}$ and defining $G(s)=F(z)=F\left(\frac{1}{s}\right).$ Using this transformation the second order ODE above becomes
$$
\left(2I_1^4G^2s^2-2I_1^4Gs^2\right)\left(\frac{d^2G}{ds^2}\right)+\left(-3I_1^4Gs^2+I_1^4s^2\right)\left(\frac{dG}{ds}\right)^2+\left(2I_1^4G^2s-2I_1^4Gs\right)\left(\frac{dG}{ds}\right)+
$$
$$+\alpha G^5-3G^4\alpha+(I_1^6s^2-2I_1^5I_3s-2I_1^5s+I_1^4I_2^2+3\alpha)G^3+$$
\begin{equation}\label{painleveV1}
+(I_1^6s^2+2I_1^5I_3s+2I_1^5s-3I_1^4I_2^2-\alpha)G^2+3I_1^4I_2^2G-I_1^4I_2^2=0.\end{equation}
Now we show that this is indeed the Painlev\' e V equation. 

Recall that the Painlev\'e V is given by 
$$\frac{d^2y}{dx^2}=\left(\frac{1}{2y}+\frac{1}{y-1}\right)\left(\frac{dy}{dx}\right)^2-\frac{\frac{dy}{dx}}{x}+\frac{(y-1)^2\left(a
y+\frac{b}{y}\right)}{x^2}+\frac{gy}{x}+\frac{dy(y+1)}{y-1}$$
where $a, b, g, d$ are parameters. 
Taking common denominator and multiplying it by $2y(y-1)x^2$ the Painlev\'e becomes
$$(2y^2x^2-2yx^2)\left(\frac{d^2y}{dx^2}\right)+(-3yx^2+x^2)\left(\frac{dy}{dx}\right)^2+(2y^2x-2yx)\left(\frac{dy}{dx}\right)-2y^5a$$
\begin{equation}\label{painleveV2}+6y^4a-(2dx^2+2gx+6a+2b)y^3+(2gx-2dx^2+2a+6b)y^2-6yb+2b=0
\end{equation}
In order to compare \eqref{painleveV1} with \eqref{painleveV2}, we divide
\eqref{painleveV1} by $I_1^4$ (assuming $I_1\neq 0$, see the Remark after the proof) and obtain:
$$ \left(2G^2s^2-2Gs^2\right)\left(\frac{d^2G}{ds^2}\right)+\left(-3Gs^2+s^2\right)\left(\frac{dG}{ds}\right)^2+\left(2G^2s-2Gs\right)\left(\frac{dG}{ds}\right)+
$$
$$+\frac{\alpha}{I_1^4} G^5-G^4\frac{3\alpha}{I_1^4}+\left(I_1^2s^2-2I_1I_3s-2I_1s+I_2^2+\frac{3\alpha}{I_1^4}\right)G^3+$$
\begin{equation}\label{painleveV3}
+\left(I_1^2s^2+2I_1I_3s+2I_1s-3I_2^2-\frac{\alpha}{I_1^4}\right)G^2+3I_2^2G-I_2^2=0.\end{equation}
Comparing \eqref{painleveV3} with \eqref{painleveV2} we get the following correspondence among parameters:
$$2b=-I_2^2,\; \;-2a=\frac{\alpha}{I_1^4}, \;\; -2d=I_1^2, \;\; g=I_1I_3+I_1.$$
These relations can be easily inverted determining $I_1, I_2, I_3, \alpha$ in terms of the parameters $a,b, d, g$.
Thus we have obtained the full Painlev\' e V. 
\begin{flushright}
$\Box$
\end{flushright}

\begin{rmk}
In the proof of the previous Theorem we have assumed that $I_1\neq 0$, hence the genericity statement. If $I_1=0$ then the system  (\ref{F1bis},\,\ref{F2bis},\,\ref{F3bis},\,\ref{F4bis},\,\ref{F5bis},\,\ref{F6bis}) reduces to a system of ODEs that can be integrated explicitly.  
\end{rmk}

\begin{rmk}
In the two-dimensional case there is only one regular non-semisimple model for a bi-flat $F$-manifold (the operator $L$ has necessarily two equal eigenvalues and one Jordan block). The computations become much easier and one can easily show that
 there exist local coordinates $\{u^1, u^2\}$ such that 
\begin{enumerate}
%\item $e, E, \circ$ are given by \eqref{canonical1}, \eqref{canonical2}, \eqref{canonical3}.
\item The Christoffel symbols $\Gamma_{jk}^{(1)i}$ for $\nabla_1$ are given by:
$$\Gamma_{22}^{(1)1}=\f{C_1}{u^2},\qquad 
\Gamma_{22}^{(1)3}=\f{C_2}{u^2},$$
while the other symbols are identically zero.  
\item The Christoffel symbols $\Gamma_{jk}^{(2)i}$ for $\nabla_2$ are uniquely determined by the Christoffel symbols of $\nabla_1$ via the following formulas:
\begin{eqnarray*}
\Gamma_{11}^{(2)1}&=&\f{\Gamma_{22}^{(1)1}(u^2)^2-2u^1}{(u^1)^2},\\
\Gamma_{11}^{(2)2}&=&\f{\Gamma_{22}^{(1)2}(u^2)^2+2u^2}{(u^1)^2},\\
\Gamma_{12}^{(2)1}&=&-\f{u^2}{u^1}\Gamma_{22}^{(1)1},\\
\Gamma_{12}^{(2)2}&=&-\f{u^2}{u^1}\Gamma_{22}^{(1)2}-\frac{2}{u^1},\\
\Gamma_{22}^{(2)1}&=&\Gamma_{22}^{(1)1},\\
\Gamma_{22}^{(2)2}&=&\Gamma_{22}^{(1)2}.\\
\end{eqnarray*}
\end{enumerate}
\end{rmk}

\subsection{Regular case and confluences of  Painlev\' e equations}
In this Section, we have shown that there exists an intimate relationship between regular bi-flat $F$-manifolds in dimension three on one hand and Painlev\'e transcendents on the other. Our analysis leads us to conclude that regular bi-flat $F$-manifolds in dimension three are characterized by continuous and discrete moduli. The discrete moduli are provided by the Jordan normal form for the operator $L$, which in turns determine which of the Painlev\' e equations control the continuous moduli. %To make an analogy with the moduli space of compact Riemann surfaces, the Jordan normal form corresponds to the genus $g$, while the Painlev\' e equations (among P$_{VI}$, P$_{V}$, and P$_{IV}$) play the role of $M_g$, the moduli space of compact Riemann surfaces, once the genus $g$has been fixed.

Furthermore, the well-known confluence of the Painlev\'e equations is associated to a corresponding degeneration of the form of the operator $L$ characterizing regular three-dimensional  bi-flat $F$-manifold. In this way, confluences of the Painlev\'e equations are mirrored in the collision of eigenvalues and the creation of non-trivial Jordan blocks according to the following diagram:

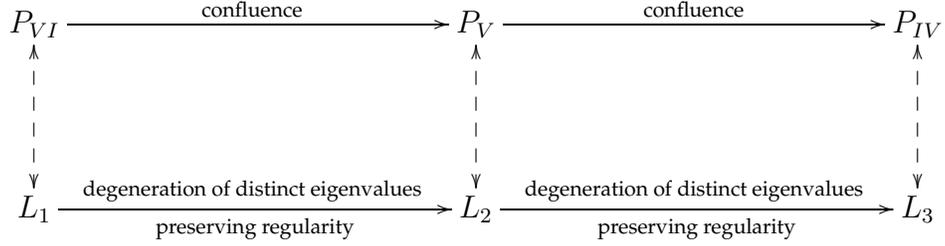
\begin{equation*}
\xymatrix{
P_{VI} \ar@{<-->}[dd] \ar[rrrrr]^{\text{confluence }}&&&&&P_{V}\ar@{<-->}[dd] \ar[rrrrr]^{\text{confluence }} &&&&&P_{IV}\ar@{<-->}[dd]\\
  &&&&&       &&&&&    \\
L_1\ar[rrrrr]^{\text{degeneration of distinct eigenvalues }}_{\text{preserving regularity}}
 &&&&& L_2 \ar[rrrrr]^{\text{degeneration of distinct eigenvalues }}_{\text{preserving regularity}}&&&&& L_3}
\end{equation*}

As an open problem, let us mention the fact that it would be interesting to extend this correspondence to include the remaining Painlev\'e transcendents on one side and possibly non-regular bi-flat $F$-manifolds on the other. 

\section{Multi-flat $F$-manifolds in the regular non-semisimple case}
\subsection{Tri-flat $F$-manifolds }
Contrary to the semisimple situation, in this Section we are going to show that in the regular non-semisimple case
there exist indeed tri and multi-flat $F$-manifolds. For simplicity we focus are attention on the case in which the Jordan normal form of the operator $L$ contains only one Jordan block with the same eigenvalues. In particular the next two theorems show that regular tri-flat and multi-flat $F$-manifolds in dimension three such that $L_p$ has three equal eigenvalues do exist and are locally represented as it follows.

\begin{thm}\label{thmtriflatregular}
Let $(M, \nabla_1, \nabla_2, \nabla_3, \circ_1, \circ_2, \circ_3,  E_1:=e, E_2:=E, E_3:=E^2=E\circ_1 E)$ be a regular tri-flat $F$-manifold in dimension three such that $L_p$ has three equal eigenvalues. Then there exist local coordinates $\{u^1, u^2, u^3\}$ such that 
\begin{enumerate}
\item $E_1:=e, E_2:=E, \circ_1=\circ$ are given by \eqref{canonical1}, \eqref{canonical2}, \eqref{canonical3}.
\item The Christoffel symbols $\Gamma_{jk}^{(1)i}$ for $\nabla_1$ are given by:
$$\Gamma_{23}^{(1)1}=\Gamma_{32}^{(1)1}=\Gamma_{33}^{(1)2}=\frac{f_1}{u^2}, \;
\Gamma_{32}^{(1)3}=\Gamma_{23}^{(1)3}=\frac{F_2\left(\frac{u^3}{u^2} \right)}{u^2},\; \Gamma_{32}^{(1)2}=\Gamma_{23}^{(1)2}=\frac{f_3}{u^2},$$
$$\Gamma_{22}^{(1)1}=\frac{F_4\left(\frac{u^3}{u^2} \right)}{u^2}, \;\Gamma_{22}^{(1)2}=\frac{F_5\left(\frac{u^3}{u^2} \right)}{u^2}, \; 
\Gamma_{22}^{(1)3}=\frac{F_6\left(\frac{u^3}{u^2} \right)}{u^2}, \; 
\Gamma_{33}^{(1)3}=\frac{f_3-F_4\left(\frac{u^3}{u^2} \right)}{u^2},$$
 where $f_1$ and $f_3$ are constants and the functions $F_2, F_4, F_5, F_6$ are given by
\begin{eqnarray}
\label{F2triflatregular}
F_2(z)&=&-f_1z^2-1,\\
\label{F4triflatregular}
F_4(z)&=&-2f_1z,\\
\label{F5triflatregular}
F_5(z)&=&-f_1z^2-2f_3z,\\
\label{F6triflatregular}
F_6(z)&=&-f_3z^2+2z
\end{eqnarray}
in the variable $z=\frac{u^3}{u^2}$ while the other symbols are identically zero.
\item The Christoffel symbols $\Gamma_{jk}^{(2)i}$ for $\nabla_2$ and the Christoffel symbols $\Gamma_{jk}^{(3)i}$ for $\nabla_3$ are uniquely determined by the Christoffel symbols of $\nabla_1$ via the procedure explained in the proof of the theorem. In particular, $\Gamma_{jk}^{(2)i}$ can be expressed in terms of $\Gamma_{jk}^{(1)i}$ via the same formulas appearing in Theorem \ref{thm1}.

\item The product $\circ_2$ is obtained via formula \eqref{nm} using $\circ_1:=\circ$ and $E$ (and analogously for $\circ_3$).
\end{enumerate}
\end{thm}

\n
\emph{ Proof}. The first part of the proof is the same as the proof given for Theorem \ref{thm1}. To determine the Christoffel symbols $\Gamma_{ij}^{(1)k}$ for the torsionless connection $\nabla_1$, the same conditions appearing in the proof of Theorem \ref{thm1} are imposed resulting in a system of algebraic equations for $\Gamma^{(1)k}_{ij}$. These symbols are in general functions of $u^1, u^2, u^3$. Furthermore, imposing the commutativity of $\nabla_1$ and $\rm{Lie}_e$ we get that $\Gamma^{(1)k}_{i,j}$ do not depend on $u^1$. 

The Christoffel symbols for the connection $\nabla_{(2)}$ are determined in terms of the Christoffel symbols for the connection $\nabla_{(1)}$ exactly like in the proof of Theorem \ref{thm1}. 

Furthermore, imposing the commutativity of $\nabla_{(2)}$ with $\rm{Lie}_E$, and using the fact that  
 $\Gamma^{(2)i}_{jk}$ are expressed uniquely in terms of $\Gamma^{(1)i}_{jk}$, we obtain a system of PDEs for the unknowns $\Gamma^{(1)i}_{jk}$, which can be solved exactly in the same way presented in the proof of Theorem \ref{thm1} (indeed the Christoffel symbols $\Gamma_{jk}^{(1)i}$ are expressed in the same way in terms of the functions $F_1, \dots F_6$ at this stage of the proof). 

Now we introduce the third connection $\nabla_{(3)}$ and we impose that it is almost hydrodynamically equivalent to $\nabla_{(1)}$ (and consequently to $\nabla_{(2)}$) and that $\nabla_{(3)} E_3=0$, where $E_{3}:=E^2=E\circ E$. In the David-Hertling coordinates $E^2$ has components $((u^1)^2, 2u^2u^1, 2u^3u^1+(u^2)^2)$. This again is enough to determine uniquely all the Christoffel symbols of $\nabla_{(3)}$ 
in terms of the Christoffel symbols of $\nabla_{(1)}$. However, at this point of the proof the Christoffel symbols for $\nabla_{(1)}$ are given in terms of functions $F_1, \dots, F_6$ of $z=\frac{u^3}{u^2}$. Therefore if we impose the commutativity of $\rm{Lie}_{E^2}$ with $\nabla_{(3)}$ (coming as always from the flatness of $\nabla_{(3)}$) we obtain a very simple system of ODEs in the functions $F_1, \dots, F_6$
In this case, the system forces $F_1$ and $F_3$ to be constants, while the other equations can be easily integrated. 
The additional constants appearing in the integration process are determined in such a way that $\nabla_{(1)}$ is flat. In this way we get the formulas for $\Gamma_{jk}^{(1)i}$ appearing in the statement of the theorem. Once the constants are chosen in this way, $\nabla_{(2)}$ and $\nabla_{(3)}$ turn out to be automatically flat and moreover the compatibility of each connection with the corresponding product is also fulfilled, as a straightforward calculation readily shows. 

\begin{flushright}
$\Box$
\end{flushright}

\subsection{An example with infinitely many compatible flat structures}
With similar computations it is possible to add further connections and try to construct  $F$-manifolds with four or more compatible flat connections. 
A very remarkable phenomenon is the following: once a quadri-flat $F$-manifold has been constructed, no new conditions arise if one tries to equip it with further flat compatible connections. 
In other words, regular quadri-flat $F$-manifolds in dimension three with operator $L$ consisting of a single Jordan block are automatically  "infinitely"-flat $F$-manifolds.

\begin{thm}\label{thm2quadriflatregular}
The data
\begin{eqnarray*} 
c^k_{ij}&=&\delta^k_{i+j-1},\\ 
E_{(0)}&=&e=\partial_{u^1},\\ 
\label{powersE}
E_{(l+1)}&=&E^l=(u^1)^l\partial_{u^1}+lu^2(u^1)^{l-1}\partial_{u^2}+\left(lu^3(u^1)^{l-1}+\f{1}{2}(l^2-l)(u^2)^2(u^1)^{l-2}\right)\partial_{u^3},\\
\end{eqnarray*}
and
\begin{eqnarray*}
&&\Gamma_{11}^{(l+1)1}=-\frac{l}{u^1},\;\;\Gamma_{11}^{(l+1)2}=\frac{lu^2(la^2 +la+a+2)}{(a+2)(u^1)^2}\\
&&\Gamma_{11}^{(l+1)3}=\frac{l((2la^2+2la+a+2)u^1u^3-(la^2+2la+a+2)(u^2)^2+(lab+2lb)u^1u^2
)}{(a+2)(u^1)^3}\\
&&\Gamma_{12}^{(l+1)1}=\Gamma_{21}^{(l+1)1}=0,\;\;\Gamma_{12}^{(l+1)2}=\Gamma_{21}^{(l+1)2}=-\frac{l(a^2+2a+2)}{(u^1)(a+2)},\;\;\Gamma_{23}^{(l+1)3}=\Gamma_{32}^{(l+1)3}=\f{a}{u^2}\\
&&\Gamma_{12}^{(l+1)3}=\Gamma_{21}^{(l+1)3}=\frac{l((la^2+a^2+2la+4a+4)(u^2)^2-2a^2u^1u^3
-(2ab+4b)u^1u^2)}{2u^2(a+2)(u^1)^2},\\
&&\Gamma_{13}^{(l+1)1}=\Gamma_{31}^{(l+1)1}=\Gamma_{13}^{(l+1)2}=\Gamma_{31}^{(l+1)2}=\Gamma_{22}^{(l+1)1}=0,\;\;\Gamma_{13}^{(l+1)3}=\Gamma_{31}^{(l+1)3}=-\f{l(a+1)}{u^1},\\
&&\Gamma_{22}^{(l+1)3}=-\frac{((la^2+3la+2l)(u^2)^2-(ab-2b)u^1u^2+2au^1u^3)}{(a+2)u^1(u^2)^2},\;\;\Gamma_{22}^{(l+1)2}=\f{a(a+1)}{u^2(a+2)}\\
&&\Gamma_{23}^{(l+1)1}=\Gamma_{32}^{(l+1)1}=\Gamma_{23}^{(l+1)2}=\Gamma_{32}^{(l+1)2}= \Gamma_{33}^{(l+1)1}=\Gamma_{33}^{(l+1)2}=\Gamma_{33}^{(l+1)3}= 0,
\end{eqnarray*}
locally define  a regular three dimensional  multi-flat $F$-manifold $(M, \nabla_{(l)}, \circ_l, E_{(l)},\,l=1,2...)$ for any value of the constants $a$ and $b$.
\end{thm}

\emph{ Proof}. The proof develops along the same lines of Theorem \ref{thmtriflatregular}, so here we just highlight the main differences. The first steps, including the determination of $\Gamma_{jk}^{(3)i}$ in terms of $\Gamma_{jk}^{(1)i}$ are the same. Imposing as in the proof of Theorem \ref{thmtriflatregular} the commutativity of $\rm{Lie}_{E_3}$ with $\nabla_{(3)}$ we obtain a simple system of ODEs in $F_i$, $i=1,\dots,6$ from which we deduce that $F_1$ and $F_3$ have to be constants. The ODEs are integrated, but this time the constants are left free at this stage of the proof. Instead we introduce the fourth connection $\nabla_{(4)}$ and as usual we impose it is almost hydrodynamically equivalent to $\nabla_{(1)}$ and that $\nabla_{(4)} E_4=0$. This is enough to express $\nabla_{(4)}$ in terms of $\nabla_{(1)}$. Furthermore we impose the commutativity of $\rm{Lie}_{E_4}$ with $\nabla_{(4)}$ which forces $f_1=f_3=0$. Some of the remaining constants appearing from the integration 
of the system of ODEs are fixed imposing that $\nabla_{(l)}$, $l=1,2,3,4$ are flat. Once this is done, the compatibility of each connection with the corresponding product is automatically satisfied and can be checked via a straightforward computation. At this point proceeding in a similar way one can construct one connection for each power of the Euler vector field (it is easy to prove by induction that the components of these vector fields are given by the formula
 \eqref{powersE}) without obtaining additional constraints.

\begin{flushright}
$\Box$
\end{flushright}

\begin{rmk}
Instead of considering the special eventual identities given by powers of the Euler vector field one can try to repeat the
 above construction considering arbitrary eventual identities. It is easy to check that these are given by
$$G_1(u^1)\partial_{u^1}+G_2(u^1, u^2)\partial_{u^2}+\left(-u^3G'_1+2u^3\f{\d G_2}{\d u^2}+G_3(u^1, u^2)\right)\partial_{u^3}$$
where $G_1(u^1),G_2(u^1, u^2),G_3(u^1, u^2)$ are arbitrary functions. It turns out (after long computations) that
 the previous construction works only for the subset of the eventual identities corresponding to the choice
\begin{eqnarray*}
&&G_1(u^1)=f(u^1),\,\,G_2(u^1, u^2)=f'u^2,\,\,G_3(u^1, u^2)=\f{(u^2)^2f''}{2}
\end{eqnarray*}
where $f$ is an arbitrary function of $u^1$. In particular, the powers of $E$ are obtained by setting $f=(u^1)^l$.  For arbitrary $f$ the formulas for the associated Christoffel symbols are
\begin{eqnarray*}
&&\Gamma_{11}^{(l+1)1}=-\frac{f'}{f},\;\;\Gamma_{11}^{(l+1)2}=-\frac{u^2((-a^2-2a-2)(f')^2+(a+2)ff'')}{f^2(a+2)}\\
&&\Gamma_{11}^{(l+1)3}=\frac{(2a^2+6a+4)(u^2)^2(f')^3+(-4a^2u^3-2abu^2-6au^3-4bu^2-4u^3)f(f')^2}{2f^3(a+2)}+\\
&&\qquad\qquad\frac{
(-2a^2-7a-6)(u^2)^2ff'f''+(2a+4)u^3f^2f''+(a+2)(u^2)^2f^2f''')}{2f^3(a+2)}\\
&&\Gamma_{12}^{(l+1)1}=\Gamma_{21}^{(l+1)1}=0,\;\;\Gamma_{12}^{(l+1)2}=\Gamma_{21}^{(l+1)2}=-\f{(a^2+2a+2)f'}{(a+2)f},\;\;\Gamma_{23}^{(l+1)3}=\Gamma_{32}^{(l+1)3}=\f{a}{u^2}\\
&&\Gamma_{12}^{(l+1)3}=\Gamma_{21}^{(l+1)3}=\frac{(a+2)^2(u^2)^2ff''-(2a^2+6a+4)(u^2)^2(f')^2+(2a^2u^3+2abu^2+4bu^2)ff'}{2u^2(a+2)f^2},\\
&&\Gamma_{13}^{(l+1)1}=\Gamma_{31}^{(l+1)1}=\Gamma_{13}^{(l+1)2}=\Gamma_{31}^{(l+1)2}=\Gamma_{22}^{(l+1)1}=0,\;\;\Gamma_{13}^{(l+1)3}=\Gamma_{31}^{(l+1)3}=-\f{(a+1)f'}{f},\\
&&\Gamma_{22}^{(l+1)3}=\f{(-a^2-3a-2)(u^2)^2f'+(abu^2-2au^3+2bu^2)f}{(a+2)(u^2)^2f},\;\;\Gamma_{22}^{(l+1)2}=\f{a(a+1)}{u^2(a+2)}\\
&&\Gamma_{23}^{(l+1)1}=\Gamma_{32}^{(l+1)1}=\Gamma_{23}^{(l+1)2}=\Gamma_{32}^{(l+1)2}= \Gamma_{33}^{(l+1)1}=\Gamma_{33}^{(l+1)2}=\Gamma_{33}^{(l+1)3}= 0.
\end{eqnarray*}
\end{rmk}                  

\section{Appendix 1}
Let us consider the system of first order partial differential equations
\begin{eqnarray}
\label{bif1}
&&\d_k\Gamma^i_{ij}=-\Gamma^i_{ij}\Gamma^i_{ik}+\Gamma^i_{ij}\Gamma^j_{jk}
+\Gamma^i_{ik}\Gamma^k_{kj}, \quad i\ne k\ne j\ne i,\\
\label{bif2}
&&e(\Gamma^i_{ij})=0,\qquad i\ne j\\
\label{bif3}
&&E(\Gamma^i_{ij})=-\Gamma^i_{ij},\qquad i\ne  j
\end{eqnarray}
for the $n(n-1)$ unknown functions $\Gamma^i_{ij}$ ($i\ne j$). In this Appendix we will prove the following theorem:
\begin{thm}
The system (\ref{bif1},\ref{bif2},\ref{bif3}) is complete, that is all the compatibility conditions 
$$
\d_l\d_k\Gamma^i_{ij}-\d_k\d_l\Gamma^i_{ij}=0, \qquad\forall k,l=1,...,n.
$$
are satisfied.
\end{thm}
\proof
First of all  it is easy to check that 
\beq\label{comp1}
\d_l\d_k\Gamma^i_{ij}-\d_k\d_l\Gamma^i_{ij}=0
\eeq
for distinct indices $i,j,k,l$.
Indeed, expanding the left hand side of \eqref{comp1} one gets
$$(\Gamma^i_{il}\Gamma^l_{lk}+\Gamma^i_{ik}\Gamma^k_{kl}-\Gamma^i_{ik}\Gamma^i_{il})\Gamma^k_{kj}+(\Gamma^k_{kl}\Gamma^l_{lj}+\Gamma^k_{kj}\Gamma^j_{jl}-\Gamma^k_{kj}\Gamma^k_{kl})\Gamma^i_{ik}$$
$$+(\Gamma^i_{il}\Gamma^l_{lj}+\Gamma^i_{ij}\Gamma^j_{jl}-\Gamma^i_{ij}\Gamma^i_{il})\Gamma^j_{jk}+(\Gamma^j_{jl}\Gamma^l_{lk}+\Gamma^j_{jk}\Gamma^k_{kl}-\Gamma^j_{jk}\Gamma^j_{jl})\Gamma^i_{ij}$$
$$-(\Gamma^i_{il}\Gamma^l_{lj}+\Gamma^i_{ij}\Gamma^j_{jl}-\Gamma^i_{ij}\Gamma^i_{il})\Gamma^i_{ik}-(\Gamma^i_{il}\Gamma^l_{lk}+\Gamma^i_{ik}\Gamma^k_{kl}-\Gamma^i_{ik}\Gamma^i_{il})\Gamma^i_{ij}$$
$$-(\Gamma^i_{ik}\Gamma^k_{kl}+\Gamma^i_{il}\Gamma^l_{lk}-\Gamma^i_{il}\Gamma^i_{ik})\Gamma^l_{lj}-(\Gamma^l_{lk}\Gamma^k_{kj}+\Gamma^l_{lj}\Gamma^j_{jk}-\Gamma^l_{lj}\Gamma^l_{lk})\Gamma^i_{il}$$
$$-(\Gamma^i_{ik}\Gamma^k_{kj}+\Gamma^i_{ij}\Gamma^j_{jk}-\Gamma^i_{ij}\Gamma^i_{ik})\Gamma^j_{jl}-(\Gamma^j_{jk}\Gamma^k_{kl}+\Gamma^j_{jl}\Gamma^l_{lk}-\Gamma^j_{jl}\Gamma^j_{jk})\Gamma^i_{ij}$$
$$+(\Gamma^i_{ik}\Gamma^k_{kj}+\Gamma^i_{ij}\Gamma^j_{jk}-\Gamma^i_{ij}\Gamma^i_{ik})\Gamma^i_{il}+(\Gamma^i_{ik}\Gamma^k_{kl}+\Gamma^i_{il}\Gamma^l_{lk}-\Gamma^i_{il}\Gamma^i_{ik})\Gamma^i_{ij}=0.$$
In order to prove that 
\begin{eqnarray*}
&&\d_i\d_k\Gamma^i_{ij}-\d_k\d_i\Gamma^i_{ij}=0\\
&&\d_j\d_k\Gamma^i_{ij}-\d_k\d_j\Gamma^i_{ij}=0
\end{eqnarray*}
for $k\ne i,j$ we observe that from \eqref{bif2} and \eqref{bif3} it follows that
\begin{eqnarray*}
\d_i\Gamma^i_{ij}&=&\frac{1}{u^j-u^i}\left(\sum_{l\ne i,j}(u^l-u^j)\d_l\Gamma^i_{ij}+\Gamma^i_{ij}\right),\\
\d_j\Gamma^i_{ij}&=&\frac{1}{u^j-u^i}\left(\sum_{l\ne i,j}(u^i-u^l)\d_l\Gamma^i_{ij}-\Gamma^i_{ij}\right).\\
\end{eqnarray*}
Using the above identities and writing $(u^i-u^j)\d_i$ and $(u^i-u^j)\d_j$ as
\begin{eqnarray*}
(u^i-u^j)\d_i&=&E-u^je-\sum_{l\ne i}(u^l-u^j)\d_l\\
(u^i-u^j)\d_j&=&-E+u^ie-\sum_{l\ne j}(u^i-u^l)\d_l
\end{eqnarray*}
respectively, we obtain
\begin{eqnarray*}
(\d_k\d_i-\d_i\d_k)\Gamma^i_{ij}&=&\frac{1}{u^j-u^i}\left(\sum_{l\ne i,j}(u^l-u^j)\d_k\d_l\Gamma^i_{ij}+2\d_k\Gamma^i_{ij}\right)-\d_i\d_k\Gamma^i_{ij}=\\
&&\frac{1}{u^j-u^i}\left(\sum_{l\ne i,j}(u^l-u^j)\d_k\d_l\Gamma^i_{ij}+2\d_k\Gamma^i_{ij}+(u^i-u^j)\d_i\d_k\Gamma^i_{ij}\right)=\\
&&\frac{1}{u^j-u^i}\left[E(\d_k\Gamma^i_{ij})-u^je(\d_k\Gamma^i_{ij})+2\d_k\Gamma^i_{ij}\right],\\
(\d_k\d_j-\d_j\d_k)\Gamma^i_{ij}&=&\frac{1}{u^j-u^i}\left(\sum_{l\ne i,j}(u^i-u^l)\d_k\d_l\Gamma^i_{ij}-2\d_k\Gamma^i_{ij}\right)-\d_j\d_k\Gamma^i_{ij}=\\
&&\frac{1}{u^j-u^i}\left(\sum_{l\ne i,j}(u^i-u^l)\d_k\d_l\Gamma^i_{ij}-2\d_k\Gamma^i_{ij}+(u^i-u^j)\d_j\d_k\Gamma^i_{ij}\right)\\
&&\frac{1}{u^j-u^i}\left[-E(\d_k\Gamma^i_{ij})+u^ie(\d_k\Gamma^i_{ij})-2\d_k\Gamma^i_{ij}\right].
\end{eqnarray*}
where we have used  the identity \eqref{comp1}. The result follow from the identities
\begin{eqnarray*}
E(\d_k\Gamma^i_{ij})&=&E(-\Gamma^i_{ij}\Gamma^i_{ik}+\Gamma^i_{ij}\Gamma^j_{jk}+\Gamma^i_{ik}\Gamma^k_{kj})\\
%&=&-E(\Gamma^i_{ij})\Gamma^i_{ik}-\Gamma^i_{ij}E(\Gamma^i_{ik})+E(\Gamma^i_{ij})\Gamma^j_{jk}+\Gamma^i_{ij}E(\Gamma^j_{jk})
%+E(\Gamma^i_{ik})\Gamma^k_{kj}+\Gamma^i_{ik}E(\Gamma^k_{kj})\\
&=&-2(-\Gamma^i_{ij}\Gamma^i_{ik}+\Gamma^i_{ij}\Gamma^j_{jk}+\Gamma^i_{ik}\Gamma^k_{kj})\\
&=&-2\d_k\Gamma^i_{ij}
\end{eqnarray*}
and
\beq\label{edgamma}
e(\d_k\Gamma^i_{ij})=e(-\Gamma^i_{ij}\Gamma^i_{ik}+\Gamma^i_{ij}\Gamma^j_{jk}+\Gamma^i_{ik}\Gamma^k_{kj})=0.
\eeq
To conclude we have to prove that
$$\d_i\d_j\Gamma^i_{ij}-\d_j\d_i\Gamma^i_{ij}=0.$$
Writing $\d_i$ as 
$$\d_i=e-\sum_{l\ne i}\d_l$$ 
and using \eqref{comp1} and \eqref{bif2}  we obtain the equivalent condition
$$\d_i(e(\Gamma^i_{ij})-e(\d_i\Gamma^i_{ij})=-e(\d_i\Gamma^i_{ij})=0.$$
Taking into account that  $e(u^j-u^i)=e(u^l-u^j)=0$ we obtain 
$$-e(\d_i\Gamma^i_{ij})=-\frac{1}{u^j-u^i}\left(\sum_{l\ne i,j}(u^l-u^j)e(\d_l\Gamma^i_{ij})+e(\Gamma^i_{ij})\right).$$
The result follows from the identities \eqref{edgamma} and \eqref{bif2}. 
\endproof

\section{Appendix 2}
In this Appendix we show how to reconstruct a solution of the system \eqref{mainsys} starting from a solution of  Painlev\'e VI equation. More precisely, we will construct solutions of the system 
\begin{equation}\label{mainsysA}
\begin{split}
\f{dF_{12}}{dz}&=-\f{(F_{12}(z)F_{23}(z)-F_{12}(z)F_{13}(z))z-F_{12}(z)F_{23}(z)+F_{32}(z)F_{13}(z)}{z(z-1)},\\
\f{dF_{21}}{dz}&=\f{(F_{21}(z)F_{23}(z)-F_{21}(z)F_{13}(z))z+F_{23}(z)F_{31}(z)-F_{23}(z)F_{21}(z)}{z(z-1)},\\
\f{dF_{13}}{dz}&=\f{(F_{12}(z)F_{23}(z)-F_{12}(z)F_{13}(z))z-F_{12}(z)F_{23}(z)+F_{32}(z)F_{13}(z)}{z(z-1)},\\
\f{dF_{31}}{dz}&=-\f{(-F_{31}(z)F_{12}(z)+F_{21}(z)F_{32}(z))z+F_{31}(z)F_{32}(z)-F_{21}(z)F_{32}(z)}{z(z-1)},\\
\f{dF_{23}}{dz}&=-\f{(F_{21}(z)F_{23}(z)-F_{21}(z)F_{13}(z))z+F_{23}(z)F_{31}(z)-F_{23}(z)F_{21}(z)}{z(z-1)},\\
\f{dF_{32}}{dz}&=\f{(-F_{31}(z)F_{12}(z)+F_{21}(z)F_{32}(z))z+F_{31}(z)F_{32}(z)-F_{21}(z)F_{32}(z)}{z(z-1)}
\end{split}
\end{equation}
 starting from solutions of
 the equation 
\begin{equation}\label{sigmaA}
\begin{split}
[z(z-1)f'']^2=&[q_2-(d_2-d_3)g_2-(d_1-d_3)g_1]^2-4f'g_1g_2.
\end{split}
\end{equation}
(where  $g_1=f-zf'+\f{q_1}{2}$ and $ g_2=(z-1)f'-f+\f{q_1}{2}$) which is related to Painlev\'e VI equation by the elementary transformation 
$$f=-\phi(z)-\f{1}{4}(d_{13}-d_{23})^2z+\f{1}{4}d_{13}(d_{13}-d_{23}).$$
%Suppose equation \eqref{sigmaA} is given, hence the parameters $q_1, q_2, d_{23}=d_2-d_3$, $d_{13}=d_1-d_3$ are known and a solution $f(z)$ is also known. 
%Once the values of $d_{23}$, $d_{13}$, $q_1$ and $q_2$ are known, one can get $d_{12}:=d_1-d_2=d_{13}-d_{23}$, and moreover, since by definition $I_5=-I_3I_4+I_1I_2I_3,$ the values of $d_1,d_2,d_3$ can be obtained from
%\beq\label{d_i}
%d_1=\frac{q_2}{d_2d_3}+\f{q_1}{d_2},\quad d_2=d_1-d_{12},\quad d_3=d_1-d_{13}.
%\eeq
%This implies that $d_1$ is a root of the cubic polynomia

Given a specific instance of equation \eqref{sigmaA} and a solution $f(z)$, 
 define $d_1$ as a root of the cubic polynomial
$$\lambda^3-(2d_{13}-d_{23})\lambda^2+(d_{13}^2-d_{13}d_{23}-q_1)\lambda+q_1d_{13}-q_2$$
and $d_2$ and $d_3$ as
$$d_2=d_1-d_{13}+d_{23},\quad d_3=d_1-d_{13}.$$
In this way the parameters $d_1,d_2,d_3,q_1,q_2$ satisfy the identity
$$q_2=-d_3q_1+d_1d_2d_3,$$
since that the values of the parameters are related to the values of the first integrals $I_i$ which are related by  a similar identity.

Notice that the constants $d_1,d_2,d_3,q_2$ are determined up to a sign, since the equation \eqref{sigmaA} is invariant under
the simultaneous substitution $$d_1\to -d_1,\quad d_2\to -d_2,\quad d_3\to -d_3,\quad q_2\to -q_2.$$ 
%Moreover, let us remark that if one performs this simultaneous change of sign, then the cubic polynomial becomes $\lambda^3+(d_{12}+d_{13})\lambda^2+(d_{12}d_{13}-q_1)-q_1d_{13}+q_2$ and the corresponding solution $d_1$ is mapped to $-d_1$ as it has to be. 
Therefore, once a root of the cubic polynomial above has been chosen, the only indetermination left is in the choice of simultaneous signs for $d_1, d_2, d_3, q_2.$
Given $d_1,d_2,d_3$, $q_1$ and $f(z)$ one can reconstruct the solution  $(F_{12}(z),F_{21}(z),F_{13}(z),F_{31}(z),F_{23}(z),F_{32}(z))$  of the system \eqref{mainsys} solving the algebraic system
\begin{eqnarray*}
&&F_{12}+F_{13}=\pm d_1,\quad F_{23}+F_{21}=\pm d_2,\quad F_{31}+F_{32}=\pm d_3,\\
&&F_{12}F_{21}=f',\qquad F_{23}F_{32}=g_1,\qquad F_{13}F_{31}=g_2.
\end{eqnarray*}
The solution is
\beq\label{F_ij}
\begin{split}
&F_{12} =\pm \f{\mu f'}{\mu d_2-g_1},\qquad F_{21} =\pm\left( d_2-\f{g_1}{\mu}\right),\qquad F_{13} =\pm\left( d_1- \f{\mu f'}{\mu d_2-g_1}\right)\\
& F_{31} =\pm\left( -\mu+d_3\right),\qquad F_{23} = \pm\f{g_1}{\mu},\qquad F_{32} =\pm\mu
\end{split}
\eeq
where $\mu$ satisfies
\beq\label{mu}
(f'-d_1d_2)\mu^2+(d_1d_2d_3+d_1g_1-d_2g_2-d_3f')\mu-d_1d_3g_1+g_1g_2=0.
\eeq
%($\mu$ has to satisfy such a constraint because for instance $F_{13}$ can be expressed in two different ways as $F_{13}=\frac{g_2}{F_{31}}=\frac{\pm g_2}{d_3-\mu}$ or as $F_{13}=\pm d_1-F_{12}=\pm (d_1-\frac{\mu f'}{\mu d_2-g_1})$ and equating the two expression one gets the polynomial above). 
Now by hypothesis the function $f$ is a solution of the equation
$$[z(z-1)f'']^2=[q_2-d_{23}g_2-d_{13}g_1]^2-4f'g_1g_2,$$
where $g_1=f-zf'+\f{q_1}{2}$ and $g_2=(z-1)f'-f+\f{q_1}{2}$. 
Defining the constants $d_1,d_2,d_3$ and the functions $F_{ij}$ as above we obtain
$$[z(z-1)f'']^2=[q_2-(d_2-d_3)F_{13}F_{31}-(d_1-d_3)F_{23}F_{32}]^2
-4F_{23}F_{31}F_{12}F_{13}F_{32}F_{21}.$$
Moreover, by construction we obtain the system that the functions $F_{ij}$ have to satisfy:
\beq\label{sysFij}\begin{split}
q_1=F_{31}F_{13}+F_{12}F_{21}+F_{23}F_{32},\qquad q_2=-d_3q_1+d_1d_2d_3,\\
d_1=\pm(F_{12}+F_{13}),\quad d_2=\pm(F_{21}+F_{23}),\quad d_3=\pm(F_{31}+F_{32}).
\end{split}
\eeq
Using these identities we obtain
$$[z(z-1)f'']^2=[F_{23}F_{31}F_{12}-F_{13}F_{32}F_{21}]^2.$$
Since the functions $F_{ij}$ are defined up to a sign, due to the form of system \eqref{sysFij}, in a neighborhood of a point $z_0\neq 0, 1$ such that $f''(z_0)\ne 0$ we can always choose the simultaneous sign of $F_{ij}$ in such a way  that the following relation holds:
$$f''=\f{F_{23}F_{31}F_{12}-F_{13}F_{32}F_{21}}{z(z-1)}.$$
In this way there is no freedom in the definition of the functions $F_{ij}$ even if we do not know {\it a priori} the right choice of the sign. Taking into account the definition of the functions $g_1$ and $g_2$ we obtain  the system 
\begin{eqnarray*}
&&(F_{12}F_{21})'=f'',\quad(F_{13}F_{31})'=(z-1)f'',\quad(F_{23}F_{32})'=-zf'',\\
&&(F_{12}+F_{13})'=0,\quad (F_{21}+F_{23})'=0,\quad (F_{31}+F_{32})'=0.
\end{eqnarray*}
It is easy to check that it is equivalent to the system \eqref{mainsysA}, just written in a different coordinate system provided that
 the jacobian determinant does not vanish. It is easy to check that this happens when $f''$ vanishes. This means that the case where $f$ is a linear function of $z$ must be treated separately. Given linear solutions of \eqref{sigmaA} the existence of corresponding
 solutions of   \eqref{mainsysA} is not automatically guaranteed. Moreover, it turns out that there are some exceptional linear solutions for which the polynomial \eqref{mu} vanishes identically. In this case families of solutions of the system \eqref{mainsysA} correspond
 to the same solution of \eqref{sigmaA}. For instance, the linear solution
 associated with tridimensional tri-flat $F$-manifolds, namely
$$f = -C_{12}C_{23}-C_{23}^2+C_{23}+zC_{12}C_{23}+\f{1}{2}(C_{12}^2+C_{12}C_{23}+C_{23}^2-C_{12}-C_{23}),$$
is related to the following one parameter family of solutions of the system \eqref{mainsysA}:
\begin{eqnarray*}
&&F_{12}=\f{C_{12}C_{23}}{F_{21}},\,F_{13}=-\f{C_{12}C_{23}-C_{12}F_{21}}{F_{21}},\,F_{23}=-F_{21}+C_{23}\\
&&F_{31}=\f{F_{21}(-1+C_{12}+C_{23})}{-F_{21}+C_{23}},\,F_{32}=-\f{C_{12}C_{23}+C_{23}^2-C_{23}}{-F_{21}+C_{23}}
\end{eqnarray*}
where
\begin{footnotesize}
\begin{eqnarray*}
&&F_{21}=\f{C(C_{12}-2)(C_{23}z+C_{12}-1){\rm hypergeom}([-C_{12}+1,C_{31}], [2-C_{12}],\f{z}{z-1})}{(z-1)\left(C(C_{12}-2){\rm hypergeom}([-C_{12}+1, C_{31}], [2-C_{12}],\f{z}{z-1})+\left(\f{z}{z-1}\right)^{C_{12}-1}\right)}+\\
&&\f{\left(C(C_{12}-1)(-C_{31}){\rm hypergeom}([2-C_{12}, 1+C_{31}], [3-C_{12}], \f{z}{z-1})+C_{23}\left(\f{z}{z-1}\right)^{C_{12}-1}(z-1)\right)z}{(z-1)^2\left(C(C_{12}-2){\rm hypergeom}([-C_{12}+1, C_{31}], [2-C_{12}],\f{z}{z-1})+\left(\f{z}{z-1}\right)^{C_{12}-1}\right)}
\end{eqnarray*}
\end{footnotesize}
with $C_{31}=1-C_{12}-C_{23}$.

\section*{Ackowledgements} 
We thank Davide Guzzetti for a useful discussion and Boris Dubrovin for very helpful comments. P.L.  is partially supported  by the Italian MIUR Research Project \emph{Teorie geometriche e analitiche dei sistemi Hamiltoniani in dimensioni finite e infinite} and by GNFM  Progetto Giovani 2014
 \emph{Aspetti geometrici e analitici dei sistemi integrabili}. The visit of A.A. at the University of Milano-Bicocca has been supported by GNFM. A.A. thanks the \emph{Dipartimento di Matematica e Applicazioni of Univesity of Milano-Bicocca} for the kind hospitality, while part of this work was done.

\end{document}